\documentclass[structabstract]{aa} 
\usepackage{graphicx}

\usepackage{txfonts}
\def\kms{\rm {km\,s$^{-1}$}}
\def\x{$\times$}
\def\etal{et~al.}
\def\cmsq{\rm {cm$^{-2}$}}
\def\cmcub{\rm {cm$^{-3}$}}
\def\vlsr{$\upsilon_\mathrm{LSR}$}
\begin{document}
\titlerunning {A spectral line survey of Orion KL in the bands 486\,--\,492 and 541\,--\,577 GHz with the Odin satellite} 
   \title{A spectral line survey of Orion KL in the bands 486-492 and 541-577 GHz with the Odin\thanks{ Odin is a Swedish-led satellite
   project
   funded jointly by the Swedish National Space Board (SNSB), the Canadian Space Agency (CSA), the National Technology Agency
   of
   Finland (Tekes) and Centre National d'Etudes Spatiales (CNES). The Swedish Space Corporation was the prime contractor and
   also is
   responsible for the satellite operation.} satellite}

   \subtitle{II. Data analysis }
\authorrunning{C.M.~Persson \etal}   
   \author{Carina M.~Persson
          \inst{1},
          A.O.H.~Olofsson \inst{1,2},
	  N.~Koning\inst{3},  P.~Bergman\inst{1,4}, P.~Bernath\inst{5, 6, 7}, J.H.~Black\inst{1},  U.~Frisk\inst{8},\\
	 W.~Geppert\inst{9}, 
	 T.I.~Hasegawa\inst{3,10}, \AA.~Hjalmarson\inst{1}, S.~Kwok\inst{3, 11},
	 B.~Larsson\inst{12}, A.~Lecacheux\inst{13}, A.~Nummelin\inst{14}, \\
	  M.~Olberg\inst{1}, Aa.~Sandqvist\inst{12}, \and E.S.~Wirstr\"om\inst{1}
          }

   \offprints{Carina Persson}

   \institute{Onsala Space Observatory (OSO), Chalmers University of Technology, SE-439 92 Onsala, Sweden,
              \email{carina@oso.chalmers.se}
	 \and LERMA, Observatoire de Paris, 61 Av. de l'Observatoire, 75014 Paris, France     
         \and Department of Physics and Astronomy, University of Calgary, Calgary, AB T2N 1N4, Canada
         \and European Southern Observatory, Alonso de Cordova 3107, Vitacura, Casilla 19001, Santiago, Chile
	 \and Department of Chemistry, University of Waterloo, Waterloo, ON N2L 3G1, Canada   
	 \and Department of Chemistry, University of Arizona, Tucson, AZ 85721, USA
	 \and Department of Chemistry, University of York, Heslington, York YO10 5DD, United Kingdom
	 \and Swedish Space Corporation, PO Box 4207, SE-171 04 Solna, Sweden
	 \and Molecular Physics Division, Department of Physics, Stockholm University AlbaNova, SE-10691 Stockholm, Sweden
	 \and Institute of Astronomy and Astrophysics, Academia Sinica, P.O.Box 23-141, Taipei 106, Taiwan, R.O.C.
	 \and Department of Physics, University of Hong Kong, Hong Kong, China
	 \and Stockholm Observatory, AlbaNova University Center, SE-10691 Stockholm, Sweden
	 \and LESIA, Observatoire de Paris, Section de Meudon, 5, Place Jules Janssen, 92195 MEUDON CEDEX, France
	 \and Computer science and engineering, Chalmers University of Technology,  SE-41296 G\"oteborg, Sweden
       }

\date{Received 2 February 2007 / accepted 17 September 2007}
 
  \abstract
   {}
   {We investigate the physical and chemical conditions in a typical star forming region, including   an unbiased search for new 
   molecules in 
   a spectral region previously unobserved.}
    { Due to its proximity, the Orion KL region offers a unique laboratory of  molecular astrophysics in a chemically rich, massive 
   star forming region. 
   Several 
   ground-based spectral line surveys have been made, 
   but due to the absorption by water and 
   oxygen, the terrestrial 
   atmosphere is completely opaque at frequencies around 487 and 557 \mbox{GHz}. 
    To cover these frequencies we   used the Odin satellite to perform a spectral line survey in the frequency ranges 486\,--\,492 
    GHz and 
    541\,--\,577 \mbox{GHz}, 
    filling the gaps between previous spectral scans.
      Odin's high main beam efficiency, $\eta_{\mathrm{mb}}$\,=\,0.9, and observations performed outside the 
       atmosphere 
       make our intensity scale very well determined.
    }
   {We observed  280 spectral lines from 38 molecules including isotopologues,  and, in addition, 64 unidentified 
    lines. A 
   few U-lines have interesting frequency coincidences
such as ND and the anion SH$^-$. The 
   beam-averaged emission is dominated by CO, H$_2$O, SO$_2$, SO, $^{13}$CO and CH$_3$OH. Species with the largest number of 
   lines are CH$_3$OH, (CH$_3)_2$O, SO$_2$, $^{13}$CH$_3$OH, CH$_3$CN and NO. 
   Six water lines are detected including
       the ground state rotational transition 1$_{1,0}$\,--\,1$_{0,1}$  of $o$-H$_2$O, its isotopologues 
       $o$-H$_2^{18}$O and $o$-H$_2^{17}$O,   the Hot Core tracing
       $p$-H$_2$O transition 
       6$_{2,4}$\,--\,7$_{1,7}$, and the 2$_{0, 2}$\,--\,1$_{1,1}$ transition of HDO. Other lines of special interest are
       the 1$_{0}$\,--\,0$_{\,0}$  transition of NH$_3$ and its isotopologue $^{15}$NH$_3$.
        Isotopologue abundance  ratios of D/H, $^{12}$C/$^{13}$C, $^{32}$S/$^{34}$S, $^{34}$S/$^{33}$S, 
	and $^{18}$O/$^{17}$O
        are estimated.
      The  temperatures, column densities and abundances in the various subregions are estimated, and we find very high 
      gas-phase
      abundances of H$_2$O, NH$_3$,  
      SO$_2$, SO, NO, and CH$_3$OH. A comparison with the ice inventory of ISO sheds new light on the origin of the abundant 
      gas-phase molecules.
       }
   {}

  
   \maketitle
%
\clearpage
\section {Introduction}

To study the important ground-state rotational transition of water  (including isotopologues), which traces shocks and heated star forming regions, is one of the main astronomy goals of the Odin satellite (Nordh \etal~\cite{Nordh} and subsequent papers in the A\&A ''Special Letters Edition: First Science with the Odin satellite'') and hence also of this spectral line survey towards the Orion KL region. The first  observations of this water line  were performed by SWAS in 1998 (NASA's Submillimeter Wave Astronomy Satellite; Melnick \etal~\cite{Melnick} and subsequent ApJ papers in that  issue). The Odin satellite 
provides a
smaller  beam than SWAS
(2.1 $\arcmin$ vs. 3.3$\arcmin\times4.5\arcmin$), and our tunable SSB receivers enable   a full  line survey in this spectral window, including the water isotopologues (H$_2^{16}$O,  H$_2^{17}$O, H$_2^{18}$O and HDO), and  a high energy $p$-H$_2$O transition.

A spectral scan offers an unbiased search for new molecules. 
It also creates opportunities to observe multiple transitions of the same species as a uniformly calibrated data set, and this can be used to calculate rotation  temperatures,  column densities, abundances, source sizes,  optical depths, and isotopic elemental abundance ratios of the observed gases. The latter are important constraints for models of the Galactic chemical evolution. These models predict the elemental abundance evolution as a function of star formation history, stellar nucleosynthesis, and the degree of mixing of the gas in the ISM (Wilson \& Rood \cite{Wilson and Rood}).

The Orion Molecular Cloud (OMC-1) is a  well known massive star forming region 
(see Genzel \& Stutzki \cite{Genzel and Stutzki} for a review), and an ideal target for spectral line surveys at millimetre and submillimetre wavelengths due to its chemical richness and proximity ($\sim$450 pc). 
The Kleinmann-Low nebula (Orion KL) is the brightest infrared region in the OMC-1   and is situated about 1$\arcmin$ NW of the Trapezium cluster.
This region enables
studies of the interaction between young massive stars and their parental molecular cloud.  Powerful outflows, shocks and turbulence cause 
a very complex and chemically structured  source, consisting of
several distinct subsources.

There are five different  components of radial velocity (e.g. Olofsson \etal~\cite{HansOlofsson81}; Olofsson \etal~\cite{HansOlofsson82}; Johansson \etal~\cite{Johansson84}; Friberg \cite{Friberg84}; Genzel \& Stutzki \cite{Genzel and Stutzki}; Wright \etal~\cite{Wright}; Schilke \etal~\cite{S01}; Beuther \etal~\cite{Beuther};   Olofsson \etal~\cite{Olofsson},  hereafter Paper~I)  within the $\sim$126$\arcsec$ Odin beam: 

\begin{itemize}
\item
\emph{The ambient medium/Extended Ridge} (ER) with  \mbox{\vlsr$\sim$8 kms$^{-1}$} in the south and an abrupt velocity shift across the  KL region to \vlsr$\sim$10 \kms~in the north. This extended emission is larger than our beam with quiescent, cool gas of narrow
line widths of  \mbox{$\Delta \upsilon\sim$3\,--\,5 \kms,} a temperature of $\sim$20\,--\,60 K, and densities of 10$^4$\,--\,$10^6$ \cmcub.
\item
\emph{The Plateau}: the out-flowing gas, centred close to IRc2 contains two  outflows (Greenhill \etal~\cite{Greenhill98}). The bipolar \emph{High Velocity Flow} (HVF) in  the SE-NW direction at \vlsr$\sim$10 \kms~reaches velocities of  150 \kms~and covers 40\,--\,70$\arcsec$. The second is a \emph{Low Velocity Flow} (LVF) in the SW-NE direction at \vlsr$\sim$5 \kms, widths of   \mbox{$\sim$18 \kms}~\mbox{(''the 18 \kms~flow'', Genzel \etal~\cite{Genzel81})}, and a size of 15\,--\,30$\arcsec$.    
The temperature and density are 100\,--\,150 K and $\sim$10$^5$ \cmcub, respectively.
\item
\emph{The Compact Ridge} (CR): a compact warm clump in the northern tip of the southern ER was first discovered by Johansson \etal~(\cite{Johansson84}), approximately 10\,--\,15$\arcsec$ south-west of IRc2 with \vlsr$\sim$8 \kms, and line widths of \mbox{$\Delta \upsilon\sim$3  \kms.}
It may be the result of an interaction between the LVF and the ER that compressed the gas to higher densities \mbox{$\sim$10$^6$ \cmcub}, temperatures of 100\,--\,150 K, and to a small size of 6\,--\,15$\arcsec$.
\item
\emph{The Hot Core} (HC): a warm  star forming region which  is heated internally, probably by one (or more) young massive protostars. The total size is  $\sim$5\,--\,10 $\arcsec$ (Hermsen \etal~\cite{Hermsen}, Wilson \etal~\cite{Wilson}) with smaller, very dense ($n\sim$10$^7$ \cmcub) clumps  (Beuther \etal~\cite{Beuther}).
It is centred only 2$\arcsec$ from IRc2, at a projected distance of  10$\arcsec$ from the CR. 
The velocity is  centred on \vlsr$\sim$3\,--\,6 \kms~with line widths of $\Delta \upsilon\sim$5\,--\,15 \kms. 
The range of temperatures obtained from inversion transitions of NH$_3$ is 165\,--\,400 K (Wilson \etal~\cite{Wilson}).
\item
\emph{Photo Dissociation Region} (PDR): the extended interface region between the molecular cloud and the foreground M42 H{\sc ii} region (Rodr\'iguez-Franco \etal~\cite{Rodriguez-Franco}, \cite{Rodriguez-Franco01}; Wirstr\"om \etal~\cite{Wirstrom}, and references therein) at velocities 8\,--\,10 \kms.
\end{itemize}
The various cloud components have been displayed  in \mbox{Fig.~6}~of Genzel \& Stutzki  (\cite{Genzel and Stutzki}), and \mbox{Fig. 7} of Irvine \etal~(\cite{Irvine87}). \mbox{Figure 1} of Greenhill \etal~(\cite{Greenhill98}) shows a model of the bipolar High Velocity Flow and the Low Velocity Flow.

The Odin satellite has
a large beam and covers high frequencies. This gives our survey the opportunity to simultaneously observe both the small, hot and dense regions, 
and the extended, cooler regions. 
Because of the complex  source structure  encompassed by large our antenna beam, we 
will compare our data with interferometric images for each species (see Paper~I for an extensive list of spectral line survey references). In this way the origin and source sizes of our detected species can be checked.

The complete submm spectrum observed by Odin together with the proposed identification of each line can be found in Paper~I. In the present paper we give a short description of our data in Sect.  \ref{The Line survey data}, and  of the different analysis methods in \mbox{Sect. \ref{section Theory}}.
In Sect.  \ref{results} we  present the results in \mbox{tables} and rotation diagrams together with spectra of typical or particularly important transitions. Tables of observed transitions
can be found as electronic Tables in the on-line material (Tables \ref{SO2 parameters }  to \ref{water parameters}).
Tables \ref{U-line parameters} and
\ref{T-line parameters} list our unidentified and tentatively identified lines.  \mbox{Sect. \ref{results}} also includes
a short analysis for each molecule. The important water  and CO lines are analysed in  \mbox{Sect. \ref{section CO} and \ref{section Water}}. 
An attempt to obtain molecular abundances in the different subregions of Orion KL and comparison with abundances in ice mantles of dust grains is found in \mbox{Sect. \ref{section Abundances}.}
We end this paper with a discussion of source 
sizes and source structure  in \mbox{Sect. \ref{section Discussion}}, followed by a short summary.


\section {The line survey data} \label{The Line survey data}

The observational method is presented in Paper I, and the data is analysed in this paper.
These data were obtained
with the Odin satellite from spring 2004 to autumn 2005 during four different runs. The spectral scan covers frequencies between \mbox{486\,--\,492} and \mbox{541\,--\,577} \mbox{GHz} and includes 280 spectral features from 38 species including isotopologues. The lines were identified using the Lovas SLAIM$\emptyset$3 molecular line catalogue\footnote{Not available on-line, but some of its content is maintained under http://physics.nist.gov/PhysRefData/}  (Lovas \cite{Lovas}), the Cologne Database for Molecular Spectroscopy\footnote{ http://www.cdms.de} (CDMS, M\"uller \etal~\cite{Muller}) and the Jet Propulsion Laboratory\footnote{http://spec.jpl.nasa.gov/} database (JPL, Pickett \etal~\cite{Pickett}).
Identifications are based not only on frequency coincidence, but also  expected abundance, line strength, width and velocity, upper state energy, and the presence of other expected transitions of the molecule.

Most lines in our survey (205 out of 280 identified lines)  are due to CH$_3$OH,  $^{13}$CH$_3$OH, (CH$_3$)$_2$O, SO$_2$, and CH$_3$CN (Table~\ref{summary Table}).
A total of 64 lines (19$\%$ of all lines)  could not be uniquely identified, although from frequency coincidences we have suggestions for a few identifications such as ND,  the  interstellar anion SH$^-$, SO$^+$, HNCO and CH$_3$OCHO
(see \mbox{Sect. \ref{results}} and Paper I). 
The spectroscopy still is sparse at higher frequencies and a number of U-lines are likely to be poorly known transitions of the identified molecules and their isotopologues, including their vibrationally or torsionally excited states.

At 557 GHz the Odin \mbox{1.1 m} mirror has a circular beam with FWHM of 2$\farcm$1. The main beam efficiency is $\eta_{\mathrm{mb}}$\,=\,0.9.  This in addition to being outside the atmosphere makes our intensity  calibration very accurate. The intensity scale is expressed in terms of antenna temperature $T^*_\mathrm{A}$. In all calculations of the column densities the main beam efficiency is properly taken into account.
The reconstructed pointing uncertainty is $<$15$\arcsec$ during most of the time.
The coordinates of \mbox{Orion KL} in our survey are \mbox{R.A. 05$^h$35$^m$14$^s$.36,} \mbox{Dec. $-$05$^\circ$22$\arcmin$29$\farcs$6} (J2000), and the  frequency scale is set in relation to a source LSR velocity of +8 \kms. The spectral resolution is 1 MHz, and the typical  rms reached  is $\sim$25 mK per 1 MHz channel.

\section {Data analysis methods -- a simplified approach} \label{section Theory}

The observed line emission  is not restricted to one single subregion in Orion KL, but may be  a complicated blend from several subregions with a complex line profile. 
Thus, when we attempt to derive column densities and abundances, we have to separate the emission into its constituent parts. 
The most simple approach
whenever several emission features are clearly present, is to use least-square fits of Gaussians to the line profiles  to separate their relative contributions. This can give a first order input to modelling attempts  including current and future knowledge of the source structure.
The resulting parameters are found in the on-line Tables and in fitted spectra (Section \ref{results}).
This is based on the assumptions that either all emissions are optically thin or that the emission subregions do not overlap each other spatially, and also that the velocity distributions are Gaussian.

The formal errors  obtained from the rotation diagram method and forward model are given in each subsection.  The formal errors obtained from the single line analysis and from the Gaussian decomposition of the lines are mostly below 20\%, with weak lines having higher formal errors. 
We estimate the
accuracy of our column densities results to be within a factor of 2\,--\,3. The uncertainties in the derived abundances can be higher because  the adopted H$_2$ column density is also uncertain (see also Sect. \ref{section Abundances}).
Details, definitions and additional uncertainties of the methods  not discussed below are found in  the on-line Section  \ref{appendix theory}.

\subsection {Single line analysis}

With the assumption of optically thin emission, neglecting the background radiation, and assuming that the source fills the antenna main  beam,  the \emph{ beam averaged upper state} column density can be calculated as 

\begin{equation} \label{Nu}
N_{ {u}}^\mathrm{\,thin}  = \frac {8  \pi k \nu_{{ul}}^2}{h c^3} \frac{1}{ A_{ {ul}}}  \int {T_{\mathrm{mb}}}  \, \mathrm{d} \upsilon,
\end{equation}
where 
$k$ is the Boltzmann constant, $\nu_{{ul}}$ is the frequency of the transition, $h$ is the Planck constant, $c$ is the speed of light,  $A_{{ul}}$ is the Einstein $A$-coefficient for the transition, and $T_\mathrm{mb}$ is the main beam brightness temperature. As customary the frequency   $\nu$ has been converted to a Doppler
velocity   $\upsilon$.

For a Boltzmann distribution, and with corrections for opacity and beam-filling, the true \emph{ total source-averaged }column density can be calculated as

\begin{equation} \label{NtotCorrected}
N_{\mathrm{LTE}}  =\frac{C_\tau}{\eta_{\mathrm{bf}}}\, \frac {8  \pi k \nu_{{ul}}^2}{h c^3} \frac{1}{ A_{{ul}}} \frac{Q(T)}{g_{{u}}}  e^{E_{{u}}/kT_{\mathrm{ex}}} \int {T_{\mathrm{mb}}}\,d \upsilon, 
\end{equation}
where 
$C_\tau$ and $\eta_\mathrm{bf}$ are  the opacity and  beam-filling correction factors,  $Q(T)$ is the partition function, $g_u$ and $E_u$ are the statistical weight and energy of the upper state, respectively, and $T_\mathrm{ex}$ is the excitation temperature for the transition.

For molecules where one or few transitions are observed, the column density is calculated using Eq. (\ref{NtotCorrected}). If no information about optical depth  or source-size is available these corrections are not taken into account, thus producing a \emph{beam-averaged} and \emph{ not opacity corrected} column density.


\subsection{Multiple line analysis}

\subsubsection{The rotation diagram method}\label{section rotation diagram method}

When we have observed a number of lines with a wide range of upper-state energies, the rotation diagram method can be used  according to

\begin{equation}\label{full solution for rotation diagram}
\ln \, \frac {N_{u}^\mathrm{\,thin}}{g_{u}} = \ln \, \frac{N_{\mathrm{tot}}}{Q(T)} - \frac {E_{u}}{kT_\mathrm{ex}}.
\end{equation}


To create a rotation diagram we plot $\ln\,(N_{u}^\mathrm{\,thin}/g_{u}$) as a function of the upper state energy $E_{u}$ in a semi-log plot.  A  least squares fit to the data will then produce a straight line with slope $-1/T_{\mathrm{ROT}}$. If we extrapolate the line to $E_{u}$ = 0 K, we obtain the total column density from the intersection  of the y-axis, $y_0$, and derive the total column density as
\begin{equation}\label{Ntotal}
N_{\mathrm{ROT}} = Q(T) \, e^{\, y_0}.
\end{equation}

To correct for beam-filling, the right hand side of  Eq. (\ref{Ntotal})  is multiplied by 1/$\eta_{\mathrm{bf}}$.
Note that this constant does not change
the rotation temperature.
However,
the optical depths  can change the slope, and therefore the rotation temperature estimated from the rotation diagram.

The error bars shown in our rotation diagrams (Section \ref {results}) include 10$\%$ calibration  error and the observed rms-noise. 

\subsubsection{The forward model ($\chi^2$-method)}

This model 
matches calculated opacity-corrected LTE integrated intensities  and beam-filling in a $\chi^2$ sense, to observed intensities vs. $E_u$ (Nummelin \etal~\cite{Nummelin98}; Nummelin \etal~\cite{Nummelin2000}; Lampton \etal~\cite{Lampton}; Bevington \cite{Bevington}).  Equation (\ref{full solution for rotation diagram})  in this case is modified to include optical depth and beam-filling corrections

\begin{equation}\label{forward model equation}
\ln \, \frac {N_{u}}{g_{u}} + \ln \, C_\tau + \ln \frac{1}{\eta_{\mathrm{bf}}}= \ln \, \frac{N_{\mathrm{tot}}}{Q(T)} - \frac {E_{u}}{kT_\mathrm{ex}}.
\end{equation}

The intensity of each transition can be calculated using Eq.~(\ref{solution}), with a specific set of free parameters $\eta_{\mathrm{bf}}$, $T_{\mathrm{ROT}}$ and $N_\mathrm{tot}$. The best fit to all the data, is obtained by finding the minimum of the reduced $\chi^2$, which is defined as

\begin{equation}\label{chi squared}
\chi^2= \frac{1}{n-p} \sum_{i=1}^{n} \left(\frac{I_i^{\mathrm{obs}}-I_i^{\mathrm{calc}}}{\sigma_i^{\mathrm{\,obs}}}\right)^2.
\end{equation}

Here $n$ is the number of data points, $p$ is the number of free parameters, $\sigma_i^\mathrm{\,obs}$ is the 1$\sigma$ uncertainty of the observed line intensity, $I_i^{\mathrm{obs}}$ and $I_i^{\mathrm{calc}}$  are the observed and calculated integrated intensities.

Note that the column density obtained with this method will be somewhat lower than that calculated from a simple rotation diagram, since solutions producing $I_i^{\mathrm{calc}}<I_i^{\mathrm{obs}}$ are favoured compared to the opposite. This can be seen from the contribution to the $\chi^2$-value from each transition giving $\chi^2_i$. 
\begin{equation}\label{chi squared limit}
\chi^2_i=  \left(\frac{I_i^{\mathrm{obs}}-I_i^{\mathrm{calc}}}{\sigma_i^{\mathrm{\,obs}}}\right)^2  \rightarrow \left(\frac{I_i^{\mathrm{obs}}}{\sigma_i^{\mathrm{\,obs}}}\right)^2 
\end{equation}
when $I_i^{\mathrm{calc}}\rightarrow0$. But if $I_i^{\mathrm{calc}}>I_i^{\mathrm{obs}}$,  $\chi^2$ will be unlimited, and this favours the lower  model intensities  (Nummelin \etal~\cite{Nummelin98}).

\begin{table*} 
\caption{Resulting column densities and rotation temperatures as well as comparison with W03, S01 and C05.} 
\label{result_table1}
\begin{tabular} { l l l l l l l l l l l}
\hline
\hline
 & & & & & \multicolumn{2}{c}{W03$^{{b,\,c}}$} & \multicolumn{2}{c}{S01$^{{b,\,c}}$} & \multicolumn{2}{c}{C05$^{{d}}$} \\
Species (No.)&Region    & $N$   & $T_{\mathrm{ex}}$&     Size$^a$ & $N$  & $T_{\mathrm{ex}}$  & $N$ &$T_{\mathrm{ex}}$ & $N$&$T_{\mathrm{ex}}$ \\
&    & [cm$^{-2}$]  & [K]      &  [\arcsec]& [cm$^{-2}$]  & [K]  &   [cm$^{-2}$]  & [K]   &   [cm$^{-2}$]  & [K]       \\

\hline
SO$_{2}$	 (31)& Total$^c$	 & 1.5\x$10^{18}$ $^{{e}}$&   103$^{{e}}$ &  8$^{\,f}$   & 1\x$10^{17}$ & 136  & 6\x$10^{16}$ & 187 & 9\x$10^{16}$ & 130\\ 
   &  CR$^g$&2.0\x$10^{17}$  $^{{h}}$ & (115)$^i$      &5$^{\,f}$     &   &    &  &   &  &  \\  
 &  LVF$^g$ &6.0\x$10^{17}$  $^{{h}}$  & 103$^{{j}}$     & 8$^{\,f}$    &   &    &  &   &  &  \\  
&  HVF$^g$&9.0\x$10^{17}$  $^{{h}}$  & 103$^{{j}}$&        8$^{\,f}$   &   &    &  &   &  &  \\  

$^{34}$SO$_{2}$ (4)   	&Total$^c$ (outflow)&   6.5\x$10^{16}$ $^{\,k}$& 103$^{{j}}$ &    8$^{\,l}$& 8\x$10^{15}$ &  156   & 8\x$10^{15}$ &  192& &\\

SO (5)	  		&Total$^c$		& 1.6\x$10^{17}$ $^{{e}}$&  132$^{e}$& 18$^f$& 3\x$10^{17}$ & (72)$^{{i}}$ & 2\x$10^{17}$ & 64 &  5\x$10^{16}$ &150 \\ 
&CR$^g$ 		&  1.7\x$10^{16}$ $^{{h}}$   & (115)$^{i}$ &  6$^{{f}}$ &   &    &  &      & & \\ 
& LVF$^g$   	&   9.3\x$10^{16}$ $^{{h}}$   & 132$^{{\,m}}$ &  11$^{{n}}$ &   &    &  &      & & \\ 
 &HVF$^g$   	 	& 8.5\x$10^{16}$ $^{{h}}$    & (100)$^{i}$  &  18$^{{n}}$&   &    &  &      & & \\  
$^{33}$SO (3)	&  Total$^c$ (outflow)	& 1.7\x$10^{15}$ $^{{o}}$&	132$^m$	   &18$^{p}$ & & & & & &\\ 
$^{34}$SO (2)&	Total$^c$ (outflow) &8.4\x$10^{15}$  $^{{o}}$  &	132$^m$	&    18$^{p}$&1\x$10^{16}$ &  89  & & & &\\ 

SiO (1)  		&	Total$^c$	&4.0\x$10^{15}$ $^{{q}}$	 &	(100)$^{i}$  & 	14$^f$	&&  & 5\x$10^{14}$ & 110 & 1\x$10^{15}$ & (150)$^i$\\ 
&	 	LVF$^g$				& 3.3\x$10^{15}$ $^{{h}}$	 &	 (100)$^{i}$	 & 	 10$^{\,n}$	&&  &  &   &    &  \\ 
&	 	HVF$^g$				&1.8\x$10^{15}$ $^{{h}}$ 	 &	(100)$^{i}$	  & 	14$^r$ 	&&  & &  &    &  \\ 
$^{29}$SiO (1)	&	Total$^c$  (outflow) & 2.0\x$10^{14}$  $^{{o}}$ 	         &	 (100)$^{i}$    &	14$^{{r}}$	&&  & & & &\\ 

H$_2$S   (1) &	 	LVF$^g$		&  4.5\x$10^{16}$ $^{{o}}$	 &	 (100)$^{i}$	&	(15)$^{i}$	&&  & 1\x$10^{16}$ &  129 & 4\x$10^{16}$    & 140  \\ 
&	 	HC$^g$			&  2.7\x$10^{16}$	  $^{{o}}$&	(200)$^{i}$	&	 	(10)$^{i}$	&&  &  &   &  &   \\ 

CH$_{3}$CN (9) &	Total$^c$ (HC)	  & 	5.0\x$10^{15}$  $^{{k}}$&  137$^{{k}}$ &	(10)$^{i}$	& 3\x$10^{15}$ &  227  & & & 4\x$10^{15}$ & 250\\ 

HC$_{3}$N (2)	&	Total$^c$ (HC)	   	& 1.8\x$10^{15}$	 $^{{o}}$ &	(200)$^{i}$	 &	(10)$^{i}$	 &  1\x$10^{15}$ &  164  & & & & \\ 

OCS	 (3)  		&	Total$^c$ (HC)	  	&  1.7\x$10^{16}$	$^{{o}}$	 &	(200)$^{i}$	&	 	(10)$^{i}$	 & 9\x$10^{16}$ &  106   & & & &\\  

NO  (1)  &Total$^c$ (HC)	&  2.8\x$10^{17\,\,o}$	&      (200)$^{i}$	 &     (10)$^{i}$  &&&  1\x$10^{17}$ &  90  &2\x$10^{17}$  & 150  \\ 

NH$_{3}$	 (1)	&	 	CR$^g$	& 4.0\x$10^{16\,\,h}$		& 	(115)$^{i}$	  & 17$^{{f}}$	&& & & & &\\ 
&	 	HC$^g$			& 1.6\x$10^{18\,\,q}$	& 	(200)$^{i}$	&	  	8$^{{f}}$	&& & & & &\\ 
$^{15}$NH$_{3}$ (1)	&	Total$^c$ (HC)	  	&3.5\x$10^{15\,\,o}$ 	&	(200)$^{i}$ 	 &	8$^{{s}}$&&  & & & &\\ 

CH$_{3}$OH   (50)  &	Total$^c$   	  &  3.4\x$10^{18\,\,e}$  &  116$^{{e}}$    & 6$^{\,f,\,t}$&  9\x$10^{16}$&  599 & 5\x$10^{16}$& 303 & &  \\ 
&	 	HC$^g$			  & 7.9\x$10^{17\,\,u}$  &  178$^u$ &  6$^{\,n}$ &  &   & & &  3\x$10^{16}$&  220\\ 
&	 	CR$^g$		  &2.4\x$10^{18\,\,u}$  &  98$^u$&  6$^{\,f}$ &  &   & & & 5\x$10^{16}$ & 160 \\ 
$^{13}$CH$_{3}$OH    (14)	 &	Total$^c$ (CR)		& 5.9\x$10^{16\,\,k}$ &   115$^{\,k}$  &    6$^{\,v}$&&  & 1\x$10^{16}$& 229 & &\\  

(CH$_{3}$)$_2$O  (37) &	Total$^c$ (CR)	 	& 1.3\x$10^{17\,\,k}$ &  112$^{k}$  &6$^{v}$&1\x$10^{16}$& 157  & 3\x$10^{16}$& 360&  2\x$10^{16}$&  160\\ 

H$_{2}$CS  (4)&	Total$^c$ (CR)	 	&	1.3\x$10^{15\,\,k}$ 	 &93$^{k}$ 	   &14$^{x}$&& & & & &\\ 

H$_{2}$CO   (3) &	 	LVF$^g$			& 4.3\x$10^{15\,\,o}$	&	 (100)$^{i}$	  & 	(15)$^{i}$	& 1\x$10^{16}$ & (166)$^{{i}}$ & 3\x$10^{15}$  &190&2\x$10^{15}$  & 155\\ 
&	 	CR$^g$		&2.0\x$10^{16\,\,q}$	&	 (115)$^{i}$	&  	14$^{f}$    	& & & & &  &   \\ 

H$_{2}^{13}$CO  (1)&	 	Total$^c$ (CR)		  &3.3\x$10^{14\,\,o}$ 	&	 (115)$^{i}$	 	&14$^{x}$  	&1\x$10^{15}$ & (166)$^{{i}}$  & & & & \\ 

HDCO (2)	  	&Total$^c$(CR)		& 2.7\x$10^{14\,\,o}$	 &	 (115)$^{i}$	 	&14$^{x}$  	&	&  & & & & \\

CS (1)  &	 	LVF$^g$		&	 3.6\x$10^{15\,\,o}$   & 	  (100)$^{i}$	&	 (15)$^{i}$	&    &  &1\x$10^{15}$  & 127  & 3\x$10^{15}$  & 100\\
&	 	HC$^g$			&2.9\x$10^{15\,\,o}$ 	     & 	(200)$^{i}$	&	 (10)$^{i}$	& &  & &   &  1\x$10^{16}$&190 \\
&N as CR$^g$			& 8.0\x$10^{15\,\,q}$ 	  & 	 (115)$^{i}$	 &	20$^{f}$	& &  & &   &  & \\
&N as ER$^g$				&4.2\x$10^{14\,\,q}$ 	  &  (60)$^{i}$	 &	...	& &  & &   &  & \\

$^{13}$CS  (1)&Total$^c$ as CR &1.3\x$10^{14\,\,o}$   &	 (115)$^{i}$	& 20$^{{y}}$ &  2\x$10^{14}$  & (120)$^{{i}}$ & & & & \\ 
&Total$^c$ as ER	 & 	7.1\x$10^{12\,\,o}$ 	 &	 (60)$^{i}$	& ... &   &  & & & & \\ 

HNC  (1)&	 LVF$^g$		& 3.6\x$10^{14\,\,o}$	&	  (100)$^{i}$	& (15)$^{i}$	& &&& & 5\x$10^{14}$ & (150)$^{{i}}$\\ 
&	 	HC$^g$			&4.4\x$10^{14\,\,o}$	&	(200)$^{i}$	& (10)$^{i}$	& && & &\\ 
&	 	ER$^g$			&1.9\x$10^{12\,\,o}$	&	  (60)$^{i}$	& ...	& && & &\\ 

CN  (2)  &PDR/ER$^g$	 	&  4.9\x$10^{13\,\,o}$	& 	 (100)$^{i}$	 &	...	& & & && &\\ 
&	 	HC$^g$			&  7.9\x$10^{15\,\,o}$ & 	(200)$^{i}$	 &	 (10)$^{i}$	& & & && &\\ 

N$_{2}$H$^{+}$ (1)	&	 	Total$^c$ (ER)	&1.0\x$10^{12\,\,o}$	&	 (60)$^{i}$	 & ... &&  & & & &\\ 

\hline
\end{tabular}
\begin{list}{}{}
\item$^{{a}}$ The column density is a source-average if a source size is given, else it is a beam-average.
$^{{b}}$HPBW is 10\,--\,12$\arcsec$ in comparison surveys, and the column densities are beam-averaged  and not corrected for opacity. $^{{c}}$From the total integrated intensity of the line(s).
$^{{d}}$Beam-filling and opacity corrected, as well as separation into components.
$^{{e}}$From an opacity-corrected rotation diagram.
$^{{f}}$From optically thick line(s). 
$^{{g}}$From Gaussian decomposition.
$^{{h}}N_\mathrm{LTE}$,  opacity corrected.
$^{{i}}$Not calculated by the authors. 
$^{{j}}$Temperature from SO$_2$ opacity-corrected rotation diagram.
$^{{k}}$From a non opacity-corrected rotation diagram.
$^{{l}}$Size from SO$_2$. 
$^{{m}}$Temperature from a SO  opacity-corrected rotation diagram.
$^{{n}}$Opacity corrected.
$^{{o}}N_\mathrm{LTE}$.
$^{{p}}$Size from SO total integrated intensity. 
$^{{q}}N_\mathrm{ISO}$.
$^{{r}}$Size from SiO total integrated intensity. 
$^{{s}}$Size from NH$_3$. 
$^{{t}}$From the forward model. 
$^{{u}}$From an opacity corrected two-component rotation diagram.
$^{{v}}$Size  from CH$_3$OH. 
$^{{x}}$Using  H$_{2}$CO calculated source size. 
$^{{y}}$Using size from CS.
\end{list}

\end{table*}

  \begin{table*} 
\caption{Column density results for water, C, CO and H$_2$.} 
\label{result_table2}
\centering
\begin{tabular} { l l l l l l l l    }
\hline
\hline
Species    & Region    & $T_{\mathrm{ex}}$& $N_{\mathrm{LTE}}$ & $N_{\mathrm{ISO}}$  &$N_{\mathrm{H_2}}$    & Size$^a$ & $\tau$   \\
 & &[K]  &   [cm$^{-2}$] &   [cm$^{-2}$] &     [cm$^{-2}$] &  [\arcsec]&         \\
\hline
C  	&	& 100	&5.6$\times10^{17}$&&  &...&   \\ 

CO   &  PDR$^b$	&	100	&1.6$\times10^{17}$&1.6$\times10^{18}$ $^c$&  2.0$\times10^{22}$	&...& \\ 
    &  LVF	&	100	& &2.5$\times10^{19}$ $^c$& 3.2$\times10^{23}$& 	(30)$^{d}$	&       \\ 
   &    HVF &	100	&1.2$\times10^{18}$&3.1$\times10^{18}$ $^e$ &3.9$\times10^{22}$   & 70$^{f}$&\\

$^{13}$CO &    PDR/ER$^b$	&	100 &5.7$\times10^{16}$		&5.4$\times10^{16}$ $^c$&	&...&  	     \\ 
   & 	 LVF &	100	&3.9$\times10^{17}$	&4.2$\times10^{17}$  $^c$&&    30  &   \\ 
   &  HVF	&	100	&5.2$\times10^{16}$	&&&    70  &  \\ 

C$^{17}$O &   PDR/ER$^b$ 	&	 100	& 2.5$\times10^{15}$	&&&   ... & 0.07 \\ 
  & 	LVF  &	100	& 2.0$\times10^{16}$	&	&&   30 & 0.1   \\ 

C$^{18}$O &   PDR/ER$^b$	&	100	& 8.9$\times10^{15}$	&9.8$\times10^{15}$ $^c$&& 	...	& 0.3  \\
 & 	LVF &	100	& 6.7$\times10^{16}$	&7.7$\times10^{16}$ $^c$	&& 	30	& 0.3  \\

H$_{2}$O   & 	Total$^g$		&	 72	&	&  1.7$\times10^{18}$ $^h$	  & &    (15)$^{d}$	 & $\sim$1100  \\ 
   & 	 CR  		&	 115	&	&  5.6$\times10^{17\,\,h}$	  & &   (6)$^{d}$	 &$\sim$860  \\ 
    	& LVF		&	72	&	& 8.7$\times10^{17\,\,h}$	  & &   (15)$^{d}$	 & $\sim$1900  \\ 
  & HVF &	72	&	8.0$\times10^{17\,\,i}$ &  8.8$\times10^{17\,\,h}$& &  70$^{{f}}$ & $\sim$910 \\ 
 &  HC		& 	200 	& 1.2$\times10^{19}$ $^j$	& 	 && 	  10	&0.3$^{k}$   \\ 

H$_{2}^{17}$O  & 	Total$^g$		&	 72	&	 1.3$\times10^{15}$ $^l$ 	 &  & &   15 	 & 0.9  \\ 
 	&     CR 	&	115 	& 4.4\x$10^{14\,\,l}$	&	& &    6	 	 & 0.7 \\ 
  & 	 LVF 		&     72	&6.7$\times10^{14\,\,l}$	&& &  	15 	&1.5  \\ 
  & 	 HVF  		&      72	& 6.8$\times10^{14\,\,l}$	&& &  	15 	&0.7  \\ 
 
H$_{2}^{18}$O   & 	Total$^g$		&	 72	&	  5.0$\times10^{15}$ $^l$	&  5.0$\times10^{15\,\,h}$ & &    15	 & 3.4  \\    
  &    CR  &	115  	&1.8$\times10^{15}$ $^l$	&  1.7$\times10^{15\,\,h}$	&&  	 6  & 2.6  \\ 
 &  LVF  &	72	& 2.7$\times10^{15}$ $^l$	& 2.6$\times10^{15\,\,h}$	&&  	15 	&  5.9  \\
 &  HVF  &72	& 2.8$\times10^{15}$ $^l$	&  2.7$\times10^{15\,\,h}$	&&  	15 	&  2.8  \\
 
HDO	&  Total$^g$				 	&	72  & 9.1$\times10^{15}$&	&&15 &  1.5$^{k}$  \\  
	   &  CR				 	&	115   & 1.8$\times10^{16}$&	&&6 &  3$^{k}$  \\  
  &  LVF				 	&	72	& 4.5$\times10^{15}$	&	&&15  &  0.3$^{k}$  \\ 
    &  HC				 	&	200	& 1.5$\times10^{16}$	&	&&10 &  0.5$^{k}$  \\ 
\hline
\end{tabular}
\begin{list}{}{}
\item$^{{a}}$The column density is a source-average if a source size is given, else it is a beam-average. $^b$The Narrow component from CO isotopologues contains emission from both PDR and ER, the CO Narrow component only from the PDR, hence the PDR column density for CO  here is divided by two as motivated in Sect. \ref{section CO}. $^{{c}}$Using C$^{17}$O together with [$^{18}$O/$^{17}$O]\,=\,3.9, [$^{16}$O/$^{18}$O]\,=\,330, and  [$^{12}$CO/$^{13}$CO]\,=\,60. $^d$Indirect size estimated using isotopologues. The full size may be larger.  $^{{e}}$Using $^{13}$CO together with [$^{12}$CO/$^{13}$CO]\,=\,60. $^{{f}}$Hjalmarson \etal~(\cite{Hjalmarson05}). $^g$From the total integrated intensity of the line. $^{{h}}$Using opacity   and beam-filling corrected $o$-H$_{2}^{17}$O together with [$^{18}$O/$^{17}$O]\,=\,3.9 and [$^{16}$O/$^{18}$O]\,=\,330. $^{{i}}$From HVF Gaussian fit of $o$-H$_{2}$O, opacity and beam-filling corrected. $^{{j}}$Using   beam-filling corrected $p$-H$_{2}$O. $^{k}$Calculated with \mbox{Eq. \ref{tau}}. $^l$Opacity and beam-filling corrected.  
\end{list}
\end{table*}


\section {Results}\label{results}

A summary of the  observed features for all the species is presented in  the on-line Table \ref{summary Table}; the  number of observed lines, 
 the range in upper state energy, and the total integrated intensity. 

A mean line-to-continuum ratio of 0.2 is reported in Paper~I. 
The largest emission comes from CO  with approximately 45\% of the total spectral line emission. The second strongest emitter is H$_2$O (13\%), followed by SO$_2$ (10\%), 
SO (7\%), $^{13}$CO (7\%) and  CH$_3$OH (4\%). The remaining species emit $\sim$14\% of the total. However, these are  \emph{beam-average} values. Since the sizes of the SO$_2$, SO and CH$_3$OH emitting regions are much smaller than the extended CO emission, the relative amount of emission will change with a smaller beam, and will in addition not be the same for the different subregions.

The resulting column densities and rotation temperatures are shown in Tables \ref{result_table1} and \ref{result_table2}, together with results from the ground-based submillimetre spectral scans by White \etal~(\cite{White}, hereafter W03) from 455 to 507 GHz, Schilke \etal~(\cite{S01}, hereafter S01) from 607 to 725 GHZ, and Comito \etal~(\cite{Comito}, hereafter C05) from 795 to 903 GHz. The number of lines used in our calculations are listed in parenthesis after the species in Table \ref{result_table1}. 

Differences may arise between the comparison surveys and ours due to the different beam-fillings. The comparison surveys  W03 and S01 mostly use beam-averaged (with HPBW of 10$\arcsec$\,--\,12$\arcsec$) and not opacity-corrected column densities,  while our results are corrected for beam-filling and optical depth, when possible. A second source of discrepancy is our effort to separate the emissions from different subregions, while the column densities in the comparison surveys are calculated from the total integrated intensity.  
Most of our column densities have therefore been calculated 
using the total integrated intensity and a source size of the main emitting component  as a first approximation and  comparison to S01 and W03. The column densities
for the different subregions,  
are also calculated when possible,
and have been corrected for opacity (if known) and  beam-filling.  In C05 beam-filling and optical depth correction are taken into account as well as separation into different subregions. 

The listed source sizes  are either
calculated with Eq.~(\ref{solution}), 
the $\chi^2$-method, 
or taken from the literature. 
The size of the ER is assumed to be larger than our 
beam, although the East-West extent of the
molecular ridge is rather limited (cf. Goldsmith \etal~\cite{Goldsmith97}).
As excitation temperatures we use the population distribution temperatures $T_\mathrm{ROT}$ obtained from rotation diagrams. These temperatures are also used for species with similar excitation conditions.  If no similar species exist, the temperatures are assumed to have the typical value for the emitting region: 100 K for the LVF, HVF and PDR,  60 K for the ER, 115 K for the CR, and 200 K for the HC. 
The rotation diagram technique as well as the forward model  have been applied to all species that have a broad upper state energy range and four lines or more; SO$_2$, $^{34}$SO$_2$, SO, CH$_3$CN, CH$_3$OH, $^{13}$CH$_3$OH, H$_2$CS, and CH$_3$OCH$_3$; to calculate $N_\mathrm{ROT}$ and $N_\mathrm{\chi^2}$.
Only in the methanol case, the forward model directly leads to a beam-filling and hence a source size.
For the other species it was impossible to discriminate between solutions for different beam size/optical depth combinations.

In  the on-line Table \ref{isotope_ratios} our estimated isotopologue abundance ratios are listed, as well as comparisons with several other studies.



\subsection {Outflow molecules}
Sulphur-bearing species are considered to be tracers of  massive  \emph{outflows} from a newly 
formed star.
The high temperatures 
caused by the intense radiation from the driving source, or by 
shocks,  
can enhance the production of 
SO$_{2}$, SO and SiO. 
The line profile of the outflow shows a characteristic triangular line shape with broad wings as seen in examples of SO in Fig.~\ref{figureSO}, SiO in Fig.~\ref{SiO and isotope}, SO$_2$ in Fig.~\ref {figureSO2}, and in a comparison of SO$_2$ and SO  in Fig. \ref{fig SO_SO2} (on-line material). The Orion outflows are also traced by other molecules like  H$_2$O and CO: see \mbox{Sect. \ref{section CO}} and \ref{section Water}, where also comparisons  of H$_2^{18}$O, SO and SO$_2$ are shown.

\begin{figure}[h] 
 \includegraphics[scale=0.5]{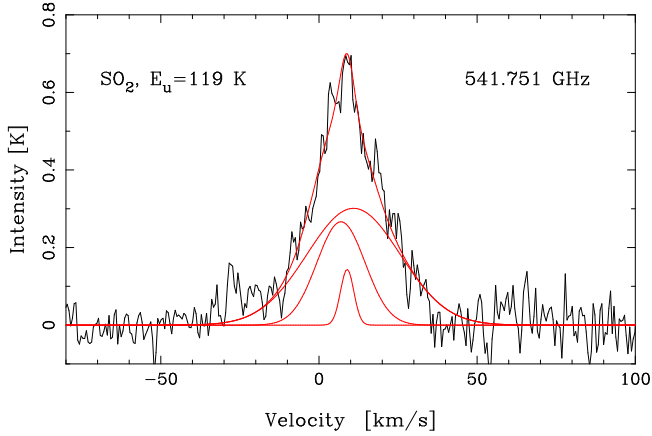} 
 \caption{The $14_{3,11}$\,--\,$13_{2,12}$ SO$_2$ transition with a three-component Gaussian fit shown together with the individual Gaussians. The line widths are 5 \kms, 18
 \kms~and 35 \kms~from the CR, 
 LVF and HVF, respectively.
 }
 \label{fig SO2 Three Gauss}
\end{figure}
\begin{figure}
   \resizebox{\hsize}{!}{\includegraphics{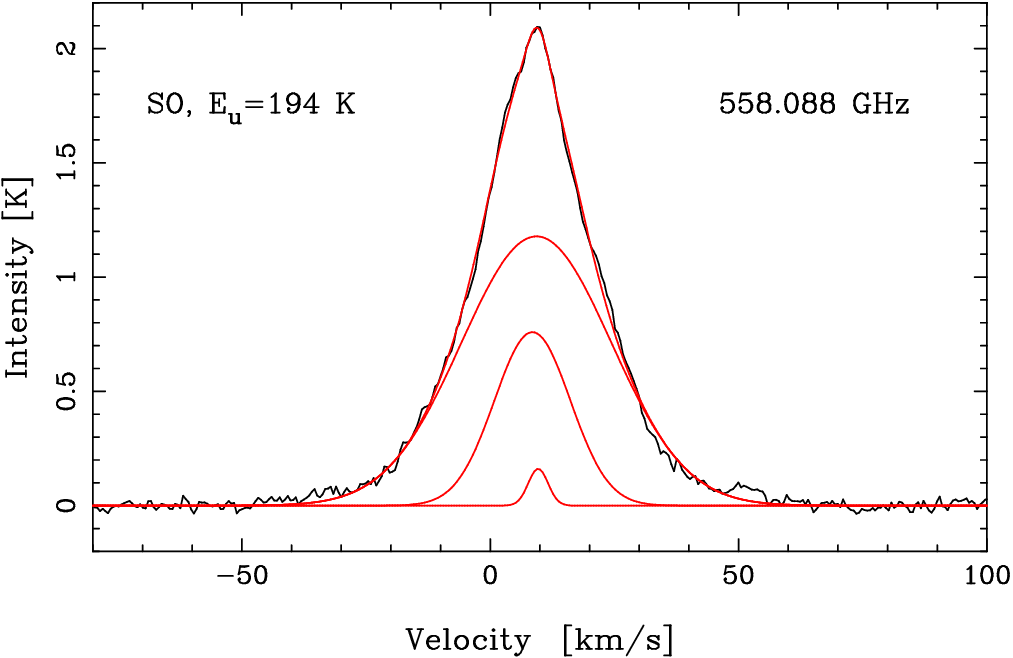}} 
 \caption{The $13_{12}$\,--\,$12_{11}$ SO transition with a three-component Gaussian fit shown together with the individual Gaussians. The line widths are 5 \kms, 18 
 \kms~and 35 \kms~from the CR, 
 LVF and HVF, respectively.}
  \label{fig SO TwoGauss}
\end{figure}
\begin{figure}
   \resizebox{\hsize}{!}{\includegraphics{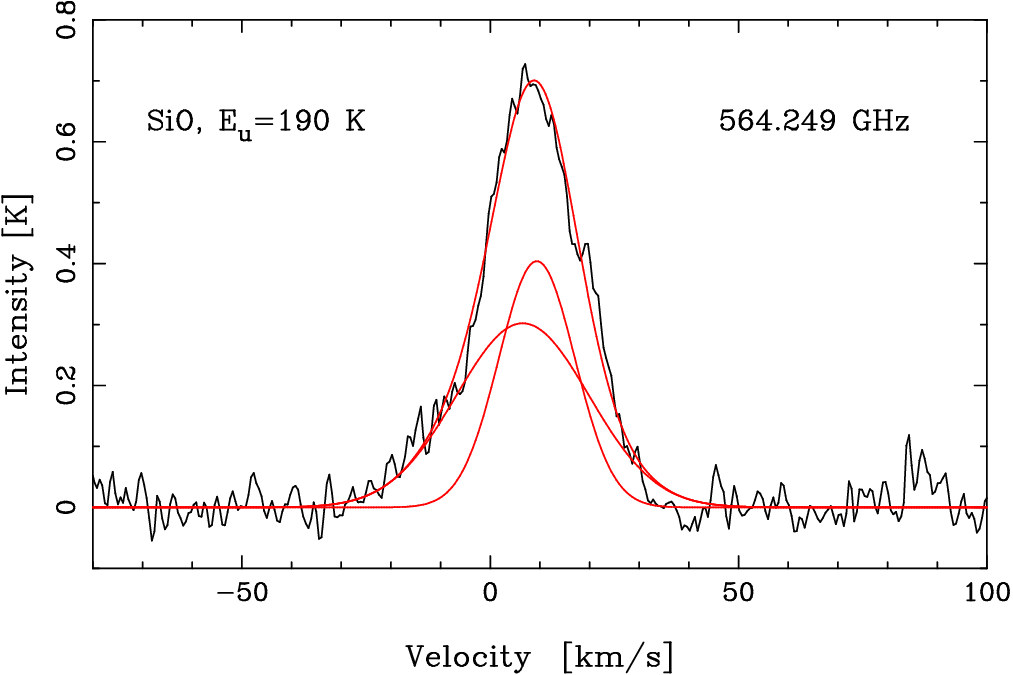}}  
 \caption{The $J$\,=\,13\,--\,12 SiO transition  with a two-component Gaussian fit shown together with the individual Gaussians. The line widths are 18 \kms~and 31 \kms~from the
 LVF and HVF, respectively.}
  \label{fig SiO TwoGauss}
\end{figure}

\subsubsection {Sulphur dioxide   (SO$_{2}$/ $^{34}$SO$_{2}$)}

We have observed 42 SO$_{2}$, and  five $^{34}$SO$_{2}$ transitions. 
Typical line profiles are shown in Fig.~\ref{figureSO2} (on-line material)  with different upper state energies.
As proposed by Johansson \etal~(\cite{Johansson84}) the complicated SO$_{2}$ and isotopologue line shapes suggest the presence of at least two velocity components, even though the emission primarily occurs in the outflow. 
Figure \ref{fig SO2 Three Gauss} shows a three-component Gaussian fit of a typical SO$_2$ line with line
widths of $\sim$5 \kms, 18 \kms~and 35 \kms~from the CR, 
 LVF and HVF, respectively. This is very similar to the Gaussian components of  SO (Fig. \ref{fig SO TwoGauss}), SiO (Fig. \ref{fig SiO TwoGauss}), and H$_2^{18}$O (Fig. \ref{3G fit to h2o18}).

Figure \ref{Figure so2 sizes} (on-line material) shows the  size of the SO$_2$ emitting region vs. energy for each transition (Eq. \ref{solution}).
The mean size is found to be 8$\arcsec$, which is consistent with the aperture synthesis mapping of Wright \etal~(\cite{Wright}). 
This size is used for beam-filling corrections.

The high line density and the broadness of the SO$_{2}$ lines result in blends between the numerous transitions as well as with other species. 
There are 31 SO$_{2}$ transitions and four  $^{34}$SO$_{2}$ transitions without blends, which are used in a rotation diagram shown in Fig. \ref{rot SO2 not opacity corrected} (on-line material)
producing $N_\mathrm{ROT}$\,=\,(3.9$\pm$0.6)$\times10^{17}$ cm$^{-2}$,
$T_\mathrm{ROT}$\,=\,(132$\pm$8) K and $N_\mathrm{ROT}$\,=\,(5.4$\pm$2.0)$\times10^{16}$ cm$^{-2}$,
$T_\mathrm{ROT}$\,=\,125$\pm$30  K for SO$_{2}$  and $^{34}$SO$_{2}$, respectively.

\begin{figure}[h] 
       \resizebox{\hsize}{!}{\includegraphics{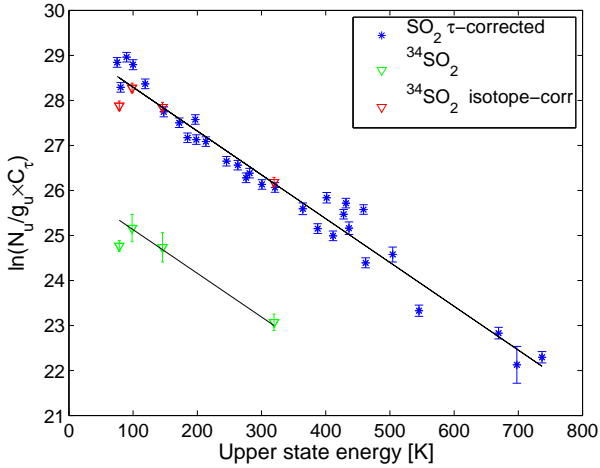}}  
\caption{Rotation diagram for SO$_2$ produces $T_{\mathrm{ROT}}$\,=\,103 K  (extended source). The $^{34}$SO$_2$ fit uses the SO$_2$ rotation
temperature.   }
 \label{rotIsoSO2}
\end{figure}

However, almost all of the SO$_{2}$ transitions are optically thick 
which lowers the  SO$_{2}$ column density. The opacity is calculated using the same excitation temperature for all transitions and the  column density obtained from the $^{34}$SO$_{2}$ rotation diagram (using an isotope ratio of 22.5, Table \ref{isotope_ratios}) and is found to be around 2\,--\,4 for most transitions. The opacity corrected rotation diagram is shown in Fig.~\ref{rotIsoSO2} together with  $^{34}$SO$_{2}$.
The column density $N_\mathrm{ROT}^{\tau-corr}$  increases to \mbox{(1.5$\pm$0.2)}$\times10^{18}$ cm$^{-2}$ and the temperature is lowered to \mbox{103$\pm$3 K}.

The isotopologue $^{34}$SO$_{2}$ is optically thin, hence no opacity correction is needed. But since the lines are weak and only four, the temperature from SO$_{2}$ is applied to the
rotation diagram, which increases the $^{34}$SO$_{2}$ column slightly to \mbox{6.5$\times10^{16}$ cm$^{-2}$.}

As a consistency check we also use Eq. (\ref{NtotCorrected}) together with the single optically thin SO$_2$ line. This 
$16_{\,3,13}$\,--\,$16_{\,0,16}$ transition has an upper state energy of 148 K, and  \mbox{$\tau\sim0.2$}.  
The column density $N_\mathrm{LTE}$ obtained  is 1.4$\times10^{18}$ cm$^{-2}$,   in agreement with $N_\mathrm{ROT}^{\tau-corr}$ and $N_\mathrm{ISO}$ from $^{34}$SO$_{2}$.

Our column densities of both isotopologues are much larger than in the comparison surveys. This can partly be caused by our beam-filling correction with a rather small size, and the non-correction for opacity in W03 and S01.
However, Johansson \etal~(\cite{Johansson84}) and Serabyn \& Weisstein (\cite{Serabyn and Weisstein}) obtain a column density of about 1$\times10^{18}$ cm$^{-2}$   (corrected for our source size) in agreement with our value.

Column densities for each subregion are estimated from the Gaussian components shown in Fig. \ref{fig SO2 Three Gauss} (Table~\ref{result_table1}).
The rarer isotopologues are too weak for a Gaussian decomposition so opacities cannot be calculated by comparison with isotopologues. 
Still, the components are likely to be optically thick and therefore  the sizes of the emitting regions are calculated with $T$\,=\,115 K for the CR,  and  \mbox{$T$\,=\,103 K} for the LVF and  HVF.
The source sizes are found to be 5$\arcsec$ for the CR and 8$\arcsec$ for both the LVF and HVF. 
These sizes correspond to optical depths of about 2\,--\,3. 
The opacity-corrected column densities become 2$\times$10$^{17}$ cm$^{-2}$, 6$\times$10$^{17}$ cm$^{-2}$, and 9$\times$10$^{17}$ cm$^{-2}$  for the CR, LVF and HVF, respectively.

The elemental  isotopic ratio of [$^{32}$S/$^{34}$S] can be estimated from a comparison of the
optically thin column densities. In this way we obtain an isotopic ratio of 23$\pm$7, in agreement with most other comparison studies listed in \mbox{Table \ref{isotope_ratios}.} 

As expected, no vibrationally excited  lines were found. The $5_{5,1}$\,--\,$4_{4,0}$ $v_2$ bending mode transition has the lowest upper state energy (822 K) of all $v_2$ lines in our spectral range. The calculated expected peak temperature of this line is 34 \mbox{mK}, with an expected line width of 23 \kms. Such weak and broad lines are marginally below our detection limit.

\begin{figure} [h]
\resizebox{\hsize}{!}{\includegraphics{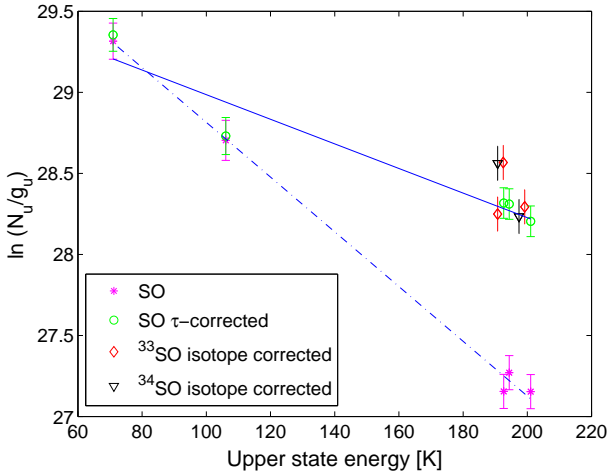}} 
\caption{Rotation diagram for SO and isotopologues (extended source). The isotopologues are plotted multiplied by respective isotopic ratio. Two fits are made: the first (dashed-dotted line) is not opacity corrected
 ($T_\mathrm{ROT}$\,=\,59 K), 
  while the second  (solid line) is corrected for opacity
 ($T_\mathrm{ROT}$\,=\,132 K).}
 \label{rotIsoSO}
\end{figure}

 \subsubsection {Sulphur monoxide  (SO/$^{33}$SO/$^{34}$SO)} \label{SO section}

Typical line profiles are shown in Fig.~\ref{figureSO} (on-line material). The line profiles of the high energy transitions show an even broader outflow emission, and with more pronounced high velocity line wings than does 
SO$_{2}$ (comparison in Fig. \ref{fig SO_SO2} in the on-line material).  
As for SO$_2$, the emission is primarily from the Plateau, and
Friberg (\cite{Friberg84}) has shown the bipolar nature of the HVF component. 
The ratios of SO$_2$ and SO emission lines vs. velocity,  also show a high degree of similarity between the line profiles except in the high velocity 
regime between  $-$30 to $-$5 \kms. At these velocities SO has stronger emission than SO$_2$.

Figure \ref{fig SO TwoGauss}  shows the 13$_{12}$\,--\,12$_{11}$ transition with a three-component Gaussian fit. 
The
broad  HVF component has a FWHM width of $\sim$35 \kms~
at \vlsr$\sim$\,9 \kms. 
The  LVF component has
widths of 18 \kms~
at \vlsr$\sim$8 \kms. 
In addition to the LVF and HVF components a third 
from the CR appears with a width of \mbox{5 \kms}~
at \vlsr$\sim$\,9\,--\,10 \kms. 

The most likely source size is 18$\arcsec$, calculated using the three optically thick SO lines.
This is in agreement with the aperture synthesis mapping by  Beuther \etal~(\cite{Beuther}), and Wright \etal~(\cite{Wright}), who find a larger source size for SO than for SO$_{2}$. The source size 18$\arcsec$ is used for beam-filling correction.

The rotation diagram in Fig.~\ref{rotIsoSO} (calculated with the total integrated intensity of the lines)  displays our  five 
SO lines.  The 
rotation temperature without any corrections is (59$\pm$2 K)  and the column density $N_\mathrm{ROT}$ is (1.5$\pm0.2$)$\times$10$^{17}$ cm$^{-2}$.  However, the three higher energy lines have  optical depth of $\sim$3, whereas the two low energy transitions are optically thin with $\tau\sim0.1$ (Eq.~\ref{tau}). We make an optical depth correction for all five transitions and plot them again in Fig.~\ref{rotIsoSO}  together with a new fit. Note that the correction is substantial for the high energy, optically thick lines (cf. Serabyn \& Weisstein \cite{Serabyn and Weisstein}).
The rotation temperature obtained is    higher than without corrections, 132$\pm$22 K,  but the resulting 
column density is only slightly higher than that found without the corrections, $N_\mathrm{ROT}^{\tau-corr}$\,=\,(1.6$\pm$0.5)$\times$10$^{17}$ cm$^{-2}$. This is
in agreement with  
the column density obtained from $^{34}$SO, $N_\mathrm{ISO}$\,=\,1.9$\times$10$^{17}$ cm$^{-2}$ (using 22.5 for [$^{32}$S/$^{34}$S]). 
The two optically thin SO transitions  gives $N_\mathrm{LTE}$\,=\,1.8$\times$10$^{17}$ cm$^{-2}$.
The column densities for both  isotopologues are calculated with the rotation temperature from SO.

The isotopologues, two $^{33}$SO and three  $^{34}$SO transitions, are optically thin with  opacities around 0.02 and 0.13, respectively. These transitions are   plotted  in Fig.~\ref{rotIsoSO}  with   the  integrated intensities  multiplied  by appropriate isotopic ratios (5.5 for [$^{34}$S/$^{33}$S], Table \ref{isotope_ratios}). As seen in Fig.~\ref{rotIsoSO}, the result of the isotopic ratio corrections is consistent with the optical depth corrected SO transitions. 

As for SO$_2$ the column densities for each SO subregion  are estimated from the Gaussian components shown in \mbox{Fig. \ref{fig SO TwoGauss}.}
Opacities cannot be calculated by comparison with the rarer isotopologues since they
are too weak for a Gaussian decomposition. However,  the components are likely to be optically thick and therefore  the sizes of the emitting regions are calculated with $T$\,=\,115 K for the CR,   \mbox{$T$\,=\,132 K} for the LVF, and   \mbox{$T$\,=\,100 K} for the  HVF.
The source sizes of the CR, LVF and HVF are found to be 6$\arcsec$, 10$\arcsec$ and 14$\arcsec$, respectively. These sizes may be larger if the opacities are low. Combining calculations of source size, optical depths, and column densities, the sizes for the LVF and HVF increase slightly to 11$\arcsec$ and 18$\arcsec$, respectively. These sizes correspond to optical depths of about 2.5 and 1.0 for respective region. 
The opacity-corrected column densities become 1.7$\times$10$^{16}$ cm$^{-2}$, 9.3$\times$10$^{16}$ cm$^{-2}$, and 8.5$\times$10$^{16}$ cm$^{-2}$  for the CR, LVF and HVF, respectively.

The elemental  isotopic ratio of [$^{32}$S/$^{34}$S] and  [$^{34}$S/$^{33}$S]  can be estimated from  comparisons of the
column densities of the isotopologues and the optically thin SO transitions. We obtain isotopic ratios of 21.0$\pm$6 and 4.9, respectively  in  agreement with most other comparison studies listed in \mbox{Table \ref{isotope_ratios}.}

\subsubsection {Silicon monoxide (SiO/$^{29}$SiO/$^{30}$SiO)}

We have observed the transition $J\!=\!13\!-\!12$ for each isotopologue, and we show the SiO and $^{29}$SiO transitions in Fig.~\ref{SiO and isotope} (on-line material). 
As for SO$_2$ and SO, the complicated line profile of SiO suggests emission from both the LVF and the bipolar HVF (present in  aperture synthesis maps of Wright \etal~\cite{Wright}),  with widths of  18 and 31  \kms~at $\upsilon_{\mathrm{LSR}}$ velocities of 9 and 7 \kms, respectively. Figure \ref{fig SiO TwoGauss} shows the two-component Gaussian fit to SiO.
The $^{29}$SiO transition is located in the high-velocity wing of $o$-H$_2$O (at \vlsr$\sim-$130  \kms). The width is
21 \kms~at a centre velocity of $\sim$9 \kms. 
The $^{30}$SiO transition is a questionable assignment due to its narrow line width of 7.5 \kms.

Comparison of the peak antenna temperatures  of SiO and $^{29}$SiO shows  that the  SiO transition 
has an optical depth of $\sim$1.0.
The source size  \mbox{(Eq. \ref{solution})} is found to be  14$\arcsec$. This is used as beam-filling correction. Using
a  LVF temperature  of   100 K (about the same temperature as the SO$_2$ rotation temperature), the total integrated intensity, 
and the simple LTE approximation,
the opacity-corrected column density is found to be  4.0$\times$10$^{15}$ cm$^{-2}$ for SiO.

The decomposition into subregions results in 
LVF and HVF source sizes  of 8$\arcsec$ and  7$\arcsec$ with temperatures of 100 K for both sources. The rather small  values are most likely due to the low opacity in these components and are therefore  only   lower limits. 
Assuming that the opacity in the LVF is about the same as for the total integrated emission, 
the LVF opacity-corrected source size increases to 10$\arcsec$. The opacity in the HVF is most likely less than in the LVF. As an upper limit  the size is assumed to be the same as for the total integrated emission, which results in an HVF opacity of about 0.4.
The sizes are consistent with Beuther \etal~(\cite{Beuther}).
Note the similarity of the SO and SiO source sizes. 
In Sect. \ref{section Water}
the H$_2^{18}$O sizes will  also be shown to be  similar. 
These source sizes are used to correct for  beam-filling, 
and the resulting LVF and HVF opacity-corrected column densities  are 3.3$\times$10$^{15}$ and 1.8$\times$10$^{15}$ cm$^{-2}$, respectively.


\subsection {Outflow and Hot Core  molecule}

The \emph{Hot Core} is a collection of warm ($\ga$200 K) and dense ($n\sim10^7$ \cmcub) clumps of gas. The 
dominating species are oxygen-free, small, saturated nitrogen-bearing molecules such as CH$_3$CN and NH$_3$.  Most N-bearing molecules are strong in the HC, and the oxygen-bearing molecules peak toward the CR (e.g. Blake \etal~\cite{Blake},  hereafter B87; Caselli \etal~\cite{Caselli93};  Beuther \etal~\cite{Beuther}).
CH$_3$OH is an exception with pronounced emissions from the HC as well as from the CR. 
In addition, high levels of deuterium fractionation are found here.  Since the HC region  probably contains one or more massive protostars it presents an ideal opportunity to study active gas-phase chemistry. And due to the high temperatures in both the HC and CR,  the gas-phase chemistry will get a 
significant contribution of molecules from grain surface chemistry through evaporation of the icy mantles caused by the intense UV radiation from  newly formed stars.


\begin{figure}
   \resizebox{\hsize}{!}{\includegraphics{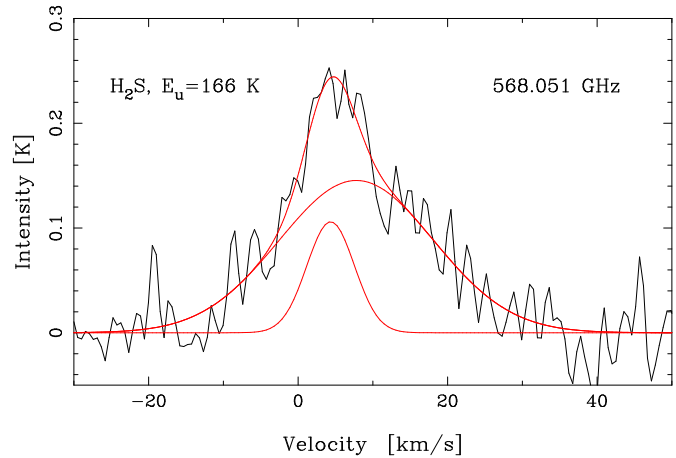}} 
 \caption{The H$_2$S 3$_{3,1}$\,--\,3$_{2,2}$ transition with a two-component Gaussian fit shown together with the individual Gaussians. The line widths are 8 \kms~and 
 24  \kms~from the HC and outflow, respectively. }
  \label{fig H2S TwoGauss}
\end{figure}

\subsubsection {Hydrogen sulphide (H$_{2}$S)}


We observe only the 3$_{3,1}$\,--\,3$_{2,2}$ transition of H$_{2}$S, with  emission from the HC and LVF, 
illustrated by a Gaussian decomposition in   Fig. \ref{fig H2S TwoGauss}.
The emission from the HC component has a width of $\sim$8 \kms~at \vlsr$\sim$ 5 \kms~between velocities $-$5 to +15 \kms. The LVF emission has a width of $\sim$24 \kms~at \vlsr$\sim$8 \kms.
The line is also shown in the bottom of Fig.~\ref{NH3 collection figure} (on-line material) together with other comparison HC molecules.

The column densities are consistent with the comparison surveys assuming typical source sizes and temperatures.

\subsection {Hot Core molecules}

\subsubsection {Methyl cyanide (CH$_{3}$CN)}

\begin{figure}[h]
 \includegraphics[scale=0.53]{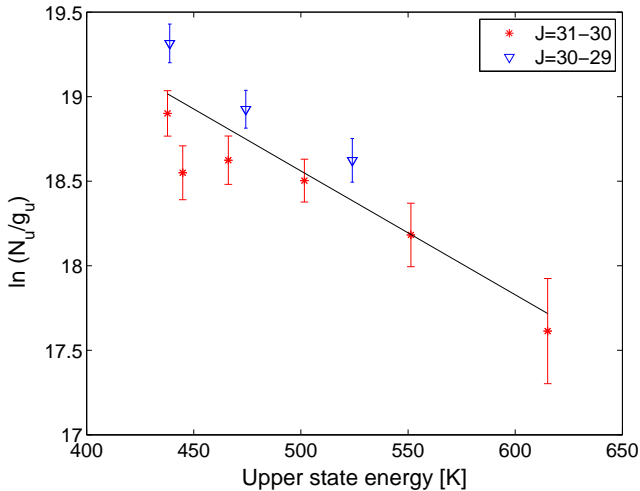}   
 \caption{Rotation diagram for CH$_3$CN  producing $T_\mathrm{ROT}$\,=\,137 K (extended source).}
  \label{rotIsoCH3CN}
\end{figure}

Previous observations of the high density tracer CH$_{3}$CN (e.g. Blake \etal~\cite{Blake etal 86}; Wilner \etal~\cite{Wilner 94}) have shown that the low-$J$ transitions in the vibrational ground state appears to be a mix of CR and HC emission, while the  high-$J$ transitions and all the vibrationally excited lines originate in the HC only.  
This is also confirmed in our survey where we  observe the  \mbox{$30_K\!-\!29_K$}  transitions with  \mbox{$K\!=\!0\!-\!9$}, and  \mbox{$31_K\!-\!30_K$} with  \mbox{$K\!=\!0\!-\!6$} and 9. These lines suggest an origin
 in the HC at $\upsilon_{\mathrm{LSR}}\sim$5\,--\,6 \kms~and widths of $\sim$8\,--\,9 \kms, also consistent with W05 and C05.
The  \mbox{30$_{4}$\,--\,29$_{4}$} ground state transition is shown in Fig.~\ref{NH3 collection figure} (on-line material).
In addition we  see a number of  weak vibrational lines from the  \mbox{$v_8$=1} bending mode with  \mbox{$30_K\!-\!29_K$} where  \mbox{$K$\,=\,0\,--\,3}. In total we observe 17 line features from this molecule. Nine of these are free from blends and are used in the rotation diagram (Fig.~\ref{rotIsoCH3CN}).
Due to the weak lines   the 
rotation temperature of 137$\pm$25 K and  the  column density of  \mbox{(5.0$\pm$3.6)$\times10^{15}$ cm$^{-2}$} 
are comparatively uncertain. The temperature is  rather low compared to  
W03 who estimate the temperature to 227 K, and C05 to 250 K.
Still, the column density agrees well with  B87, Sutton \etal~(\cite{Sutton}, hereafter S95), W03, and C05.

Wilner \etal~(\cite{Wilner 94})
find an opacity of the HC emission of at most a few for the main lines. This could explain the rather low [$^{13}$C/$^{12}$C] ratios in Blake \etal~(\cite{Blake etal 86}) and Turner (\cite{Turner 91}). Sutton \etal~(\cite{Sutton 86}) suggests significant opacity from their statistical equilibrium calculations, if the HC is as small as 10$\arcsec$. This would give even higher column densities in the HC.

The partition function is calculated as recommended in  
 Araya \etal~(\cite{Araya}).

\subsubsection {Cyanoacetylene (HC$_{3}$N)}	

Two transitions of cyanoacetylene are seen, of which $J\!=\!54\!-\!53$ is a blend with CH$_3$OH.
The transition $J\!=\!60\!-\!59$ is shown in Fig.~\ref{NH3 collection figure} (on-line material). 
HC emission is here evident at $\upsilon_{\mathrm{LSR}}\sim$5\,--\,6 \kms~and a line width of $\sim$10 \kms.
The column density is calculated with the simple LTE approximation
and is in agreement with W03.


\subsubsection {Carbonyl sulphide (OCS)}

We have identified three transitions from carbonyl sulphide, \mbox{$J\!=\!47\!-\!46$,} \mbox{$J\!=\!45\!-\!44$,} and  \mbox{$J\!=\!46\!-\!45$}
(shown in the on-line Fig.~\ref{NH3 collection figure}). The emission has its origin in the HC with $\upsilon_{\mathrm{LSR}}\sim$6 \kms, and a width of $\sim$6 \kms.

The estimated column density 
is about five times lower than found by both W03 and  S95.

\subsubsection {Nitric oxide (NO)}

We observe three features with $^2\Pi_{3/2}$, $J$\,=\,11/2\,--\,9/2,  and \mbox{$^2\Pi_{1/2}$, $J$\,=\,11/2\,--\,9/2} from both e and f species, which are composed of twelve non-resolved  hyperfine   transitions.
No separation into components is  possible due to blends between the hyperfine transitions and other species. 
Our estimated rotation temperature from our  transitions  with  upper state energies of 84 K and   232 K is found to be 75 K. This is highly uncertain due to the severe blends in both  low energy transitions. Since 
S01 and C05 observed HC emission we therefore use a typical HC temperature of 200 K and source size of 10$\arcsec$. The resulting column density (using the high energy line) is 
in agreement with S01 and C05.


\subsection {Hot Core and  Compact Ridge molecules}

The \emph{Compact Ridge} is a more quiescent region as compared to the Hot Core. Here we find high abundances of   oxygen-bearing species such as CH$_3$OH, (CH$_3$)$_2$O and HDO (B87; Caselli \etal~\cite{Caselli93}; Beuther \etal~\cite{Beuther}).
As in the HC, the evaporation of the icy mantles  in the warm CR will release molecules produced by grain surface chemistry into
the gas-phase.



\subsubsection {Ammonia   (NH$_{3}$/$^{15}$NH$_{3}$)}

The symmetric top ammonia molecule is a valuable diagnostic because its complex energy level structure covers a very broad range of critical densities and temperatures (see Ho \& Townes \cite{Ho and Townes} for energy level diagram and review). 

Many observations have been made of the NH$_{3}$ inversion lines at cm wavelengths since the first detection by Cheung \etal~(\cite{Cheung68}).
The upper state energy of the lowest metastable inversion lines   are 24 K and 64 K comparable to \mbox{28 K} for the 
rotational ground state transition 1$_{0}$\,--\,0$_{\,0}$ at 572 GHz.
The critical density is very different though, and
is 3.6$\times10^7$ \cmcub~(calculated for 20 K) for the rotational ground state transition, 
and about 10$^3$ \cmcub~for the inversion lines. 
The non-metastable inversion lines also trace higher excitation and density regions. 
Comparison of all 
these transitions could therefore give valuable information about both high- and low density and temperature regions.
The previous low quantity of observations of rotational transitions is due to the fact that they fall into the submillimetre and infrared regimes, which are
generally not accessible from the ground and therefore has to be observed from space.

Observations of both metastable and non-metastable inversion lines (e.g. Batrla \etal~\cite{Batrla83}; Hermsen \etal~\cite{Hermsen}, \cite{Hermsen88b}; Migenes \etal~\cite{Migenes89}) have shown NH$_{3}$ in the HC, CR, ER and LVF regions. The existence of an outflow component was  however questioned by Genzel \etal~(\cite{Genzel82}) since the hyperfine satellite lines could cause the broadness of the line if the opacity is large.

The rotational ground state transition was first and solely detected twenty-four years ago with the Kuiper Airborne Observatory (Keene \etal~\cite{Keene}). Note that the Kuiper Airborne Observatory had a similar beam size (2$\arcmin$) to that of Odin  (2$\arcmin.$1).  Using  Odin, sensitive observations have been  made recently for example towards  Orion KL and the Orion Bar (Larsson \etal~\cite{Larsson}), the $\rho$ Oph A core (Liseau \etal~\cite{Liseau}), Sgr B2 (Hjalmarson \etal~\cite{Hjalmarson05}), as well as the molecular cloud S140. The resulting NH$_3$ abundance in the Orion Bar is 5\x10$^{-9}$ (Larsson \etal~\cite{Larsson}).

In this spectral survey we have observed the rotational ground state 1$_{0}$\,--\,0$_{\,0}$ transitions of NH$_{3}$ and $^{15}$NH$_{3}$, which are shown in Fig.~\ref{NH3 collection figure} (on-line material). We show the NH$_{3}$ transition twice to emphasize the  line wings. Our peak temperature agrees to within 5\% with  Larsson \etal~(\cite{Larsson}) who used a rather different Odin observational setup, demonstrating the excellent calibration of the Odin data. The vibrational transition $\nu_2=1$ of this line at 466 GHz has previously been observed by Schilke \etal~(\cite{Schilke92}).

\begin{figure}
   \resizebox{\hsize}{!}{\includegraphics{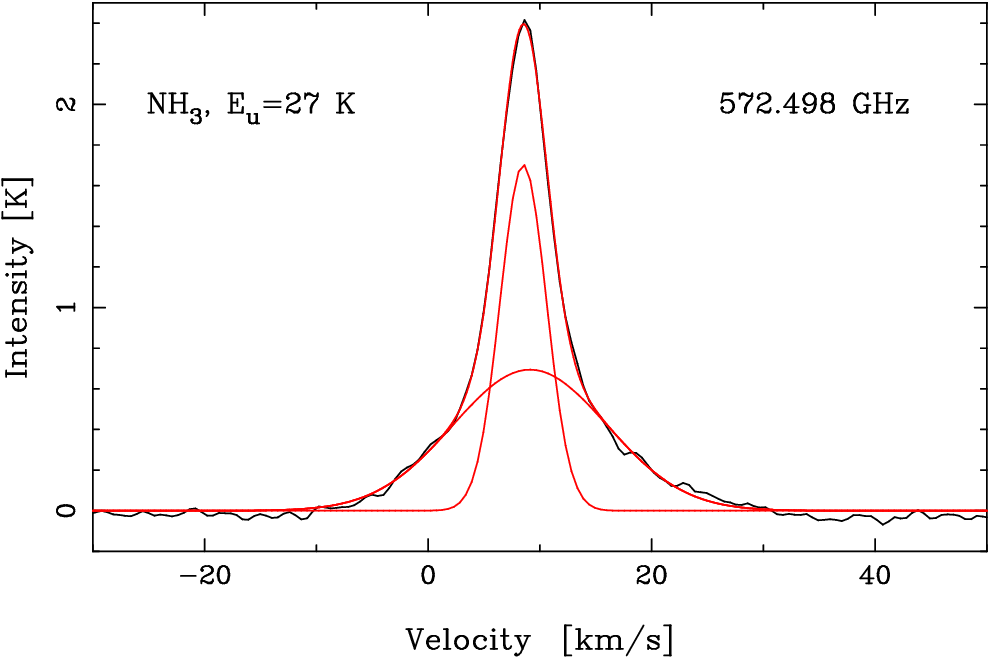}}  
 \caption{The NH$_3$ transition with a two-component Gaussian fit shown together with the individual Gaussians. The line widths are 5~\kms~and 16~\kms~ from the CR and 
 HC/outflow, respectively.}
  \label{fig NH3 Three Gauss}
\end{figure}

Fig. \ref{fig NH3 Three Gauss} shows our two-component Gaussian fit to the NH$_{3}$ line which has pronounced  features of the CR and a broad component.
The 
line widths are 
5 and 16~\kms~at LSR  velocities  8,  and 9~\kms~for the CR and broad components, respectively.  The CR emission   was also observed by Keene \etal~(\cite{Keene}), while the broader  component clearly seen in our Odin data, was only marginally present in their lower signal-to-noise data.
Our $^{15}$NH$_{3}$ spectrum  shows only evidence of the HC  component (cf. Hermsen \etal~\cite{Hermsen85}), with a width of \mbox{$\sim$7 \kms}~at \mbox{$ \upsilon_{\mathrm{LSR}}\approx7$~\kms.} 
The width of the  broad NH$_{3}$ component   may seem  too  large to have an origin in the HC. However, the broadness of the line may be  caused by opacity broadening  (\mbox{Eq. \ref{opacity broadening};} cf. Phillips \etal~\cite{Phillips79}). 
From \mbox{Eq. \ref{tau_iso_comp}}
combined with an assumed $^{14}$N/$^{15}$N isotope ratio of 450 (Table \ref{isotope_ratios}), we estimate optical depths of $\sim$100 and $\sim$0.3 in the  NH$_{3}$ and $^{15}$NH$_{3}$ HC lines, respectively. According to \mbox{Eq. \ref{opacity broadening}} this will 
broaden the
optically thick NH$_{3}$ emission line   by  approximately 2.6 times from a line width comparable to 
the optically thin $^{15}$NH$_{3}$ HC emission to a 
resulting width of $\sim$17~\kms.
This is very close to the width of our Gaussian HC component, 16~\kms. 
However, the high opacity in this component will cause the line profile to be flat topped with little or no line wings. 
Hence our broad Gaussian  component not only 
contains the opacity broadened HC emission but 
also the outflow component seen by e.g. Wilson \etal~(\cite{Wilson 1979}) and Pauls \etal~(\cite{Pauls 1983}), in our 
spectrum visible as pronounced line wings. Alternatively it could be that the HC emission is hidden by optically thick NH$_3$ LVF emission just as in case of water (cf. Section~\ref{section p-h2o from HC}).


The NH$_{3}$ source sizes of the CR and HC regions are  found to be 17$\arcsec$ and 8$\arcsec$, respectively, and are used as beam-filling corrections. The rather large CR size as compared to the 6$\arcsec$ mean source size obtained for CH$_3$OH, might be due to the low upper state energy of 27 K for NH$_{3}$. Figure \ref{SourceSize 2comp}  shows the decreasing methanol source size with upper state energy, where the lowest methanol transitions with upper state energies of  40\,--\,100 K reach a source size of  $\sim$11$\arcsec$. Hermsen \etal~(\cite{Hermsen88b}) find source sizes of 15$\arcsec$ and 6$\arcsec$ for the CR and HC, respectively, in agreement with our calculations. VLA maps by Migenes \etal~(\cite{Migenes89}) also show that the HC emission is clumped on 1$\arcsec$ scales.

Hermsen \etal~(\cite{Hermsen88b}) find a HC temperature of 160$\pm$25~K and
a CR temperature above 100 K. In addition Wilson \etal~(\cite{Wilson}) detect
an even hotter HC component with a temperature of about 400~K. Using  HC and CR temperatures of 200 K and 115 K, respectively, we find
a NH$_{3}$  HC column density 
  (calculated from the optically thin $^{15}$NH$_{3}$ line) of
1.6$\times10^{18}$~cm$^{-2}$. Our comparison surveys have no observations of this molecule, but our result agrees with
Genzel \etal~(\cite{Genzel82}), who report column densities of NH$_{3}$ that reach  5$ \times10^{18}$~cm$^{-2}$ from the HC, with size 10$\arcsec$ and temperatures about 200 K.  Their observations also confirmed increasing line width with increasing optical depth.   Hermsen \etal~(\cite{Hermsen88b}) and Pauls \etal~(\cite{Pauls 1983}) find  values  of 1$\times10^{18}$ cm$^{-2}$ for the HC. 

The optically thick NH$_{3}$  CR column density  is found to be 3.4$\times10^{15}$ cm$^{-2}$.
Optical depth broadening is used to
estimate the opacity in this component.  Batrla \etal~(\cite{Batrla83}) found an
intrinsic velocity width of 2.6 \kms~ by ammonia inversion lines observations.
From a comparison with the observed line width, the opacity is estimated to be about 12 in the CR component.  
The opacity-corrected CR column density then becomes 4.0$\times10^{16}$ cm$^{-2}$. This is in agreement with the estimation of 
Hermsen \etal~(\cite{Hermsen88b}) who find a column density in the range 8$\times10^{15}$\,--\,8$\times10^{16}$ cm$^{-2}$ from the metastable (6,6) inversion line.


\subsubsection {Methanol (CH$_{3}$OH/$^{13}$CH$_{3}$OH) } \label{methanol}

Methanol is an organic asymmetric top molecule with many energy levels (see energy level diagram in Nagai \etal~\cite{Nagai}), and behaves like two different species labelled A and E for  symmetry reasons.

We have observed 76 methanol lines of which 42 are from the $v_t$=1 state, which is the first excited vibrational state of the torsional motion of the CH$_3$ group relative to the OH group. 
In  the on-line Fig.~\ref{figureCH3OH} we have collected a number of 
examples of typical line profiles of CH$_{3}$OH, with  different upper state energies and  $A$-coefficients.
The rarer isotopologue $^{13}$CH$_{3}$OH is seen with 23 lines, of which two are vibrationally excited. Three typical line profiles are shown in the on-line Fig.~\ref{figure13CH3OH}.

\begin{figure}[h] 
       \resizebox{\hsize}{!}{\includegraphics{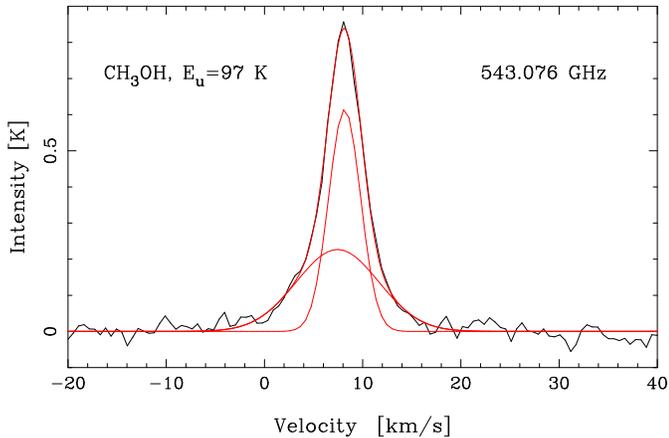}} 
\caption{Methanol line with a two-component fit shown together with the individual Gaussians. The line widths are 4~\kms~and 10~\kms~ from the CR and HC, respectively.}
 \label{2componentsCH3OH}
\end{figure}

The CH$_{3}$OH lines show evidence of two velocity components. One narrow, likely from the CR, with  a line width of  \mbox{$\sim$3--4 \kms,} and average velocity $ \sim$8 \kms. The other broader component with a probable origin in the HC has a line width of $\sim$6\,--10 \kms, and average velocity $ \sim$7 \kms~(see  an example of a two-component Gaussian fit in Fig.~\ref{2componentsCH3OH}). This is consistent with the findings of   Menten \etal~(\cite{Menten}), S95, C05, and also of Beuther \etal~(\cite{Beuther}) who locate the methanol emission to the HC as well as the CR in their SMA aperture synthesis maps. 
According to recent CRYRING storage ring measurements (Geppert \etal~\cite{Geppert}; Millar \cite{Millar}) 
the dissociative recombination of  a parent ion \mbox{CH$_3$OH$_2^+$ + e$^-$ $\rightarrow$ CH$_3$OH + H} is so slow 
that gas-phase formation of methanol is unable to explain the abundance of this molecule, even
in dark clouds where it is rare. Instead we have to rely on efficient hydrogenation 
reactions on grain surfaces, and subsequent release of the methanol into gas-phase. In this scenario the presence of very large amounts of CH$_3$OH in the compact, heated CR and HC sources  is indeed expected.

\begin{figure}[h] 
 \resizebox{\hsize}{!}{\includegraphics{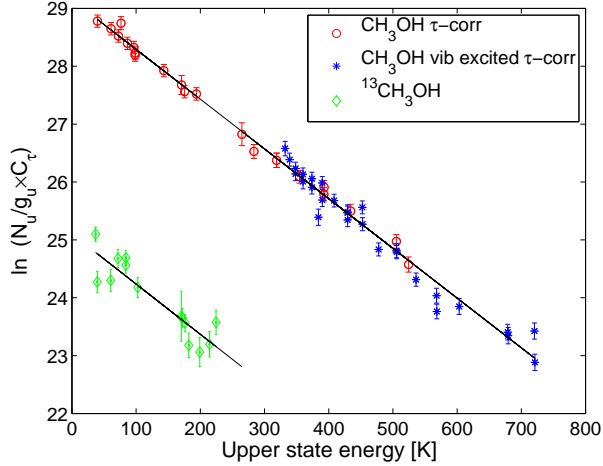}}  
\caption{CH$_3$OH opacity corrected rotation diagram producing $T_\mathrm{ROT}$\,=\,116 K (extended source).  The opacity is calculated with varying source-sizes with energy. The
 $^{13}$CH$_3$OH rotation diagram gives a temperature of 115 K.}
 \label{rotIsoCH3OH}
\end{figure}

The  isotopologue $^{13}$CH$_{3}$OH  show only evidence of the narrow component, which can be well fitted by a single Gaussian with $\Delta \upsilon\approx 3$ \kms~at $\upsilon_{\mathrm{LSR}}\approx8$~\kms. This is also consistent with Menten \etal~(\cite{Menten}) and Beuther \etal~(\cite{Beuther}).  The slight broadening of the CH$_{3}$OH narrow components compared with  that of  $^{13}$CH$_{3}$OH might be caused by optical depth broadening. The $^{13}$CH$_{3}$OH from the HC is expected to be well below our detection limit.
There are
14 $^{13}$CH$_{3}$OH lines free from blends with an upper state energy range between \mbox{37\,--\,225 K,}
which are used in a rotation diagram (Fig.~\ref{rotIsoCH3OH}).  No corrections for optical depths are needed 
since the $^{13}$CH$_{3}$OH transitions are optically thin.
The rotation temperature is found to be \mbox{115$\pm$16 K,} and the column density   (5.9$\pm$1.5)$\times 10^{16}$ cm$^{-2}$. 

If we exclude blended and very weak lines we have  50  CH$_{3}$OH lines  with an upper state energy range from 40 to 721~K. The large number of lines and the wide temperature range make methanol well suited to be used in a rotational diagram. However, one difficulty 
that may occur  with this method is that the optical depths may vary
considerably between the CH$_3$OH transitions. In the rotation diagram seen in Fig.~\ref{Rot diagram for non-tau-corrected CH3OH and 3 optically thin ch3oh and 13ch3oh} (on-line material) the lines are plotted  (using the total integrated intensity) without any attempt to correct for optical depth or beam-dilution. As can be seen there is a large scatter of the CH$_3$OH lines.
Three transitions with  upper state energies of 77, 171 and 265 K lie clearly  very high above the others  due to their low transition probability and low opacity.
A separate fit of these three lines is made and the resulting beam-filling corrected column density becomes (2.6$\pm$0.4)$\times 10^{18}$ cm$^{-2}$. This is about 3 times higher than the resulting column density from all the lines, (9.3$\pm$1.1)$\times10^{17}$ cm$^{-2}$.
This  indicates that opacity correction needs to be included in the rotation diagram.

Using the forward model, which includes opacity and beam-filling correction (see Fig. \ref{rot Albert CH3OH} in the on-line material), we find a column density  of (1.3$\pm$0.1)$\times10^{18}$ cm$^{-2}$~in a source size of 6\arcsec. 
(This size is used as beam-filling correction in all calculations of the column densities above.)
The scatter in the rotation diagram is  reduced and approaches the column density obtained from the three (assumed) optically thin lines. 
However, since this method has a tendency to underestimate the column density we proceed with opacity correction of the traditional rotation diagram. 
We note that most of the low energy lines seem to be optically thick (opacities between $\sim$1\,--\,6) and most of the high energy  lines seem to be optically thin (opacities between $\sim$0.3\,--\,1.5).
The rotation temperature would be too high if not opacity corrected.

An additional complication is that   the extent of the emitting regions may be different for lines of different energy (cf. Menten \etal~\cite{Menten86}). This is affecting our estimation of the opacity since we need a total column density (corrected for beam-filling) in the calculations. 
The on-line Fig.~\ref{SourceSizeCH3OH} shows
that the source size of the low-energy lines varies between \mbox{5\,--\,12$\arcsec$,} whereas the size of the high-energy lines is almost constant (about 6$\arcsec$), based on Eq.  (\ref{source-size}) at \mbox{$T_{\mathrm{ROT}}$\,=\,120 K.}

In Fig. \ref{rotIsoCH3OH} we show the opacity corrected rotation diagram. The opacity is calculated using the column density obtained from the three optically thin lines   corrected for different beam-fillings for each transition,  and the same excitation temperature for all lines (120 K). The scatter in the plot is even more reduced than in the forward model and the resulting column density  becomes (3.4$\pm$0.2)$\times 10^{18}$ cm$^{-2}$. This is much higher than in our comparison surveys, but consistent with  Johansson \etal~(\cite{Johansson84}), Menten \etal~(\cite{Menten86}),  and S01 
using the $^{13}$CH$_{3}$OH column density  (all corrected for our source size).

The 
rotation temperature is 116$\pm$2 K with opacity correction which is  the same as produced by 
the $^{13}$CH$_3$OH rotation diagram (115$\pm$16 K)  and the optically thin fit \mbox{(120$\pm$10 K).} The forward model produces a slightly higher temperature (\mbox{136$\pm$4 K}), which suggests that the  opacity correction is too low with this method.

There is also a possibility that the high- and low energy lines are emitted from different regions even though our rotation diagram does not indicate a change of rotation temperature. Using the Gaussian decomposition of the 27 strongest lines, we note that
the integrated intensity of the low energy lines is dominated by the narrow CR component, and the high energy lines by the broad HC emission.

When  calculating the opacity of the components we   again take into account the varying source size with energy.    However,  Fig.~\ref{SourceSize 2comp} shows that the pronounced variation in size is only true for the narrow   component. The broad component seems to have approximately the same size  as the energy increases.
This again supports the scenario in which the narrow component arises in
the CR,  which is denser and hotter in the central parts. Hence only the central parts have the ability to emit the high energy lines. The broad component keeps the small size across the 
transition energy range, supporting  an origin in the HC. This source is small and hot and thus can emit all transitions throughout the whole region.
The opacity of the CR component is found to be higher than in the HC component which is about 1 or less.

\begin{figure}[h] 
 \resizebox{\hsize}{!}{\includegraphics{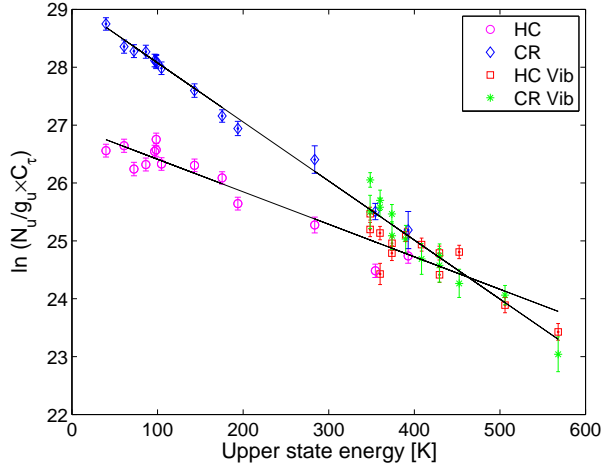}} 
 \caption{Two component CH$_3$OH opacity corrected rotation diagram  producing $T_\mathrm{ROT}$\,=\,178 K and 98 K  
for the HC and CR, respectively  (extended source).  The opacity is calculated with the same parameters together with varying source-size with energy. }
 \label{rotCH3OH_2comp}
\end{figure}

\begin{figure}[h] 
       \resizebox{\hsize}{!}{\includegraphics{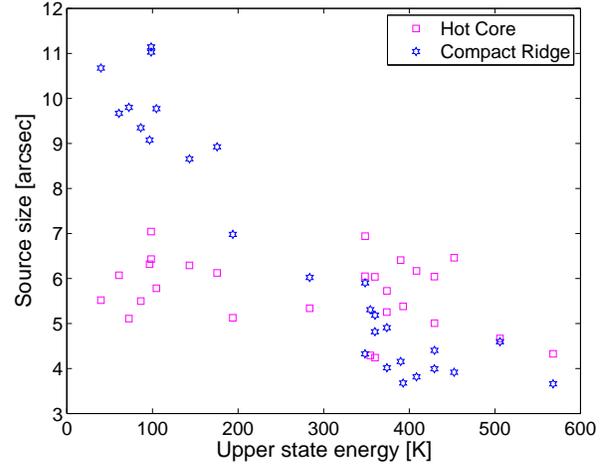}}  
 \caption{Source size variation with energy for the CR and HC components of CH$_3$OH (opacity-corrected with 
$N_\mathrm{CR}$\,=\,1.7$\times10^{18}$ cm$^{-2}$, $T_\mathrm{CR}$\,=\,120 K and $N_\mathrm{HC}$\,=\,7.9$\times10^{17}$ cm$^{-2}$, $T_\mathrm{HC}$\,=\,200 K )}.
 \label{SourceSize 2comp}
\end{figure}

Plotting the components   in a opacity corrected rotation diagram (Fig.~\ref{rotCH3OH_2comp}),  produces  column  densities and rotation temperatures for each region: 
$N_\mathrm{ROT}$\,=\,(2.4$\pm$0.2)$\times 10^{18}$ cm$^{-2}$, $T_\mathrm{ROT}$\,=\,98$\pm$2 K and (7.9$\pm$1.0)$\times 10^{17}$ cm$^{-2}$, $T_\mathrm{ROT}$\,=\,178$\pm$11 K for the CR and HC, respectively. Both column densities are much higher than in our comparison surveys, but agrees well with S95 (corrected for our source size). The calculated temperatures are lower than in the comparison surveys, but
the high apparent rotation temperatures may be caused by high opacity.
Hollis \etal~(\cite{Hollis 83}) found that  the ground-state transitions originate in a 90 K region, while the torsionally excited transitions come from a 200 K region.



The isotopic ratio of $^{12}$C/$^{13}$C can be estimated from
the ratio of the optically thin  CH$_{3}$OH  and $^{13}$CH$_{3}$OH  column densities, and is found to be 57$\pm$14. This is consistent with previous estimates \mbox{(Table \ref{isotope_ratios})}.

\subsection {Compact  Ridge molecules}


\subsubsection {Dimethyl ether ((CH$_{3}$)$_2$O)}

This molecule is affected by two internal rotors which are the origin of the fine structure lines of the AA, AE, EE and EA symmetries (Groner \etal~\cite{Groner98}). The emission only shows characteristics of the CR with narrow widths of 3\,--\,4 \kms~at $\upsilon_{\mathrm{LSR}}$ velocities of $\sim$6\,--\,8 \kms.

\begin{figure}[h] 
       \resizebox{\hsize}{!}{\includegraphics{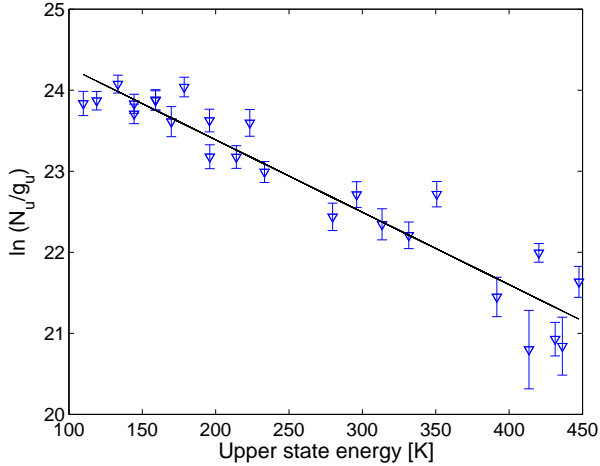}}  
\caption{Rotation diagram for (CH$_3$)$_2$O  producing $T_\mathrm{ROT}$\,=\,112 K  (extended source).}
 \label{rotIsoCH3OCH3}
\end{figure}

Since we cannot resolve these fine structure transitions, we treat them as one single line. The statistical weights and the partition function   are changed accordingly.
We observe 47 quartets out of which 37 are free from blends and hence can be used in the rotation diagram shown in Fig.~\ref{rotIsoCH3OCH3}. 
The resulting beam-filling corrected column density is (1.3$\pm$0.3)\x$10^{17}$ \cmsq~and the  rotation temperature is 112$\pm$8 K, which is higher than in the comparison line surveys (Table \ref{result_table1}). 
The adopted source size is the same as obtained for CH$_3$OH, since these molecules most likely have a rather similar origin  in the CR. This is also verified when calculating the source size with Eq. \ref{source-size}, assuming an opacity larger than unity. For a temperature of 112 K we find a CR size of 5\,--\,6$\arcsec$. This is also indicating optically thick lines which could increase the column density even further.

\subsubsection {Thio-formaldehyde (H$_{2}$CS)} \label{subsubsection h2cs}

\begin{figure}[h] 
       \resizebox{\hsize}{!}{\includegraphics{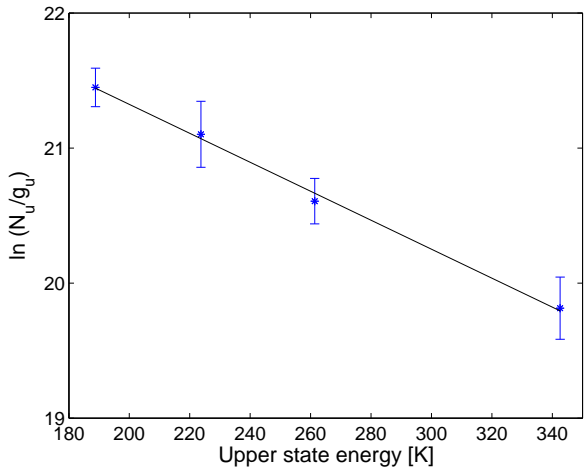}}  
 \caption{Rotation diagram for H$_2$CS producing $T_\mathrm{ROT}$\,=\,93 K  (extended source).}
 \label{rotH2CS}
\end{figure}

Five transitions of the CR-emitting H$_{2}$CS are observed,  of which the 16$_{3,13}$\,--\,15$_{3,12}$ transition is a blend with a U-line. The line profile of the 14$_{1,13}$\,--\,13$_{1,12}$ transition is shown in Fig.~\ref{H2CO} (on-line material). 
The four lines with no blends are used in the rotation diagram shown in Fig.~\ref{rotH2CS}, producing a rotation temperature, very similar to that of CH$_3$OH, $T_\mathrm{ROT}$\,=\,(93$\pm$4) K. 
The resulting beam-filling corrected column density is (1.3$\pm$0.2)\x$10^{15}$ \cmsq, with a source size of 14$\arcsec$ guided by our calculations for the H$_2$CO optically thick CR emission (see Sect. \ref{subsubsection h2co}). 

A comparison of the  H$_{2}$CS and the optically thin  H$_{2}^{13}$CO results in a molecular abundance ratio of H$_{2}$CO/H$_{2}$CS$\sim$15. This is lower than the quoted [O/S] ratio of 35 (on-line Table \ref{isotope_ratios}) from Grevesse \etal~(\cite{Grevesse}). From the comparison of H$_2$O and H$_2$S in Sect.~\ref{section Water}
we obtain a similar  value of $\sim$20.


\subsubsection {Thioformyl cation (HCS$^{+}$)}

The thioformyl cation  previously has not been seen either by W03 nor S01, and here we only observe the $J\!=\!13\!-\!12$ transition as a visible blend with $^{33}$SO. Due to the blend we cannot analyse this transition further, but
this emission is most likely emitted in the same small hot and dense CR source as CH$_3$OH and (CH$_3$)$_2$O. 

 \subsection {Outflow and Compact Ridge molecule}

\subsubsection {Formaldehyde (H$_{2}$CO/H$_{2}^{13}$CO/HDCO)}	 \label{subsubsection h2co}

We detect three transitions from each of H$_2$CO and HDCO, and one transition from H$_{2}^{13}$CO. Since the energy range is  small (106\,--\,133 K), no rotation diagram can be made.
The  8$_{1,8}$\,--\,7$_{1,7}$ transition of H$_{2}$CO is shown in Fig.~\ref{H2CO} (on-line material), together with the same transition of H$_{2}^{13}$CO and the 9$_{1,9}$\,--\,8$_{1,8}$ transition of HDCO.
The H$_2$CO 8$_{0,8}$\,--\,7$_{0,7}$ transition shows a blend with Hot Core NS at 576.720 \mbox{GHz}. The 8$_{1,7}$\,--\,7$_{1,6}$ transition of H$_{2}$C$^{18}$O  is tentatively found at 571.477 GHz.

The H$_{2}$CO lines show two velocity components. Figure \ref{fig H2CO Two Gauss} shows a two-component Gaussian fit. The narrow component from the CR has widths of \mbox{$\sim$5 \kms}~at $\upsilon_{\mathrm{LSR}}\sim$8.5 \kms, and the broader component from the LVF has widths of $\sim$19 \kms~at $\upsilon_{\mathrm{LSR}}\sim$8 \kms.  H$_{2}^{13}$CO and HDCO show only  emission from the CR with similar widths and \mbox{LSR} velocities as for the narrow H$_{2}$CO component.
Comparison of the CR component of the H$_{2}$CO \mbox{8$_{1,8}$\,--\,7$_{1,7}$} transition  with the same H$_{2}^{13}$CO transition, results in optical depths of  $\sim$6.6 and $\sim$0.1, respectively (using  [$^{12}$C/$^{13}$C]\,=\,60). 

Since the CR component is optically thick in H$_{2}$CO, this
source size is calculated with Eq. \ref{source-size} and is found to be as large as 14$\arcsec$ for a temperature of 115 K, in agreement with Mangum \etal~(\cite{Mangum90}).
The LVF source size  becomes 10$\arcsec$, which might be caused by a low opacity. Hence a LVF size of 15$\arcsec$ is used for beam-filling correction.
The resulting CR and LVF 
column densities are 3.0$\times$10$^{15}$~cm$^{-2}$ and 4.3$\times$10$^{15}$~cm$^{-2}$, respectively. With the use of the optically thin H$_{2}^{13}$CO the CR column density increases to 2.0$\times$10$^{16}$~cm$^{-2}$,  in agreement with Turner (\cite{Turner90}), Mangum \etal~(\cite{Mangum90}), and S95.

Since H$_2$CO is optically thick we cannot calculate the [$^{12}$C/$^{13}$C] elemental ratio.
But with the use of the optically thin H$_{2}^{13}$CO and HDCO, the 
abundance ratio of D/H is estimated to $\sim$0.01, which implies a high deuterium fractionation in the CR.
Turner (\cite{Turner90}) derived a ratio of HDCO/H$_2$CO\,=\,0.14$^{+0.12}_{-0.07}$  and  D$_2$CO/HDCO\,=\,2.1$^{+1.2}_{-0.5}$\x10$^{-2}$ for the CR. These large abundance ratios were interpreted as a result of active grain surface chemistry.

\subsection {Outflow, Hot Core and Compact/Extended  Ridge molecules}
\subsubsection {Carbon monosulfide (CS/$^{13}$CS)}  

Fig. \ref{fig CS Three Gauss} shows a three-component Gaussian fit to the observed
$J$\,=\,10\,--\,9 transition of
CS.  Emission is seen from a narrow component, the HC and the LVF at LSR velocities   9, 7  and 10~\kms~with widths  4, 9 and 18 \kms, respectively.  The narrow component may have an origin either from the ER or the CR, hence the column density is calculated with both alternatives. 
The CS line is also compared  to  H$_2$CS and isotopologues of H$_2$CO  in Fig.~\ref{H2CO} (on-line material). 

The $^{13}$CS $J\!=\!12\!-\!11$ transition is observed  with emission from a narrow (ER or CR) component, 
but is blended with a $^{34}$SO$_2$ transition. This makes the Gaussian fit with a width of 5 \kms, at LSR velocity 7 \kms~approximate. Comparison of  peak antenna temperatures of the isotopologues (using a $^{12}$C/$^{13}$C ratio of 60) suggests that the narrow CS component is optically thick ($\tau \sim 6-12$).
The  source size of an ER component is calculated with Eq. \ref{source-size} and is found to be 30$\arcsec$ for at temperature of 60 K. This suggests that either the emission of this  component is rather extended and clumpy (see Sect. \ref{section Discussion}), or has  an origin in the CR. For a typical CR temperature of 115 K, we find a size of 20$\arcsec$. 
 
\begin{figure}
       \resizebox{\hsize}{!}{\includegraphics {7225fg15.eps}}  
 \caption{ The H$_2$CO 8$_{1,8}$\,--\,7$_{1,7}$ transition with a two-component Gaussian fit shown together with the individual Gaussians. The  line widths are 5 
 \kms~and 19 \kms~from the CR and LVF, respectively.}
 \label{fig H2CO Two Gauss}
\end{figure}


\begin{figure}
       \resizebox{\hsize}{!}{\includegraphics {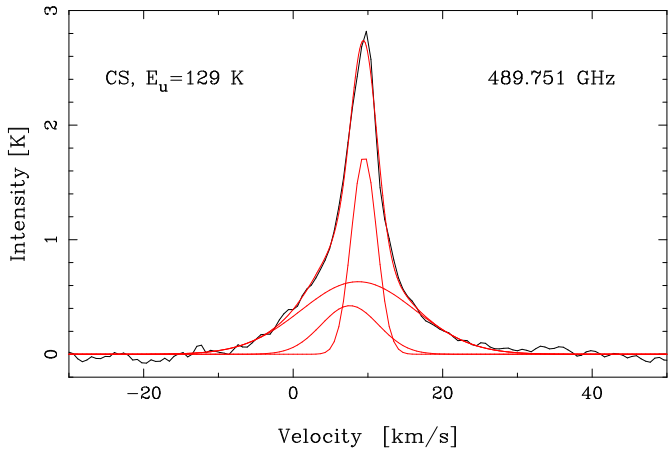}}  
 \caption{ The CS $J$\,=\,10\,--\,9 transition with a three-component Gaussian fit shown together with the individual Gaussians. The line widths are 4 \kms, 9 \kms, and 
 18 \kms~ from   the CR/ER, the HC and LVF, respectively.}
 \label{fig CS Three Gauss}
\end{figure}

The resulting column densities are listed in Table \ref{result_table1}. The column density of the LVF agrees well with B87, S95, S01 and C05, and the HC column also agrees with S95, but is lower than found in C05.
The narrow component from either the CR och ER is more difficult to compare. Our ER column agrees rather well with B87, but is an order of magnitude lower than found by S95.  As  CR emission it agrees with S95. The differences may arise due to opacity, beam-sizes and energy levels.

\subsubsection {Hydrogen isocyanide (HNC)}		

\begin{figure}
       \resizebox{\hsize}{!}{\includegraphics {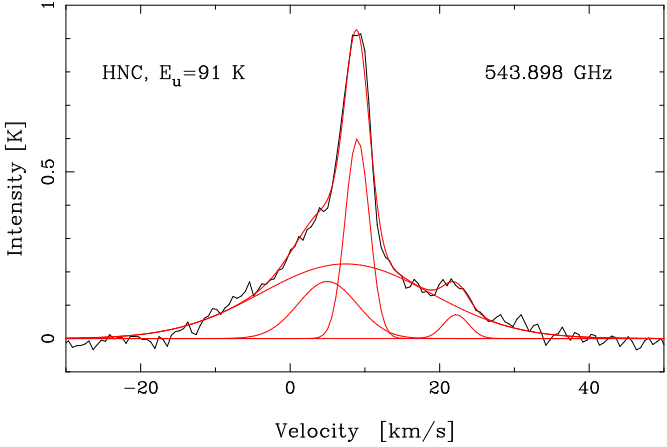}}  
 \caption{ The HNC $J$\,=\,6\,--\,5 transition and a U-line at 543.873 GHz with a four-component Gaussian fit shown together with the individual Gaussians. The line widths are
  4~\kms~for the U-line, 4~\kms, 9~\kms, and 27~\kms for the HNC ER, HC and outflow components, respectively.
}
 \label{fig HNC Three Gauss}
\end{figure}

Figure \ref{fig HNC Three Gauss} shows a four-component Gaussian fit of the HNC  \mbox{$J\!=\!6\!-\!5$} transition with an upper state energy   of 91 K,
and a U-line  seen in the red-ward LVF line wing  at a velocity of 22~\kms.
As for CS, three velocity components, from the ER, HC, and LVF, are clearly seen 
at $\upsilon_{\mathrm{LSR}}$= 9, 6 and 7 with widths of 4, 9 and 27 \kms, respectively.  The sizes and temperatures for the subregions are taken to be representative of typical  values (see Table \ref{result_table1}).
The HNC line  is also shown in Fig.~\ref{HNCcollection} (on-line material).

\subsection {PDR/Extended  Ridge and Hot Core molecule}

\subsubsection { The cyanide radical (CN)}

\begin{figure}
       \resizebox{\hsize}{!}{\includegraphics {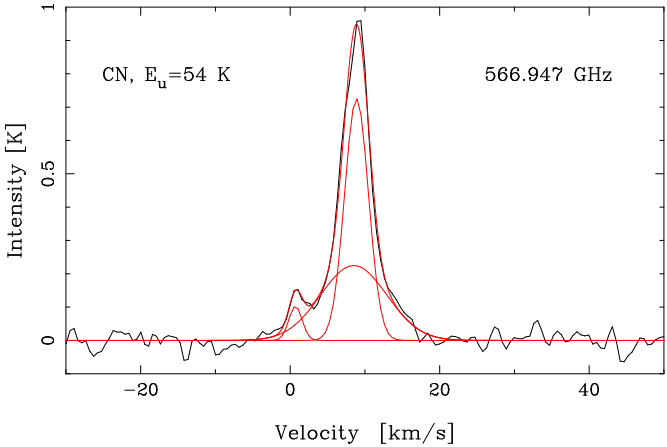}}  
 \caption{CN   with a three-component Gaussian fit of five unresolved hyperfine-structure lines shown together with the individual Gaussians. The widths are 4~\kms~and 11~\kms~for the PDR/ER and HC components of the three strong hyperfine-structure lines, respectively, and 2~\kms~for the weak transitions at a velocity of 0~\kms.}
 \label{fig CN Three Gauss}
\end{figure}

The main CN emission has its origin in the PDR/ER region and the HC  (Rodr\'iguez-Franco \etal~\cite{Rodriguez-Franco}) at $\upsilon_{\mathrm{LSR}}\sim9$ and 8.5~\kms~with widths of $\sim$4 and $\sim$10~\kms, respectively. 
In total we have observed three lines with 8 non-resolved hyperfine structure features. Figure  \ref{HNCcollection} (on-line material) shows one of the $N\!=\!5\!-\!4$ transitions, consisting of 
three non-resolved hyperfine structure lines, with two additional ones in the line wing at a velocity of 1~\kms. The same transitions are shown in  Fig.~\ref{fig CN Three Gauss} with a three-Gaussian fit of the five transitions.
No rotation diagram is made
since the upper state energy of 54 K is the same for all transitions.

Using   Gaussian fits, the LTE approximation and typical temperatures and source sizes, 
the column densities for the HC and PDR/ER  regions are estimated to be
7.9x10$^{15}$~\cmsq, and 4.9x10$^{13}$~\cmsq~for the HC and PDR/ER, respectively.
Our comparison surveys have no observations of CN, but Rodr\'iguez-Franco \etal~(\cite{Rodriguez-Franco01}) obtained column densities by CN mapping, ranging from \mbox{10$^{13}$~\cmsq}~in the Trapezium region to 10$^{14}$\,--\,10$^{15}$~\cmsq~in the Ridge region. 
S95 find $N$\,=\,1\x10$^{15}$~\cmsq~with a 14$\arcsec$ beam, and B87 also find the same CN column density 
with a 30$\arcsec$ beam. 

 \subsection { Extended  Ridge molecule}

\subsubsection {Diazenylium (N$_{2}$H$^{+}$)}

The diazenylium transition $J\!=\!6\!-\!5$ is shown in Fig.~\ref{HNCcollection} (on-line material). The width of $\sim$5~\kms~at \mbox{$\upsilon_{\mathrm{LSR}}\approx9$~kms$^{-1}$} indicates an ER origin of the emission, in agreement
with mapping of the $J$\,=\,1\,--\,0 transitions by Womack \etal~(\cite{Womack90}) and
Ungerechts \etal~(\cite{Ungerechts97}). The column density of \mbox{1.0$\times$10$^{12}$ cm$^{-2}$,} that we calculate using the simple LTE approximation, is much lower
than that found by Ungerechts \etal~(\cite{Ungerechts97}), 8.4$\times$10$^{12}$~cm$^{-2}$.


\subsection {Unidentified line features}

We observe 64 unidentified line features. Tentative assignments have been given to 26 lines, such as the first tentative detections of ND, and of the anion SH$^-$ (see Fig. \ref{TentativeLines}, Tables \ref{U-line parameters}, and \ref{T-line parameters}  in the on-line Material and Table 3 in Paper I).  There are 28 \mbox{U-lines}, i.e. clearly detected lines, and 36 \mbox{T-lines}, which means that they are only marginally visible against the noise or in a blend.
The tentative assignments also include the species SO$^+$,  CH$_3$CHO, CH$_3$OCHO, SiS, HNCO,  H$_2$C$^{18}$O, and a high energy HDO line. For details see \mbox{Paper {\sc{I}}}.



The strongest U-line is found at 542.945 GHz with a peak intensity of 140 mK. The line appears to  show emission from two components, probably the CR and the HC (see Fig. \ref{TentativeLines} in the on-line material).

\section {Carbon monoxide   (CO/$^{13}$CO/C$^{17}$O/C$^{18}$O),  carbon (C) and H$_2$ column densities} \label{section CO}

We have observed the $J\!=\!5\!-\!4$ transition of CO, $^{13}$CO, C$^{17}$O, and C$^{18}$O (Fig.~\ref{fig CO}).  The CO line is the most intense single line in our 42 GHz wide band.  
The FWZP (Full Width Zero Power) of CO is approximately 230  \kms, as compared to
120  \kms~reported in Wirstr\"om \etal~(\cite{Wirstrom}, hereafter W06), a result of our much lower noise level.
Since W06 also used Odin but with another observation mode, we can again demonstrate  the high accuracy of the Odin calibration with a comparison of the amplitudes, which
agree    within less than five percent.

As  pointed out in W06  it is clear that  CO $J\!=\!5\!-\!4$ has emission from at least three different components -- LVF, HVF  and a narrow component. The high brightness temperature of the last component  suggests that this emission originates in the extended
 and warm PDR, whereas the narrow components from 
the optically thin isotopologues have approximately equal
 emission from the PDR and the colder ER gas behind it. 
We observe all three  components in the CO and $^{13}$CO emission, but only the narrow ER/PDR component  and the LVF
for the C$^{17}$O 
and C$^{18}$O isotopologues  (W06).
The Gaussian components are given in  Tables \ref{result_tableCO} and \ref{CO parameters } in the on-line material, and agree well with W06, especially when our higher signal-to-noise ratio is 
taken into account,  which enables us to see line wings that were previously unobserved. 

A summary of the resulting column densities,  estimated optical depths, used source sizes and temperatures
is found in Table \ref{result_table2}, and also in more detail in the on-line
Table \ref{result_tableCO}. Here also column densities calculated from all isotopologues are given together with the parameters of the Gaussian fits. Note that the column density for the CO narrow component (calculated from C$^{17}$O) is lowered by a factor of two, since this component only has emission from the PDR, while the isotopologues have approximately equal emission from both the PDR and ER. For detailed arguments see W06.


The only observed atomic species in this survey is the \mbox{$^3P_1$\,--\,$^3P_0$} transition of C at 492.1607 GHz. It shows a narrow line profile from an extended emission with a width of 4.5 \kms, at   LSR velocity $\sim$9 \kms. Due to the loss of orbits during this observation, the noise level here is 200 mK, as compared to our average level of 25 mK in the rest of the spectral survey. This makes it impossible to distinguish a possible broad emission 
 in this transition.
Our beam-averaged column density of C is 5.6\x$10^{17}$ \cmsq. 
Tauber \etal~(\cite{Tauber}) find a lower limit for a beam-averaged C column density of $\sim$7\x$10^{17}$ \cmsq~ (beam size 17$\arcsec$) in the Orion bar.
Ikeda \etal~(\cite{Ikeda}) find a column density very similar to ours (6.2\x$10^{17}$ \cmsq) from observations of the 492.1607 GHz transition with the Mount Fuji submillimetre-wave telescope towards the Orion KL position, in a HPBW of 2$\arcmin$.2. The optical depth was estimated to be 0.2. B87 find $N\ga$7.5\x$10^{17}$ \cmsq~with a 30$\arcsec$ beam towards the Orion KL region. 
Plume \etal~(\cite{Plume}) presented maps of the same transition,
obtained with the SWAS satellite, resulting in an average column density of 2\x$10^{17}$ \cmsq. 

When estimating abundances we need comparison column densities of H$_2$ for each subregion (results also given in Table \ref{result_table2}). This is provided by  C$^{17}$O for the PDR/ER  and LVF components,
using [CO]/[H$_2$]=8\x$10^{-5}$ (e.g. Wilson \& Matteucci \cite{Wilson  Matteucci }),  an isotope ratio [$^{18}$O]/[$^{17}$O]\,=\,3.9 (Table \ref{isotope_ratios}), together with [$^{16}$O]/[$^{18}$O]=330 (Olofsson \cite{Olofsson thesis}). The latter  value was found from high S/N S$^{18}$O observations of molecular cloud cores. This is somewhat lower than the usually quoted value of 560  (Wilson \& Rood \cite{Wilson and Rood}),  valid for the local ISM and estimated from  H$_2$CO surveys in 1981 and 1985. A likely  
explanation for the lower value is a  
local enrichment of $^{18}$O relative 
to $^{16}$O by  the ejecta from massive stars.
For the  HVF component we use $^{13}$CO since C$^{17}$O has no HVF emission.

The resulting H$_2$ column density from the LVF is  3.2$\times10^{23}$ \cmsq.
This is close to the limits given by Masson et al (\cite{Masson}) (3\,--\,10)$\times10^{22}$ \cmsq, 
as well as 1$\times10^{23}$ \cmsq~by Genzel \& Stutzki (\cite{Genzel and Stutzki}).
Wright  \etal~(\cite{Wright 2000}) find a beam-averaged  H$_2$ column density of 2.8$\times10^{23}$ \cmsq~from observations of the  28.2 $\mu$m  H$_2$ 0--0 S(0) line for a temperature of 130 K  (beam size 20$\arcsec\times33\arcsec$).

Our resulting HVF H$_2$ column density is 3.9$\times10^{22}$ \cmsq, in agreement with the Genzel \& Stutzki value of 5$\times10^{22}$ \cmsq. In contrast, Watson \etal~(\cite{Watson}) found that the HVF column of warm shock heated H$_2$ is only 3$\times10^{21}$ \cmsq, a result based upon their KAO observation of high-$J$ CO lines.

Tauber  \etal~(\cite{Tauber}) reported an average  H$_2$ column density of $\sim$3$\times10^{22}$ \cmsq~from the Orion Bar (calculated from  $^{13}$CO mapping). This is in agreement with  our total narrow component, which we find to be 4$\times10^{22}$ \cmsq.
When we calculate the abundances in Sect.  \ref{section Abundances}, we divide this value by two, since there are about equal contributions from the PDR and ER to the C$^{17}$O emission (W06). Our value is also consistent with the results of Goldsmith  \etal~(\cite{Goldsmith97}) convolved with the 2$\arcmin$.1 Odin beam.




\begin{figure}
       \resizebox{\hsize}{!}{\includegraphics[angle=-90]{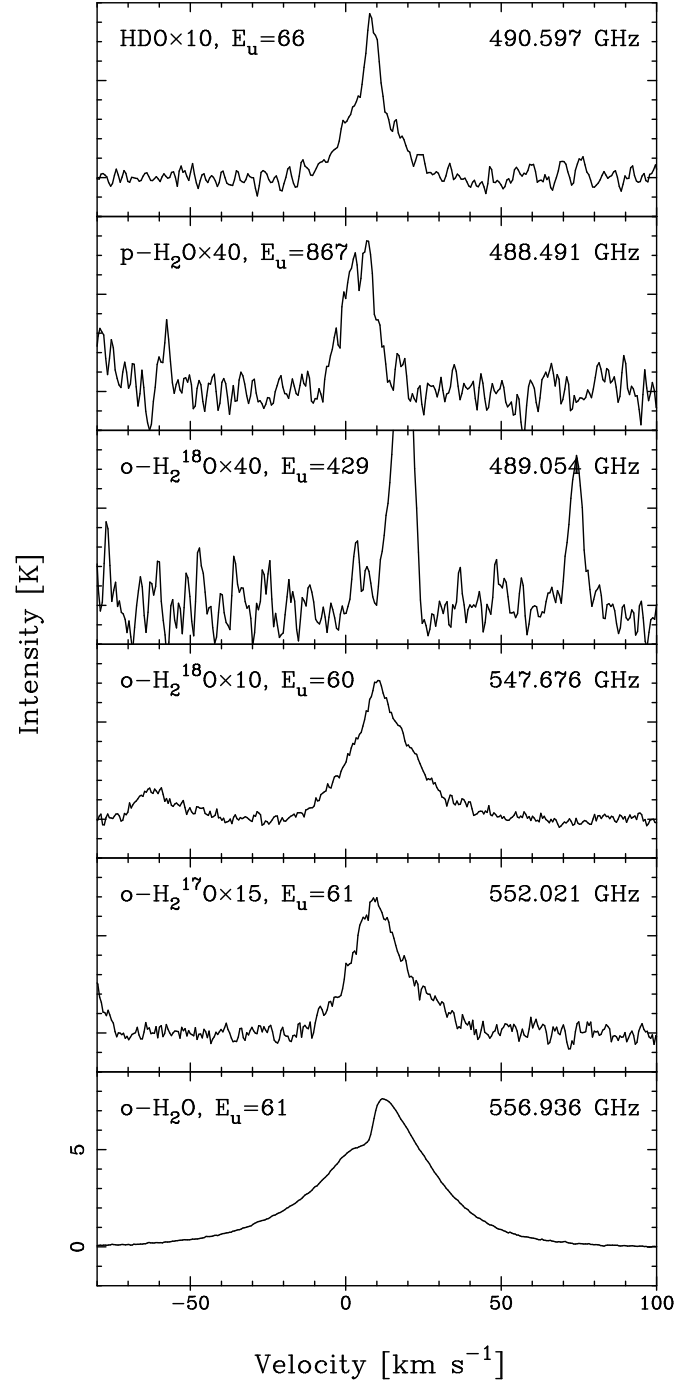}}   
 \caption{Water and isotopologues. The  $o$-H$_2^{17}$O is reconstructed spectra with blending lines subtracted. An  intensity scale factor is given after the molecular species.}
  \label{fig vatten_Nobel}
\end{figure}

\section {Water  ($o$-H$_2^{16}$O,  $p$-H$_2^{16}$O, $o$-H$_2^{17}$O, $o$-H$_2^{18}$O, HDO)}\label{section Water}

\subsection {Correcting the water emission lines for blends}\label{section Blends}

Because of the large number of methanol and sulphur dioxide  lines observed, they cause the most common blends in all lines. Since we are particularly interested in water, we attempt to reconstruct the water isotopologues without blends. We use observed transitions in our survey with similar parameters ($E_u$, $A$-coefficient and $g_u$), and scale them  with the parameters of the blending lines  before removal from the water isotopologue line of interest. The molecular line parameters of the blending transitions are found in the on-line Tables (\ref{SO2 parameters },  \ref{34SO2 parameters } and  \ref{CH3OH parameters 1 }).

Two lines are blended with  the $o$-H$_2^{18}$O  $1_{1,0}$\,--\,$1_{0,1}$ ground state rotational transition. 
The $^{34}$SO$_2$ $21_{3,19}$\,--\,$20_{2,18}$  transition is blended with  the  red wing, and 
in the blue wing there is an overlapping methanol line, $15_{1,15}$\,--\,$15_{2,14}$, $v_t$=1. 
However, since the simultaneous observations of $o$-H$_2^{17}$O show that the H$_2^{18}$O  transition is optically thick  even in the line wings (see next section), we do not attempt to remove these blending transitions.

The $o$-H$_2^{17}$O ground state rotational transition is, however, optically thin in the line wings, and we therefore   remove  three blending lines.
In  Fig.~\ref{H217O with blending lines before removal} (on-line material) we show two of the blending lines together with the  $o$-H$_2^{17}$O line. 
In the blue \mbox{$o$-H$_2^{17}$O} wing the blending SO$_2$ transition $26_{\,6,20}$\,--\,$26_{\,5,21}$ is overlapping. 
In the red wing there are two blends. One from the $6_{\,6,1}$\,--\,$7_{5,3}$ methanol transition shown, and one  from the very weak  SO$_2$ $34_{1,33}$\,--\,$34_{0,34}$ transition.

\subsection{Water analysis}  

We have observed the \mbox{1$_{1,0}$\,--\,1$_{0,1}$} ground state rotational transition   of $o$-H$_2$O and its isotopologues $o$-H$_2^{18}$O and $o$-H$_2^{17}$O, which mainly show emission from the Plateau. A very weak feature at 489.054 GHz is tentatively identified as the HC-tracing \mbox{4$_{2,3}$\,--\,3$_{3,0}$} transition of $o$-H$_2^{18}$O with an upper state energy of 429 K. The HC-tracing $p$-H$_2$O transition \mbox{6$_{2,4}$\,--\,7$_{1,7}$} with an upper state energy of 867 K,  is also observed, as well as the \mbox{2$_{0,2}$\,--\,1$_{1,1}$} HDO transition showing emission from the CR, HC and LVF.~Figure \ref{fig vatten_Nobel} shows all detected water lines after removal of some blends in H$_2^{17}$O as discussed in the previous section.

The optical depths, column densities, 
assumed source sizes and excitation temperatures are found in Table  \ref{result_table2}, and   in more detail in Table \ref{result_tableWater} (on-line material), together with the
  parameters of the Gaussian fits.


Both the $o$-H$_2^{18}$O  and  $o$-H$_2^{17}$O ground state rotational transition
show features of a weak, narrow component from the CR, 
a broad stronger component from the LVF, and with HVF emission mainly in the red wing. Figures \ref{3G fit to h2o17} and \ref{3G fit to h2o18} (on-line material)  show  three-component Gaussian fits  to the water isotopologues.
The emission from the ER and PDR is considered to be very low since 
the water mainly  will be frozen onto the dust grains at the rather low temperatures in this region. 

The $o$-H$_2$O line is very optically thick at all velocities as seen in Fig.~\ref{Extra fig h2o_over_h2o18_zoom50_narrow}  (on-line material) which displays the ratio of \mbox{$o$-H$_2$O} and \mbox{$o$-H$_2^{18}$O.}
The excitation conditions for these two isotopologues are therefore very different. A similar figure of the ratio of $o$-H$_2^{18}$O and $o$-H$_2^{17}$O   (Fig.~\ref {fig h2o18overh2o17_zoom_narrow} in the on-line material) shows an almost constant ratio of 1.5 for velocities between $-$10 and +30 km s$^{-1}$. This demonstrates that the two line profiles are almost identical, and that the  $o$-H$_2^{18}$O emission is rather optically thick at all velocities since  [$^{18}$O/$^{17}$O]\,=\,3.9. 
By comparison of column densities from the total integrated intensities, the optical depths for  $o$-H$_2^{17}$O and  $o$-H$_2^{18}$O, are 0.9 and 3.4, respectively.
The small changes of the ratio with velocity as seen in Fig.~\ref {fig h2o18overh2o17_zoom_narrow} also are consistent with our decomposition into Gaussian components. The LVF is optically thick in both isotopologues,  
whereas the HVF and CR have lower optical depths causing increase of the ratio at their emission velocities.


\begin{figure}
   \resizebox{\hsize}{!}{\includegraphics[angle=-90]{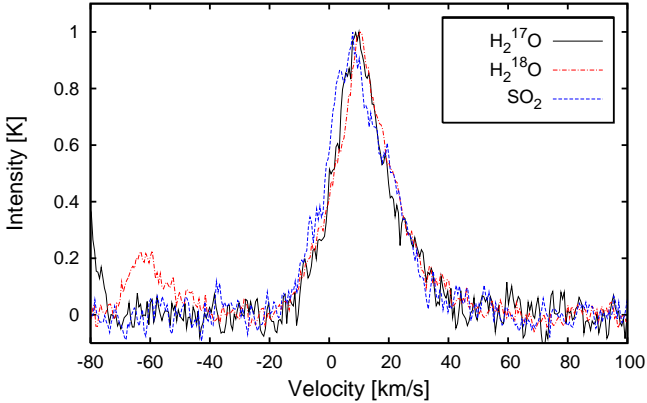}}  
\caption{Comparison of H$_2^{18}$O, with H$_2^{17}$O, and the 19$_{3, 17}$\,--\,18$_{2, 16}$\  SO$_2$ transition, all normalised with respective peak temperature. The high 
degree of similarity of all three line profiles suggests an origin in the same gas and velocity fields, mainly from the LVF with additional
emission from HVF in the red wings.}
  \label{fig H218O_H217O_SO2}
\end{figure}

\begin{figure}
   \resizebox{\hsize}{!}{\includegraphics[angle=-90]{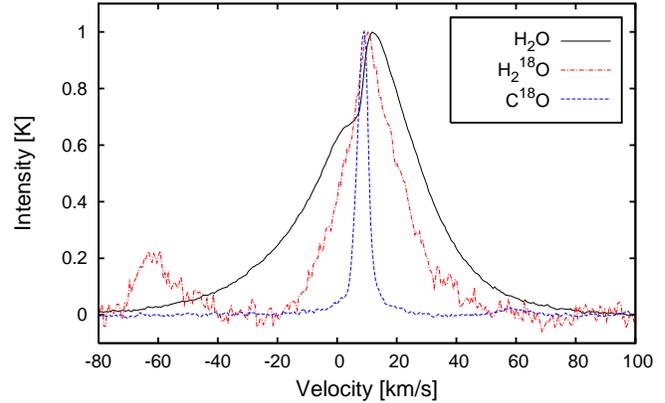}}  
 \caption{Comparison of the optically thick self-absorbed  $o$-H$_2$O with  $o$-H$_2^{18}$O and C$^{18}$O $J$\,=\,5\,--\,4, all 
 normalised with respective peak intensity. The LVF self-absorption of  $o$-H$_2$O in the blue wing is seen when compared to   
 $o$-H$_2^{18}$O, which 
 shows LVF emission in both wings and HVF emission mainly in the red wing. }
  \label{fig H2O_H218O_C18O}
\end{figure}

\subsubsection{{\emph{Ortho}}-H$_2$O from the Low- and High Velocity Flow and the Compact Ridge}

The similarity of the line profiles is also illustrated in Fig.~\ref{fig H218O_H217O_SO2} showing a comparison of  $o$-H$_2^{17}$O,  $o$-H$_2^{18}$O and the  19$_{3, 17}$\,--\,18$_{2, 16}$  SO$_2$  transition, all normalised to their respective 
peak temperature. The  remarkable similarity of the line profiles  suggests a very similar origin and chemistry of the water isotopologues and SO$_2$:   the LVF and with additional HVF emission mainly in the red wings.

The resulting   column densities (found in Table \ref{result_table2} and in the on-line Table \ref{result_tableWater}) are calculated with the simple LTE approximation and for an $ortho/para$ ratio of 3.
As a first approximation of the column density  we have  used the full integrated intensity of the lines, assuming the Plateau to be the main emitting source (with $T_\mathrm{ex}$\,=\,72 K and a source size of 15$\arcsec$, see below). We have also calculated the column densities for the different subregions using the Gaussian components.
In addition, the   \mbox{$o$-H$_2^{17}$O} and  \mbox{$o$-H$_2^{18}$O} column densities  are  calculated with
and without optical depth corrections. Since the  \mbox{$o$-H$_2$O} transition is highly optically thick, we calculate the column density from  \mbox{$o$-H$_2^{17}$O}  and  \mbox{$o$-H$_2^{18}$O}.
With  isotope ratios of [$^{18}$O]/[$^{17}$O]=3.9 and  [$^{16}$O]/[$^{18}$O]=330 (Table \ref{isotope_ratios}), we determine the   opacity-corrected column density of H$_2$O to be  1.7$\times10^{18}$ \cmsq. The opacity-corrected LVF and HVF column densities, obtained from the Gaussian fits of the isotopologues, are 8.7$\times10^{17}$ \cmsq~and 8.8$\times10^{17}$ \cmsq, respectively.
These calculations assume LVF and HVF source sizes for the isotopologues of 15$\arcsec$ (cf. Olofsson  \etal~\cite{Olofsson03}), which is the same extent as the submillimetre HDO 
emission from the LVF mapped by Pardo  \etal~(\cite{Pardo}). As excitation temperature for both LVF and HVF we use 72 K as found by Wright \etal~(\cite{Wright 2000}).
We also calculate the H$_2$O  HVF column density from the Gaussian fit to  H$_2$O itself, and with opacity-correction (calculated from the isotopologues) almost the same value is obtained as from the isotopologues. 
The size of the H$_2$O HVF is assumed to have an extent of 70$\arcsec$ (Olofsson  \etal~\cite{Olofsson03},  Hjalmarson  \etal~\cite{Hjalmarson05}). 

The opacity-corrected column density for the H$_2$O  CR is 5.6$\times10^{17}$ \cmsq.
For the CR we use the temperature and size obtained from our CH$_3$OH rotation diagram of 115 K and 6$\arcsec$. This is also consistent with our calculation of the excitation temperature from the optically thick  H$_2^{18}$O CR component, if a source size of 6$\arcsec$ is assumed. 

\begin{figure}
       \resizebox{\hsize}{!}{\includegraphics[angle=-90]{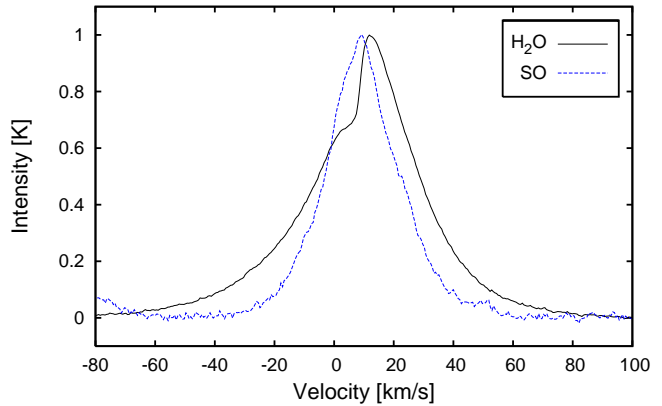}} 
 \caption{Comparison of the 13$_{13}$\,--\,12$_{12}$ SO transition and $o$-H$_2$O, both normalised with respective peak 
 intensity. Both species 
 show emission from the LVF and the HVF, but the water LVF is self-absorbed in the blue-shifted emission. }
  \label{fig SO_and_H2O}
\end{figure}

The $o$-H$_2$O line  has a central asymmetry that suggests strong self-absorption in the blue LVF by lower excitation foreground gas. The steep change in the self-absorption occurs in the velocity range of 2 to 12 \kms.
Fig.~\ref{fig H2O_H218O_C18O} compares the   self-absorbed \mbox{$o$-H$_2$O} transition  both with $o$-H$_2^{18}$O, and with the narrow emission from the C$^{18}$O $J$\,=\,5\,--\,4 line, all normalised to their respective peak temperature.
The LVF self-absorption of $o$-H$_2$O in the blue wing is seen when compared to  $o$-H$_2^{18}$O, which 
shows LVF emission in both wings, and HVF emission mainly in the red wing.
Fig.~\ref{fig SO_and_H2O} shows a similar comparison between $o$-H$_2$O and an optically thick SO transition at 559.320~GHz. Both species display emission from LVF and HVF, although the blue water LVF emission is self-absorbed.

\begin{figure}
       \resizebox{\hsize}{!}{\includegraphics[angle=-90]{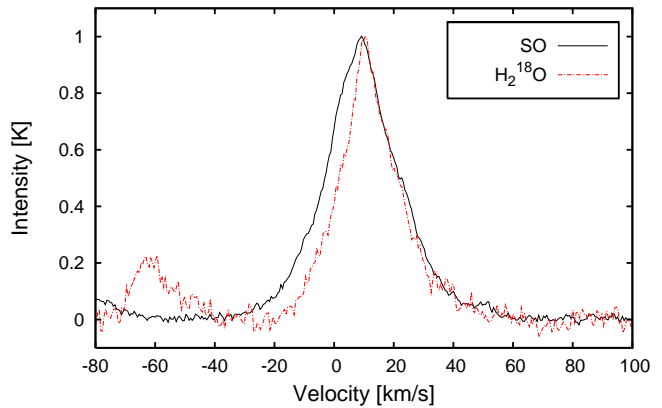}} 
 \caption{Comparison of the 13$_{13}$\,--\,12$_{12}$ SO transition,  and  $o$-H$_2^{18}$O normalised with 
 respective peak 
 intensity. 
}
  \label{fig SO_H218O}
\end{figure}

In Fig.~\ref{fig SO_H218O} we show  the same SO 13$_{13}$\,--\,12$_{12}$ transition again, but this time compared to 
$o$-H$_2^{18}$O.
In the blue wing 
SO has excess emission as compared to the water emission, whereas 
the  red  wings of SO and $o$-H$_2^{18}$O are almost identical.
This might be caused by shock chemistry in the red HVF which is pushing into the ambient molecular cloud (Genzel \& Stutzki \cite{Genzel and Stutzki}),  thereby producing a high water abundance. In the blue HVF, which is leaving the molecular cloud, no such shock chemistry seems to be present. The water abundance in this part of the HVF is likely due to evaporation from  icy dust grains, which produces less water than shock chemistry. In contradiction to this, the SO emission is symmetric in both wings, suggesting that shock chemistry is not required to produce high SO abundances. 

\begin{figure}
       \resizebox{\hsize}{!}{\includegraphics[angle=-90]{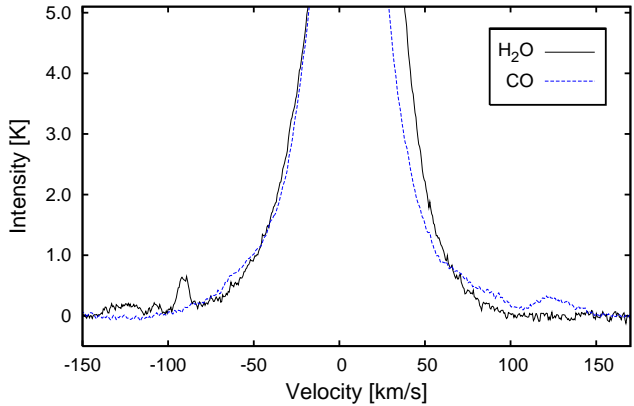}} 
 \caption{Comparison of the broad HVF emission in the line wings of CO and $o$-H$_2$O scaled 2.5 times. Note the large 
 velocity scale.}
  \label{fig CO_H2OLVFcompzoom5}
\end{figure}

\begin{figure}
   \resizebox{\hsize}{!}{\includegraphics{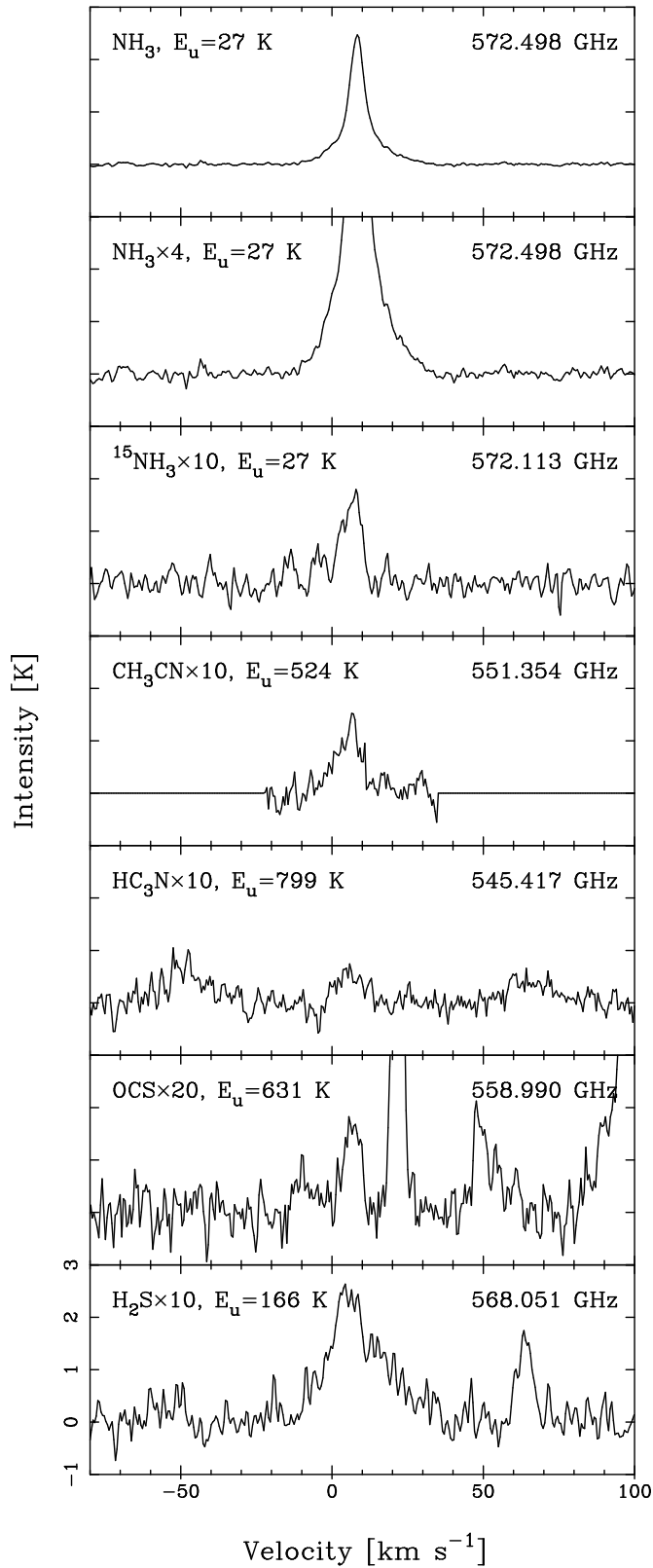}}  
 \caption{Top figure shows the NH$_3$ line, the second figure shows four times magnified NH$_3$ line wings. The NH$_3$ line
 profile shows
 emission from the CR and HC. The $^{15}$NH$_3$, 
 CH$_3$CN, HC$_3$N and OCS transitions suggest emission from the HC, and  H$_2$S from both the HC and LVF.  An  intensity 
 scale factor is given after the 
 molecular species.}
 \label{NH3 collection figure}
\end{figure}

\begin{figure} 
       \resizebox{\hsize}{!}{\includegraphics[angle=-90]{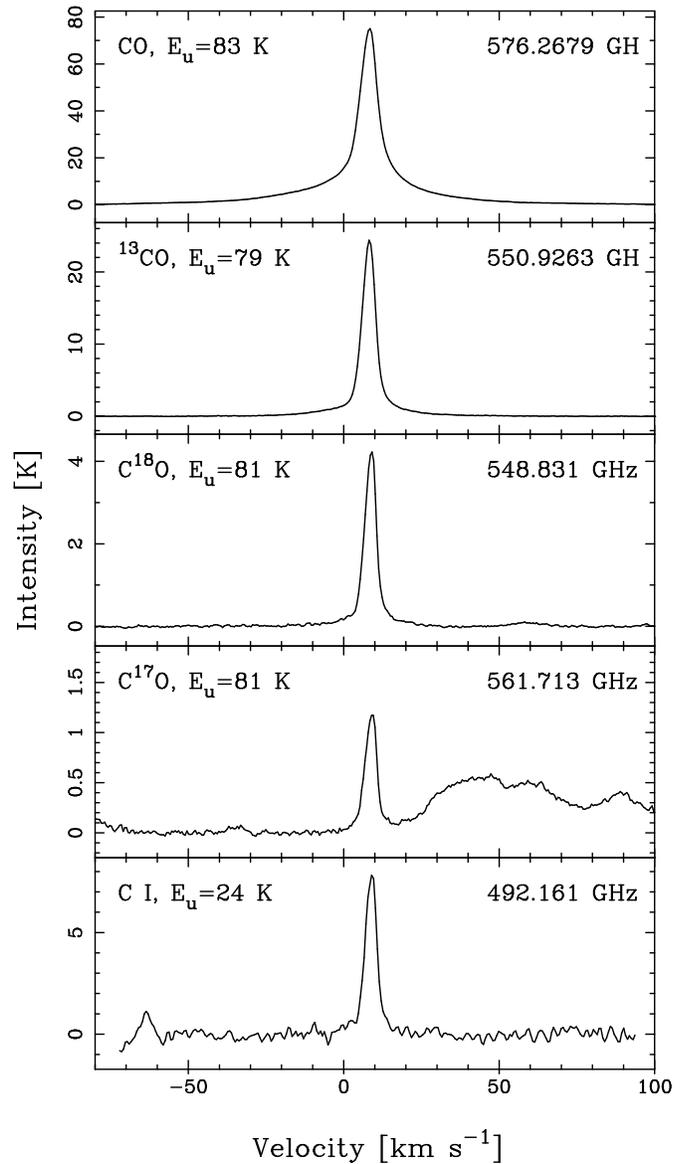}}  
 \caption{CO,  isotopologues and atomic C. Note the different intensity scales.}
 \label{fig CO}
\end{figure}

\begin{figure}
    \resizebox{\hsize}{!}{\includegraphics[angle=-90]{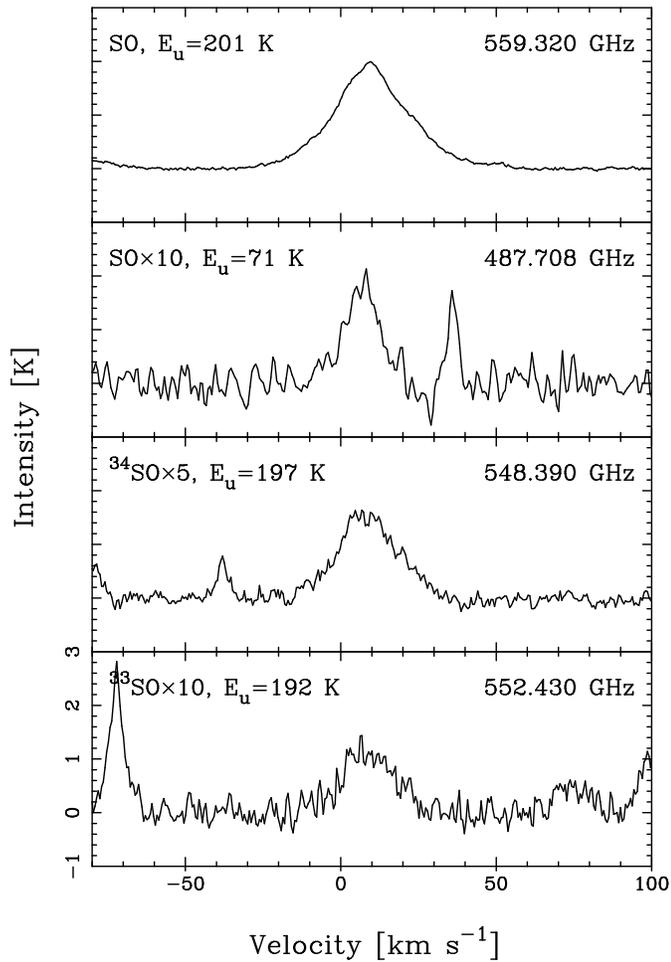}}  
 \caption{SO and isotopologues.  The SO line in the top panel is optically thick, and shows clear HVF line wings. The low-energy SO transition is optically thin.
 An  intensity scale factor is given after the molecular species.}
  \label{figureSO}
\end{figure}

\begin{figure}
   \resizebox{\hsize}{!}{\includegraphics{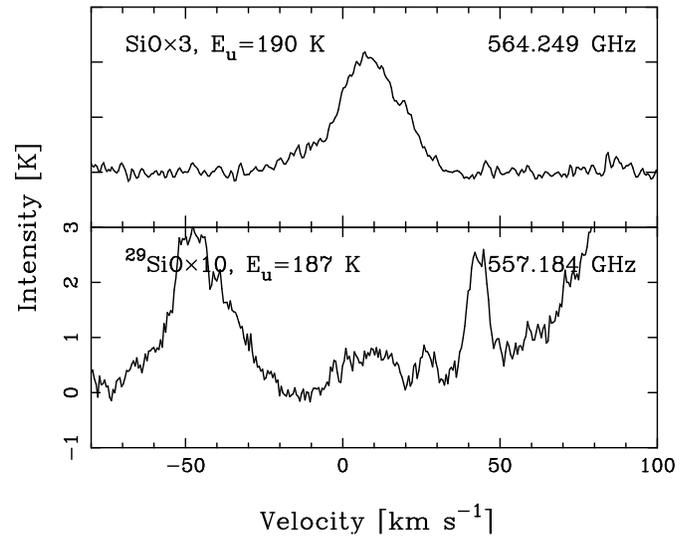}}  
\caption{The $J$\,=\,13\,--\,12 transition for both SiO and $^{29}$SiO. The SiO line is optically thick and also 
exhibits pronounced HVF line wings. A blend from CH$_3$OH and  $^{13}$CH$_3$OH is visible at 19.6 \kms~in the SiO line.  An  intensity scale factor is given after the molecular species.}
  \label{SiO and isotope}
\end{figure}

\begin{figure}
    \resizebox{\hsize}{!}{\includegraphics{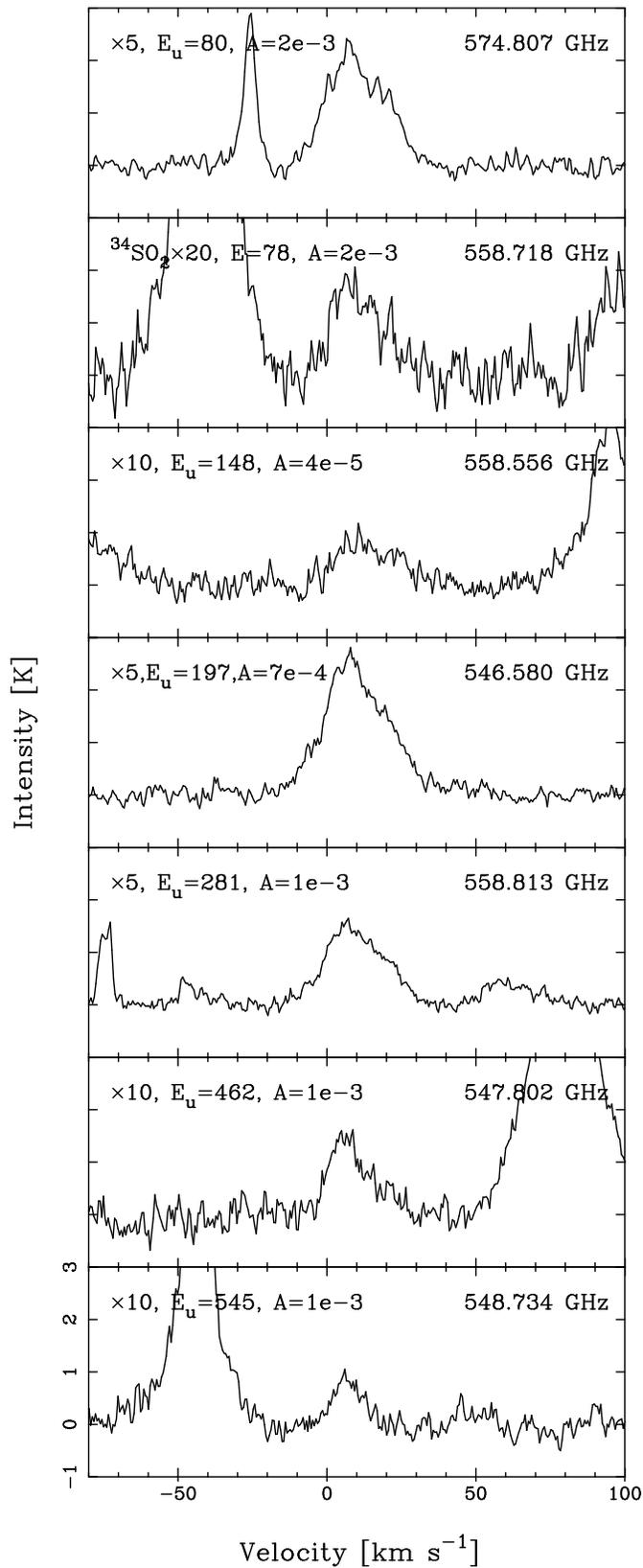}}  
 \caption{SO$_2$ with different upper state energy levels and A-coefficients.
 The $^{34}$SO$_2$ line is the same transition as the SO$_2$ transition 
 with $E_u$=80 K. 
An  intensity scale factor is given in the top left corners.}
  \label{figureSO2}
\end{figure}

\begin{figure} 
    \resizebox{\hsize}{!}{\includegraphics{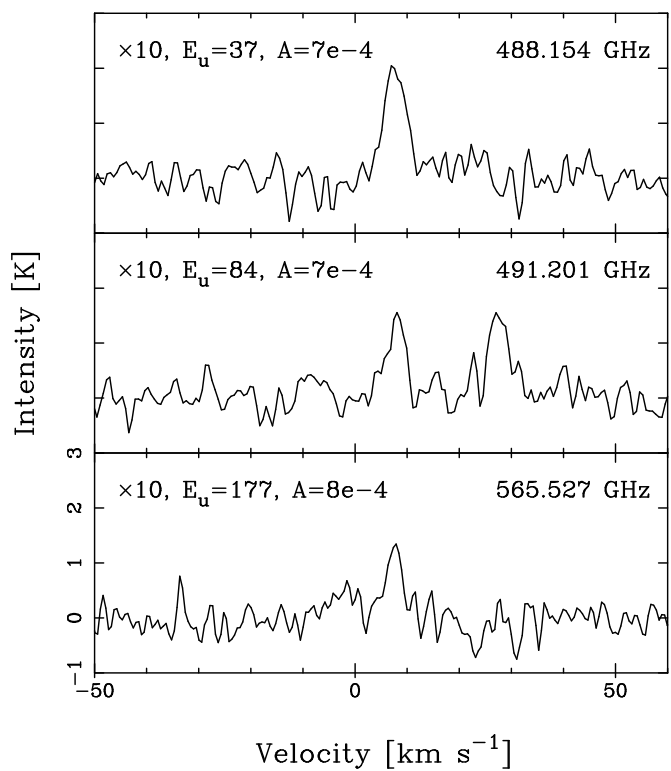}}  
 \caption{$^{13}$CH$_3$OH with emission from the CR. An  intensity scale factor is given in the top left corners, upper state 
 energies given in K and $A$-coefficients in s$^{-1}$.}
 \label{figure13CH3OH}
\end{figure}

\begin{figure}
   \resizebox{\hsize}{!}{\includegraphics{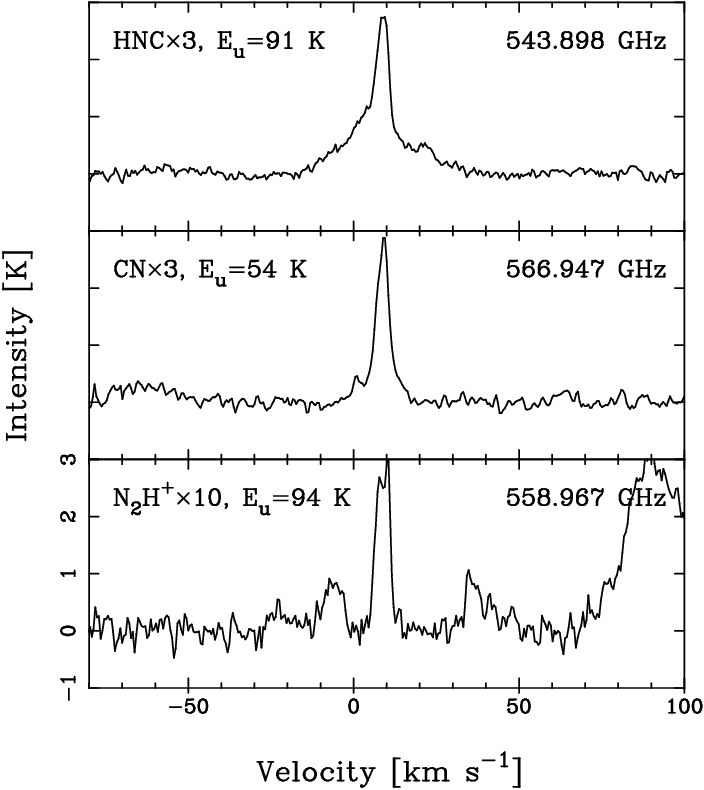}} 
 \caption{The line profile of HNC shows emission from the ER, HC and LVF,  CN from the PDR/ER and HC, and  
 N$_2$H$^+$ from the ER. The line at v$\approx-$8 \kms~next to N$_2$H$^+$ is OCS, which is also shown in 
 \mbox{Fig. \ref{NH3 collection figure}}. 
An  intensity scale factor is given after the molecular species.}
  \label{HNCcollection}
\end{figure}

The similarity of the broad HVF emissions  from CO and H$_2$O is shown in Fig.~\ref{fig CO_H2OLVFcompzoom5}.  The FWZP of the broad component  is  $\sim$230  \kms~for $o$-H$_2^{16}$O, 50  \kms~for the isotopologues, and 35 \kms~for HDO.

\subsubsection{{\emph{Para}}-H$_2$O from the Hot Core} \label{section p-h2o from HC}

In the main $o$-H$_2$O  1$_{1,0}$\,--\,1$_{0,1}$ line spectrum (Fig. \ref{fig vatten_Nobel}) possible  emission from the HC and CR would be blended with the much stronger and broader component from the LVF. Melnick  \etal~(\cite{Melnick}) concluded that the HC contributes negligibly to the water emission within the SWAS beam, and the CR would contribute less than 5\,--\,10\%. The highest energy levels in the ISO data presented by Lerate  \etal~(\cite{Lerate}) and Cernicharo  \etal~(\cite{Cernicharo}) may have a contribution from the HC, but  those authors remark that the large far-IR line-plus-continuum opacity  probably would hide most of this emission.


However, our detected optically thin $p$-H$_2$O 6$_{2,4}$\,--\,7$_{1,7}$ transition with upper state energy 867 K, clearly reveals the  emission from the HC. The width of the line is 12 \kms~at $\upsilon_{\mathrm{LSR}}\sim$4.4 \kms, between velocities $-$8 and 15 \kms. 
When column densities are calculated using the simple LTE approximation,
we find values in the range (3.7\,--\,12)$\times10^{18}$ \cmsq~for a temperature range of 200\,--\,500 K and a typical source size of 10$\arcsec$. 
Since no clear HC emission is seen from the ground rotational state transition in   H$_2^{17}$O, either the temperature in the HC is high enough  to result in the negligible  emission of this transition from the HC,  or the H$_2^{17}$O transition has an optically thick LVF emission which is blocking possible HC emission. If the temperature is 500 K, the emission from the HC would only be about 10\% of the total (if no optical depths are taken into account), and therefore would  be hidden in the stronger and broader LVF  emission even if H$_2^{17}$O is optically thin. However, if the temperature in the HC is about 200 K, the  H$_2^{17}$O LVF emission has to have an optical depth larger than unity. Since our estimated value is about 1.5, this is consistent with a HC temperature of  200 K and a column density of HC H$_2$O 1.2$\times10^{19}$ \cmsq.  This is in agreement with the
only previous observation of H$_2$O emission from the HC made by S01. They detected the vibrationally excited \mbox{1$_{1,0}$\,--\,1$_{0,1}$} $v_2=1$ transition and found a HC column density of 3$\times10^{19}$ \cmsq~with a temperature of 200 K and a water abundance of 1$\times10^{-5}$. Gensheimer \etal~(\cite{Gensheimer}) estimated the HC H$_2^{18}$O column density to be 2.7$\times10^{16}$ \cmsq~from H$_2^{18}$O mapping of the quasi-thermal \mbox{3$_{1,3}$\,--\,2$_{2,0}$} transition with the IRAM telescope (12$\arcsec$ beam). This column density translates to 
1.3$\times10^{19}$ \cmsq~for H$_2$O (corrected for source size and isotopologue ratio differences), in excellent agreement with our result, even though their observations suffered from severe blends.
Their HC water mapping also
showed that the emission from both H$_2^{18}$O and HDO were unresolved by their beam, and the HDO mapping showed sizes of 6\,--\,8$\arcsec$.

Our tentative detection of the optically thin HC emitting \mbox{$o$-H$_2^{18}$O} \mbox{4$_{2,3}$\,--\,3$_{3,0}$}
 transition,  with an upper state energy of 430 K, results in a HC  H$_2$O   column density of about 4$\times10^{19}$ \cmsq.

 \subsubsection{Comparisons of outflow column densities}
 
Our column density results for the Plateau agrees excellently with those of Wright  \etal~(\cite{Wright 2000}), who 
observed 19 pure rotational lines in absorption using the Short Wavelength Spectrometer (SWS) on board the Infrared Space Observatory (ISO), with a beam size of 14$\arcsec\times20\arcsec$ to 20$\arcsec\times30\arcsec$. Their rotation diagram, including a generalised curve-of-growth method, results in a rotation temperature of 72 K, and a beam-averaged column density (from the total emission) of 1.5$\times10^{18}$ \cmsq. They also conclude that the observed water arises from an outflow centred near IRc2.

Lerate  \etal~(\cite{Lerate}) observed more than 70 far-IR pure rotational H$_2^{16}$O lines, and 5    H$_2^{18}$O lines, with the ISO LWS (Long Wavelength Spectrometer) between $\sim$43 and 197 $\mu$m (beam size about 80$\arcsec$). Their rotation diagram of the total emission from H$_2^{18}$O results in a  beam-averaged column density of (2\,--\,5)$\times10^{14}$ \cmsq, and a rotation temperature of 60 K. For a 15$\arcsec$ outflow source, their result translates to  (1.8\,--\,4.7)$\times10^{18}$ \cmsq~for H$_2^{16}$O, also in good agreement with our value. In the analysis by
Cernicharo  \etal~(\cite{Cernicharo}) of this data set,
they concluded that most of the water emission and absorption arises from an extended flow of gas with velocity 25$\pm5$ \kms, with an inferred kinetic gas temperature of 80\,--\,100 K.

\subsubsection{HDO from the Compact Ridge, Low Velocity Flow and the Hot Core}

The deuterated species HDO is observed in the  \mbox{2$_{0,2}$\,--\,1$_{1,1}$}  transition with $E_u$\,=\,66 K, and 
shows  evidence of CR, LVF, and also HC  emission as is observed by W03. Figure \ref{3G fit to hdo} shows a 
three-component Gaussian fit.
Pardo \etal~(\cite{Pardo}) reported detections of the  \mbox{2$_{1,2}$\,--\,1$_{1,1}$} and  \mbox{1$_{1,1}$\,--\,1$_{0,0}$} lines in the 850\,--\,900 \mbox{GHz} spectral region, which seem to trace the Plateau gas and not the HC. Their conclusion is that the HC component is hidden behind the optically thick HDO LVF in their transitions, which is supporting our  analysis of the non-detection of HC emission in the ground state rotational water transitions.


A T-line at 559.816 GHz  tentatively is identified as the high energy (580 K) HDO transition 6$_{2,4}$\,--\,6$_{2,5}$.

Our estimated  column density of HDO, assuming that  the main emission  originates in the outflows, is 9.1\x$10^{15}$ \cmsq, which is about the same value as reported by Lerate  \etal~(\cite{Lerate}),  $8.5\times10^{15}$  \cmsq~(corrected for source size).  The column density reported by Pardo  \etal~(\cite{Pardo}) is higher, 5$\times10^{16}$   \cmsq, calculated from an  LVG model with a source size of 15$\arcsec$, and opacities of 3.7 and 6.7 for their two
 lines.

We  also calculate the column densities for the CR, LVF and HC separately, which are found to be 1.8$\times10^{16}$  \cmsq, 
4.5$\times10^{15}$  \cmsq, and 1.5$\times10^{16}$  \cmsq, respectively. With correction for source-size, this is 2\,--\,4 times lower than found in Olofsson (\cite{HansOlofsson84}) by mapping the 1$_{1,0}$\,--\,1$_{1,1}$ transition with a 47$\arcsec$ beam (at Onsala Space Observatory); 3$\times10^{16}$  \cmsq, 
1$\times10^{16}$  \cmsq, and 7$\times10^{16}$  \cmsq, respectively. 

\subsubsection{Molecular abundance ratios}

The  [D/H] ratio calculated from from HDO/H$_2$O is 0.005, 0.001 and 0.03 in the LVF, HC, and CR, respectively (Table \ref{isotope_ratios}). The CR ratio may be compared to the  HDCO/H$_2^{13}$CO ratio from which we obtain a similar [D/H] value of 0.01 (see Sect. \ref{subsubsection h2co}).
Lerate  \etal~(\cite{Lerate})  found  [D/H] values in the range  0.004\,--\,0.01. 

The column densities for H$_2$O and  H$_2$S from the  LVF  are also used to estimate the molecular abundance ratio of O/S to  $\sim$20 (Table \ref{isotope_ratios}). Using the H$_2$CS/H$_2^{13}$CO ratio we obtain a similar value of $\sim$15 (see Sect. \ref{subsubsection h2cs}). 


\section {Molecular abundance estimates} \label{section Abundances}

\subsection{Gas-phase abundances from the Odin spectral line survey}

Our estimated abundances for each subregion are summarised in Table \ref{abundances},  together with comparison abundances mostly from B87 and S95. We find very high gas-phase
abundances of H$_2$O, NH$_3$,  SO$_2$, SO, NO, and CH$_3$OH. Note that both our  LVF and HC abundances are source averages. 
S95 use beam-averaged abundances (with a HPBW of 14$\arcsec$), and
B87 use  a HC size of 10$\arcsec$ and a Plateau size of 20$\arcsec$,  while we separate the LVF and HVF emissions with a slightly smaller size for the LVF (15$\arcsec$). The beam-averaged CR abundances in both B87  and S95 are corrected to our  source-sizes to allow an easier comparison.

  \begin{table} 
\caption{Derived abundances  and comparisons.} 
\label{abundances}
\begin{tabular} {l l l l l}
\hline
\hline
Region &Species 	&	X$^{\,g}$ & B87 X$^{\,h}$ & S95 X$^{\,i}$ \\
&&  [$\times10^{7}$] &  [$\times10^{7}$]&  [$\times10^{7}$]\\

\hline
\hline
LVF$^{\,a}$     &H$_{2}$O  &  29	&	     \\  
&HDO  			&    0.15  	&	0.17   \\ 
 &SO$_{2}$		   & 20 	&	5.2   &  1.3 \\
&SO 	 			 &   3.1 &	 5.2 & 2  \\ 
&SiO 				&     0.11 	&	0.28 & 0.08  \\  
&H$_2$S  			&     1.5 	&	0.98   \\  
&H$_{2}$CO	 	&     0.14 			&	0.31  & 1.1  \\  
&CS 	 			&     0.12 		&	0.22  &  0.04 \\  
&HNC		 		&     0.012 		&	   \\ 

\hline
HVF	&H$_{2}$O/Total H$_2^{\,b}$ 	&  220	&	200-300$^{\,j}$   \\  
		&H$_{2}$O/Hot H$_2^{\,c}$	&  2900	&	2000-5000$^{\,k}$     \\ 
		& SO$_2$				& 225	& \\ 
		& SO					& 21	& \\ 
		& SiO					& 0.45   & \\ 

\hline
HC$^{\,d}$ 	 &H$_{2}$O 			&    120 	&	140$^{\,l}$  \\ 
&HDO			 	& 0.15	 &	 0.5  &  0.14  \\  
&H$_2$S 	 	&   0.27 	&	0.3$^{m}$     \\  
&CH$_{3}$CN  			&   0.05 	&	0.078  &  0.04  \\  
&NH$_{3}$	 	 	&   16	&	10$^{\,n}$   \\  
&HC$_{3}$N 	 		&   0.018	&	0.016  &  0.018   \\  
&OCS		  	 	&   0.17  	&   &  1.1	  \\  
&CH$_{3}$OH 	  		&   7.9$^{\,o}$	&	1-10$^{\,p}$   & 1.4  \\  
&HNC  				&   0.0044  	&	   \\  
&CS  				&     0.029 	&&	  0.06 \\  
& CN				&	0.079		&&	 0.008	\\ 
& NO				&    2.8		&  2.0$^{\,m}$\\   
\hline


CR$^{\,e}$	  &H$_{2}$O   &   28     &   \\  
&	HDO		  &  0.87	&	&   0.93    \\ 
&      NH$_3$       &   2.0      & \\ 
&CH$_{3}$OH 	 		&   120$^{\,o}$	&	30  & 22  \\   
&(CH$_{3}$)$_2$O 	  	&   6.5 	&	2.5  & 1.0  \\  
&H$_{2}$CO 	 		&     1.0	&	0.6  &  0.46\\  
&HDCO 	 			&    0.014	&	 \\  
&H$_{2}$CS 	 		&    0.065 	&     0.06	  &  0.014   \\  
&SO$_2$				& 10     & \\ 
&SO					& 0.85       & \\ 
& CS as CR			& 0.40   &  &    0.1 \\ 
\hline


ER$^{\,f}$ 
&CS 	as ER 		&   0.21 	&	0.025      & 0.11 \\   
&HNC  		&   0.001&	0.005 \\  
&N$_2$H$^+$  	&    0.0005	&	\\ 
\hline

PDR$^{\,f}$ 	 &H$_{2}$O 			   	&$\ga$1.1$^{q}$	&0.33$^{r}$  \\ 
 		 &CN 		&  0.02   	& 0.03	  \\ 
 		 &NH$_3$ 		&  0.05$^{s}$   	& 	  \\
\hline
\end{tabular}
\begin{list}{}{}
\item[$^{{a}}$]LVF $N$(H$_2$)\,=\,3$\times10^{23}$ cm$^{-2}$ (this work). $^{{b}}$HVF total $N$(H$_2$)\,=\,4$\times10^{22}$ cm$^{-2}$ (this work). $^{{c}}$HVF hot $N$(H$_2$)\,=\,3$\times10^{21}$ cm$^{-2}$ (Watson \etal~\cite{Watson}). $^{{d}}$HC $N$(H$_2$)\,=\,1$\times10^{24}$ cm$^{-2}$ (B87). 
$^{{e}}$CR $N$(H$_2$)\,=\,2$\times10^{23}$ cm$^{-2}$ (Wilson  \etal~\cite{Wilson86}; Goldsmith  \etal~\cite{Goldsmith97}). 
$^{{f}}$For both the ER and PDR: $N$(H$_2$)\,=\,2.0$\times10^{22}$ cm$^{-2}$  (this work, see Sect. \ref{section CO}). 
$^{{g}}$Source averages with sizes given in Tables \ref{result_table1} and \ref{result_table2}. 
$^{{h}}$Source averages from Blake \etal~\cite{Blake}.  
$^{{i}}$Source averages from Sutton \etal~\cite{Sutton}.
$^{{j}}$Plateau abundance (both LVF and HVF), Cernicharo   \etal~(\cite{Cernicharo}). 
$^{{k}}$Wright \etal~(\cite{Wright}).
$^{{l}}$Gensheimer \etal~(\cite{Gensheimer}).
$^{{m}}$C05.
$^{{n}}$Hermsen  \etal~(\cite{Hermsen}).  
$^{o}$Estimated from the CH$_3$OH two-component rotation diagram.
$^{{p}}$Menten \etal~(\cite{Menten}).
 $^{{q}}$W06.
$^{{r}}$Melnick \etal~(\cite{Melnick}).
$^{{s}}$Larsson \etal~(\cite{Larsson}) towards the Orion Bar.
\end{list}
\end{table}

A large source of uncertainty in these calculations is the adopted H$_2$ column densities. Whenever possible we have used our own calculated H$_2$ column densities (for the ER,  LVF, HVF, Table \ref{result_table2}). For the HC and CR we have adopted  \mbox{$N$(H$_2$)\,=\,1$\times10^{24}$   \cmsq} (calculated for a HC size of 10$\arcsec$ in B87), 
and 
\mbox{$N$(H$_2$)\,=\,2$\times10^{23}$   \cmsq} (Wilson  \etal~\cite{Wilson86}; B87; Goldsmith  \etal~\cite{Goldsmith97}), respectively. 
In case of the HVF we also compare the (shocked) water column with the column density of the hot (shocked) H$_2$, 
where \mbox{$N_\mathrm{HVF}$\,=\,3$\times10^{21}$   \cmsq} (Watson \etal~\cite{Watson}).
When we calculate the abundances we assume that the derived H$_2$ column density
spatially coincides with  the emission from the species of interest.

Most of our abundances, listed in Table \ref{abundances}, are in very good agreement with  B87 and S95.
However, there are 
a few  exceptions -- a 4\,--\,15 times  higher 
abundance for SO$_2$  in the LVF, and \mbox{2\,--\,5} times higher abundances in general 
in the CR  than in B87 and S95. 
The differences may arise because our observed transitions probe higher density and more compact regions in the CR.
Since we cannot discriminate between CR or ER emission for CS, the  abundance is calculated with both alternatives. It turns out that the CS abundance is about the same for either source of emission.  However, the high ER abundance of CS compared to that of B87 is to a large extent due to the very   
different H$_2$ column densities used in our survey and B87, \mbox{$N$(H$_2$)\,=\,2$\times10^{22}$   \cmsq} and  \mbox{$N$(H$_2$)\,=\,3$\times10^{23}$   \cmsq}, respectively. 

Our HVF H$_2$O  abundance, as compared to the total H$_2$ density in the flow, is 3$\times10^{-5}$, in agreement with
the water abundance in the Plateau estimated by Cernicharo  \etal~(\cite{Cernicharo}). If we compare the
HVF water abundance to the hot shocked H$_2$ it is   consistent with 
Wright  \etal~(\cite{Wright 2000}) and Melnick \etal~(\cite{Melnick}),  who estimate the shocked Plateau water
abundance to be (2\,--\,5)$\times10^{-4}$.

The HC H$_2$O abundance is in agreement with the mapping of H$_2^{18}$O towards the Orion Hot Core with the IRAM telescope
(12$\arcsec$ beam)  by Gensheimer \etal~(\cite{Gensheimer}). Their estimated abundance is 1.4$\times10^{-5}$. S01 estimate the water abundance to be $\sim$1.0$\times10^{-5}$ from their observation of the vibrationally excited 
H$_2$O \mbox{1$_{1,0}$\,--\,1$_{0,1}$} $v_2$=1 transition, also in accordance with our value.


High water abundances in high temperature regions,  for example in 
outflows,  PDRs, and hot cores  are consistent with both water
and deuterated water forming on grains at low 
temperatures,
and subsequently evaporating from the grain surfaces at high temperatures above $\sim$90 K. At temperatures above $\sim$400 K, easily reached in shocks, neutral-neutral reactions produce even higher water abundances (cf. Neufeld \etal~\cite{Neufeld95}). 
Hence, our
high  abundance
in the HC of  H$_2$O can be the result of evaporation from grain surfaces, which also applies to CH$_3$OH and NH$_3$, as is discussed in more detail in Sect. \ref{subsection: Gas-phase vs grain surface abundances}. The water abundance in the CR and LVF is about the same, and lower 
than in the HC. This might be a natural consequence of a lower temperature in these regions with less evaporation from grain surfaces, which also applies to the PDR region. This is also the cause of the non-detection of water in the ER which has a temperature below the sublimation temperature. 
The highest water 
abundance is found in the HVF, which is suggestive of an even more efficient  production in shocks.


 \begin{table} 
\caption{Relative abundances ratios and comparisons with ice abundances.} 
\label{ice abundances}
\centering
\begin{tabular} {l l l l l l  }
\hline
\hline  
Abundance ratio		 & LVF	&	HC	&	CR	 	& \multicolumn{2}{c}{Ice abundances$^{{a}}$}  \\
& \multicolumn{3}{c}{$N$/$N$(H$_2$O)$\times$100} &Orion IRc2& W33A\\
\hline

CH$_3$OH/H$_2$O 	&	 	&	7 	&	430	 &	10	 	& 11\,--\,17 \\
 
HDO/H$_2$O 		& 	0.52	&	0.13	&  	3.2	 &	 	 &  0.3 \\

HDCO/H$_2$CO		& 		& 		& 	1.2	& 		& 	 \\

NH$_3$/H$_2$O 	&	 	& 	13	&  	7	&   	&  15	\\  
 
SO$_2$/H$_2$O 	&   	69	&  		 & 36   	&   	& $\sim$1.6	\\

H$_2$CO/H$_2$O 	&	 0.5  	 & 	&	3.6 	&  	& $\sim$3	\\

SO/H$_2$O             	&   	11	&   		 &    3.0	&  		&  	\\

OCS/H$_2$O		&	 	 &   0.14	&	  	&  $<$0.2	 	&  0.2 	\\

CS/H$_2$O		&	0.41	&	0.02	&	1.4	&	 	&	 	\\	

H$_2$S/H$_2$O   	&  	5.2	&   	0.22	&   	&  	& \\

\hline
\end{tabular}
\begin{list}{}{}
\item$^{{a}}$Gibb \etal~(\cite{Gibb00}) and (\cite{Gibb04}).
\end{list}
\end{table}

\subsection{Gas-phase vs grain surface abundances} \label{subsection: Gas-phase vs grain surface abundances}

Ratios of our observed gas-phase column densities in the Orion LVF, HC and CR sources (as derived from Tables \ref{result_table1} and \ref{result_table2}) and the water column density
 is found in Table \ref{ice abundances}.
These ratios are compared with the corresponding grain-surface abundance ratios towards two of the sources observed by ISO (Gibb \etal~\cite{Gibb00}, \cite{Gibb04}), Orion IRc2 and the embedded high-mass protostar W33A.

Some suggestions from these comparisons are:
\begin{itemize}
\item
Both  CH$_3$OH and  H$_2$O are very abundant in the dense and warm HC, and their gas-phase abundance ratio is very similar to that  in the grain-surface ice of Orion IRc2 and W33A. This strongly points at a dominant production via hydrogenation on cold  grain surfaces and subsequent evaporation in the warm and hot cores (cf. Brown \etal~\cite{Brown88}; Caselli \etal~\cite{Caselli93}; Stantcheva and Herbst \cite{Stantcheva and Herbst 04}; Garrod \& Herbst \cite{Garrod and Herbst06}; Chang
\etal~\cite{Chang Cuppen and Herbst 07}), especially so since gas-phase production of CH$_3$OH has been shown to be very inefficient 
(cf. Geppert \etal~\cite{Geppert}; Millar \cite{Millar}; Garrod \etal~\cite{Garrod Wakelam and Herbst 07}).
In the CR the CH$_3$OH/H$_2$O ratio is about 60  times higher than in the HC and  in ices, caused  by  the four times decrease of the water abundance, and the  15  times higher methanol abundance as compared to the HC. This may suggest that H$_2$O is consumed in the formation of CH$_3$OH in accordance with the recent laboratory study of methanol formation from electron-irradiated mixed H$_2$O/CH$_4$ ice at 10 K by Wada \etal~(\cite{Wada Mochizuki and Hiraoka  06}).
\item
The rather similar gas-phase HDO/H$_2$O abundance ratios in the LVF and HC   compared with the ice  ratio in W33A   most likely suggest efficient deuteration reactions on grain surfaces as the cause of the high water deuteration level. This is supported by our previous conclusions about the grain surface origin of the high water abundances. 
The much higher HDO/H$_2$O ratio found in the CR, as well as the similar HDCO/H$_2$CO ratio, at least partly is caused by a decreasing H$_2$O abundance -- possibly a result of H$_2$O consumption in the efficient grain-surface  formation of CH$_3$OH in this source.
\item
Likewise, the  gas-phase abundance ratios  of  NH$_3$ and  H$_2$O in the   CR and HC are similar to the W33A ice abundance ratio, again strongly suggesting that both these  abundant  species originate primarily from 
hydrogenation on cold  
grain surfaces with subsequent evaporation (cf. Stantcheva and Herbst \cite{Stantcheva and Herbst 04}; Garrod  \etal~\cite{Garrod Wakelam and Herbst 07}).
\item
The high SO$_2$/H$_2$O abundance ratio observed in the LVF is contrasted with a low ice ratio in W33A. The latter ratio is 
most likely explained by rather inefficient gas-phase formation of SO$_2$ and subsequent adsorption onto already icy grain mantles formed by efficient hydrogenation on the cold grains. The high gas-phase H$_2$O abundance in the LVF may directly result from evaporation caused by the strong radiation from the nearby LVF driving source. This heating also could release S and Si atoms from the grains. Subsequent gas-phase reactions, based upon undepleted elemental abundances, then could lead to the elevated abundances of SO$_2$, SO, H$_2$S and SiO observed in the LVF. 
These abundances are several orders of magnitude higher than those in quiescent clouds where the abundances of S and Si appear to be depleted (B87;  Irvine \etal~\cite{Irvine87}). 
In this scenario the ISO observations of OH at high abundance in the LVF (Goicoechea \etal~\cite{Goicoechea}) are important.
Low velocity shocks also may play a role here (cf. Mitchell \cite{Mitchell84}; Pineau des For\^ets \etal~\cite{Pineau des Foretes93}).
\item
The very similar CR H$_2$CO/H$_2$O gas-phase and W33A ice abundances  most likely just tells us that both abundance ratios have the same main origin.
\item
The similarity of the HC gas-phase OCS/H$_2$O abundance ratio and the corresponding ratio in the W33A ice, also might hint at a grain surface origin of OCS. However, the comparatively low OCS abundance is accommodated by current ion-molecule reaction models and the ice content then likely is a result of adsorption.

\end{itemize}

\section {Discussion -- source sizes \& source structure } \label{section Discussion}

We have in our column density and molecular abundance calculations in all cases treated the various Orion KL subregions, probed by the large Odin antenna beam, as homogeneous sources having specified average temperatures, densities and equivalent circular sizes and beam-filling factors. 

However,   the \emph{High Velocity Outflow} is known to be bi-polar with a FWHP size of 60\,--\,70\arcsec, as estimated from simultaneous Odin mapping of the H$_2$O and CO \mbox{$J$\,=\,5\,--\,4} brightness distributions (Olofsson  \etal~\cite{Olofsson03},  Hjalmarson  \etal~\cite{Hjalmarson05}). 
As seen from the radiative transfer equation, the H$_2$O  excitation temperature corresponding to a size of 70\arcsec~is only 26 K, while a temperature of 72 K as found by Wright \etal~(\cite{Wright 2000})  corresponds to a size of  only 32\arcsec. This indicates a  very clumpy  H$_2$O  brightness distribution,  and that this source is filled with approximately one fourth of water emission. Similar results are obtained investigating the size-temperature relation of 
CO. In fact, the appearance of the HVF may be similar to the clumpy, finger-like emission seen in shock-excited H$_2$ (Salas \etal~\cite{Salas})\footnote{See also www: http://subarutelescope.org/Science/press-release/99$\emptyset$1/OrionKL\_3$\emptyset\emptyset$.jpg.}. 

The \emph{Low Velocity Outflow} has a NE-SW elongated structure, roughly orthogonal to the HVF, which also must have small scale structure (Genzel \etal~\cite{Genzel81}; Greenhill \etal~\cite{Greenhill98}). If we guide ourselves by the optically thick HDO lines observed by Pardo \etal~(\cite{Pardo}) and use a HPBW size of 15\arcsec~for H$_2^{18}$O,
the corresponding excitation temperature is only 40 K. The size corresponding to 72 K is 10$\arcsec$
which suggests that this source is filled with about one half of radiating gas-phase water. Similar results are found for all optically thick outflow species.

The \emph{Compact Ridge}  and \emph{Hot Core} size-temperature relations are more consistent with the assumed values, although we know from Beuther \etal~(\cite{Beuther}) that both sources are very clumpy. This might be caused by their much smaller size as compared to the outflows, where the clumping affects larger scales. As seen from the CH$_3$OH Fig.   \ref{SourceSizeCH3OH} (on-line material) and Fig. \ref{SourceSize 2comp}, the source size of the CR also varies with the upper state energies of the lines. This might indicate considerable temperature and density variations within the CR.

Considering the uncertainties discussed above and in \mbox{Sect. \ref{section Theory}} and \ref{uncertainties appendix} (Appendix), the striking agreement with B87 and S95 strengthens our confidence of our results.



\section {Summary}
  We present  first results from a spectral line survey
  towards Orion KL in a frequency range inaccessible from the ground covering  487\,--\,492 and 
 542\,--\,577 \mbox{GHz}.
       
      Some of the results from this survey:
      
   \begin{enumerate} 
      \item 
We detect   a total of  280 lines from 38 different molecular species.
       \item
      In addition we detect 64 unidentified lines,  which represents 19$\%$ of the total. Some tentative assignments of a few of 
      them have been made such as
      the  interstellar anion SH$^-$,  ND, SO$^+$, and CH$_3$OCHO.  
       \item
       The total \emph{beam-averaged} emission in our survey is dominated by CO, \mbox{$o$-H$_2$O}, SO$_2$, SO, 
       $^{13}$CO and  
       CH$_3$OH. Species 
       with 
       the
       largest number of lines are CH$_3$OH, (CH$_3$)$_2$O, SO$_2$, $^{13}$CH$_3$OH, CH$_3$CN and NO.
       \item
       Six water lines are detected, including
       the ground state rotational transition 1$_{1,0}$\,--\,1$_{0,1}$  of $o$-H$_2$O, and its isotopologues 
       $o$-H$_2^{18}$O and $o$-H$_2^{17}$O, which shows emission from the Low- and High Velocity Flow and the Compact 
       Ridge.  
       Hot Core emission from water is observed 
       from the $p$-H$_2$O transition 
       6$_{2,4}$\,--\,7$_{1,7}$ with an upper state energy 867 K, and from a
       weak line feature at 489.054 GHz  identified as the 4$_{2,3}$\,--\,3$_{3,0}$
       transition of $o$-H$_2^{18}$O with an upper state energy of 430 K. 
       We have also observed
       the HDO  
       2$_{0,2}$\,--\,1$_{1,1}$ transition from the Low Velocity Flow, Compact Ridge and the Hot Core,  and have a tentative 
       detection 
       of
       the high energy transition (581 K) 6$_{2,4}$\,--\,6$_{2,5}$ of HDO.
       \item
       We detect the 1$_{0}$\,--\,0$_{\,0}$  transitions of NH$_3$ and the isotopologue $^{15}$NH$_3$. The main 
       isotopologue 
       shows emission from both the Hot Core, LVF and Compact Ridge, while the rarer  isotopologue only exhibits 
       emission
       from the Hot Core.
       \item
       Isotopologue abundance ratios of D/H, $^{12}$C/$^{13}$C, $^{32}$S/$^{34}$S, $^{34}$S/$^{33}$S
      and $^{18}$O/$^{17}$O
       are calculated, as well as the molecular abundance ratio of O/S, all in agreement with previous findings.
      \item
      Different methods are used to obtain rotation temperatures and column densities. For eight different species with at least
      four lines and a sufficient energy range in the transitions, 
      the rotation diagram method and the forward model are applied. 
      The
       LTE approximation is used for all the other species. 
       \item
       Abundances are estimated for the observed species from the different subregions, and we find very high gas-phase
      abundances of H$_2$O, NH$_3$,  
      SO$_2$, SO, NO, and CH$_3$OH.
      An important fact here is that all our abundance determinations, including those for water vapour, are based upon the same 
      methodology.
       \item
       A comparison of our estimated gas-phase abundances with the ice inventory of ISO is shedding new light on the chemical 
       origins of 
       H$_2$O, CH$_3$OH, NH$_3$ and SO$_2$ in the various Orion KL subregions.
       \item
       The line density in our survey is 4\,--\,20 per GHz, with a mean of 8 per GHz. This is comparable with larger telescopes ($\sim$10 
       \mbox{GHz$^{-1}$}), 
       showing
       the excellent performance of the Odin satellite.
   \end{enumerate}


\begin{acknowledgements}
Generous financial support from the Research Councils and Space Agencies in Sweden, Canada, Finland and France is gratefully acknowledged.
We sincerely thank Frank Lovas for a CD containing his molecular spectroscopy database SLAIM$\emptyset$3, 
and are very grateful to the dedicated scientists supporting the molecular spectroscopy database 
the Cologne Database for Molecular Spectroscopy (CDMS) and the Jet Propulsion Laboratory (JPL) for making the difficult but absolutely necessary molecular spectroscopy available on the Internet.
We also thank the  referees whose constructive comments   led to significant 
improvements of the paper.
\end{acknowledgements}

\Online

\section{Data analysis methods} \label{appendix theory}

\subsection {Single line analysis}

With the assumption of optically thin emission, neglecting the background radiation, and assuming that the source fills the antenna main  beam,  the beam averaged upper state column density is calculated as

\begin{equation} \label{Nu appendix}
N_{ {u}}^\mathrm{\,thin}  = \frac {8  \pi k \nu_{{ul}}^2}{h c^3} \frac{1}{ A_{ {ul}}}  \int {T_{\mathrm{mb}}}  \, \mathrm{d} \upsilon, 
\end{equation}
where 
$k$ is the Boltzmann constant, $\nu_{{ul}}$ is the frequency of the transition, $h$ is the Planck constant, $c$ is the speed of light,  $A_{{ul}}$ is the Einstein $A$-coefficient for the transition, and $T_\mathrm{mb}$ is the main beam brightness temperature. As customary the frequency axis $\nu$ has been converted to a 
velocity axis $\upsilon$ using the speed of light.

The total column density of each species can then be found assuming LTE (Local Thermodynamic Equilibrium), where the excitation temperatures, $T_{\mathrm{ex}}$, for all the energy levels are the same. The molecular population of each level is then given by the Boltzmann equation, 
which also defines $T_{\mathrm{ex}}$

\begin{equation}\label{Boltzmann}
N_{u}=N_{\mathrm{tot}} \frac{g_{u}}{Q(T)} \, e^{-E_{u}/kT_\mathrm{ex}},
\end{equation}
where $g_{u}$ is the statistical weight of the upper state, and $Q(T)$ is the partition function, which only depends on temperature and molecular constants and hence differs for different kinds of species.
The $A$-coefficients, statistical weights, partition functions and upper state energy levels are available via the  databases JPL, CDMS,  Leiden\footnote{http://www.strw.leidenuniv.nl/$\sim$moldata/} (Sch\"oier et al. \cite{Schoier}) or SLAIM$\emptyset$3. For a few molecules, e.g. (CH$_3)_2$O, we  calculate the $A$-coefficients using line strengths found in SLAIM$\emptyset$3 as

\begin{equation}\label{Acoeff eq}
A_{ul}=\frac{16 \, \pi^ 3\,  \nu_{ul}^3\,  \mu^2}{3 \, \epsilon_0 \,h\, c^3} \frac{S_{ul}}{g_u}
\end{equation}
where $S_{ul}$ is 
the rotational part of the  line strength, and $\mu$ is the molecular dipole moment.

From Eq. (\ref{Boltzmann}) and (\ref{Nu}), we obtain \emph{the beam-averaged total column density }

\begin{equation} \label{Ntot}
N_{\,\mathrm{tot}}^{\,\mathrm{thin}} = \frac {8  \pi k \nu_{{ul}}^2}{h c^3} \frac{1}{ A_{{ul}}} \frac{Q(T)}{g_{{u}}}  e^{\,E_{{u}}/kT_{\mathrm{ex}}} \int{T_{\mathrm{mb}}}\,\mathrm{d} \upsilon
\end{equation}
assuming optically thin emission.

The solution of the radiative transport equation, neglecting  background radiation and with a constant source function, is

\begin{equation}\label{solution}
T^*_\mathrm{A} =T_\mathrm{b}  \, \eta_{\mathrm{mb}}\, \eta_{\mathrm{bf}}=J(T_\mathrm{ex})  \, (1-e^{-\tau})  \, \eta_{\mathrm{mb}}\, \eta_{\mathrm{bf}},
\end{equation}
where the beam-filling factor
$\eta_{\mathrm{bf}} =\theta_\mathrm{s}^2 / (\theta_\mathrm{s}^2+\theta_\mathrm{mb}^2)$, assuming that both the source brightness distribution and the antenna response are circularly symmetric and Gaussian. The
radiation temperature, $J(T_\mathrm{ex}$) is 

\begin{equation}
J(T_\mathrm{ex})  = \frac{h \nu}{k} \frac{1}{e^{\,h\nu/kT_\mathrm{ex}}-1} \approx T_\mathrm{ex},
\end{equation}
where the approximation is valid only if h$\nu$ $\ll$ k$T_{\mathrm{ex}}$. For the appropriate temperatures in this region,  $T\approx$ 100\,--\,200 K, and  frequencies $\sim$550 \mbox{GHz}, the radiation temperature will differ from $T_\mathrm{ex}$ by  approximately 
10\,--\,15\%. Accordingly, we use $J(T_\mathrm{ex}$) and not $T_\mathrm{ex}$ in our calculations.

If the emission is optically thick ($\tau\gg$1) Eq.  (\ref{solution}) simplifies to

\begin{equation}\label{source-size}
T^*_\mathrm{A} =J(T_\mathrm{ex})\, \eta_{\mathrm{mb}}\,\eta_{\mathrm{bf}} =J(T_\mathrm{ex})\,\eta_{\mathrm{mb}}\, \frac{\theta_\mathrm{s}^2 }{ \theta_\mathrm{s}^2+\theta_\mathrm{mb}^2}.
\end{equation}

Using this equation  the beam-filling, and hence the approximate source size, can be determined without relying on correctly calculated optical depths. The variation in $T^*_\mathrm{A}$ that is seen mainly will be due to
variations in the  beam-filling factor, but there can also be variations in $T_\mathrm{ex}$, both along the path and between different states, which cannot be determined.

Since molecular clouds are known to be  clumpy down to very small scales, it is indeed likely that the source does not fill the beam. 
This is especially true in case of the large Odin beam, and the total column density from Eq.  (\ref{Ntot}) will  be too low by a factor of 1/$\eta_\mathrm{{bf}}$. 
Variations in beam-filling between the different transitions of the same molecule also are probable. Hence, when we are comparing other observations (e.g. interferometry
mapping) with ours to estimate the beam-filling, by necessity we have to compare transitions with similar parameters tracing the same gas.

The optical depth at the centre of the line can be calculated, assuming LTE  and  a Gaussian line profile, using

\begin{equation} \label{tau}
\tau_{\mathrm{max}} =\sqrt{\frac{\ln \,2}{16\, \pi^3} }\frac{c^3}{ \nu_{{ul}}^3\,\Delta \upsilon} \, A_{{ul}} \, N_\mathrm{tot} \,\frac{g_{u}}{Q(T)}\,e^{-E_{u}/kT_\mathrm{ex}}\, (e^{\,h\nu_{{ul}}/kT_\mathrm{ex}}-1) \\
\end{equation}

\noindent
where $\Delta \upsilon$ is the width of the line,  and $E_{u}$ is the upper state energy. However, this approach demands knowledge of the total column density.
If we have observations of isotopologues, and if the  isotopologue abundance ratio $R$ is known, then the optical depths can be determined by means of Eq. (\ref{solution}). If the  excitation temperatures are about the same for both isotopologues,  and the optical depth of isotopologue one is larger than isotopologue two by a factor $ R$, we then will have
\begin{equation}\label{tau_iso_comp}
\frac{T_\mathrm{A, 1}}{T_\mathrm{A, 2}}
 = \frac { (1-e^{-\tau_1}) \, \eta_{\mathrm{\,bf, 1}}}{ (1-e^{-\tau_2}) \, \eta_{\mathrm{\,bf, 2}}}= \frac { (1-e^{-R\,\tau_2}) \,
 \eta_{\mathrm{\,bf, 1}} }{ (1-e^{-\tau_2})  \, \eta_{\mathrm{\,bf, 2}}}.
\end{equation}
This will give the mean optical depths of both species.
If the emissions are co-spatial the beam-filling factors cancel.

Once a transition has become
optically thick, the value of $N_{u}$/$g_{u}$ cannot increase any further. 
The derived total column density will then be too low
and needs an
optical depth correction factor $C_\tau$, which will be one or larger
\begin{equation}
C_\tau=\frac{\tau}{1-e^{-\tau}}.
\end{equation}

With corrections for optical depth and beam-filling the \emph{true total source averaged column density} will be

\begin{equation} \label{NtotCorrected appendix}
N_{\mathrm{tot}}  =\frac{C_\tau}{\eta_{\mathrm{bf}}}\, \frac {8  \pi k \nu_{{ul}}^2}{h c^3} \frac{1}{ A_{{ul}}} \frac{Q(T)}{g_{{u}}}  e^{E_{{u}}/kT_{\mathrm{ex}}} \int {T_{\mathrm{mb}}}\,d \upsilon = \frac{C_\tau}{\eta_{\mathrm{bf}}}\,N_{\,\mathrm{tot}}^{\,\mathrm{thin}}.
\end{equation}

If no information about optical depth is available this correction is not taken into account.
For the excitation temperature in Eq. (\ref{source-size}) and (\ref{NtotCorrected appendix}), we use an adopted temperature that fits the species (Table \ref{result_table1}), or the calculated rotation temperature  $T_\mathrm{ROT}$ (see Sect. \ref{section rotation diagram method}). 

Note the importance of the partition function $Q(T)$, and the statistical weight $g_{u}$ in Eq. (\ref{NtotCorrected appendix}). 
If the statistical weights include the $ortho$/$para$ ratio whenever these molecular sub-divisions exist,  the partition function
must include them. It is also important to use the same  statistical weights $g_u$ in Eq.  (\ref{NtotCorrected appendix})  when  the partition function $Q(T)$ is calculated.

\subsection {Additional limitations} \label{uncertainties appendix}
If there are
deviations from the basic assumptions, errors will occur in the calculations for all models (cf. Goldsmith and Langer \cite{Goldsmith Langer}). These effects can to some extent be corrected for, like optical depth effects and beam-filling, as discussed above.  In addition we have:
  \begin{enumerate}
 
  \item
  \emph{Several excitation temperatures/Non-LTE.}
  The population distribution may not be characterised by a single rotational temperature. The temperature can vary due to density,
  excitation gradients along the line-of-sight, IR flux, subthermal excitation etc. Since our beam is very large and encompasses a 
  variety 
  of different conditions, we may expect emission from several different sources, each with different temperature.
  \item
 \emph{Adopted excitation temperature.}
  Whenever we have too few lines to use the rotation diagram method or forward model we use the LTE approximation for a single line. 
  The excitation temperature is then
 adopted from a rotation temperature of a similar species in our survey, or taken from the literature. This can create errors if the adopted temperature is not appropriate.
  \item
 \emph{Source structure.}
   The beam-filling correction are calculated assuming a homogenous source (see also~Sect. \ref{section Discussion}). 
   Clumping and substructures can introduce errors in the beam-filling corrections. Vastly different opacitites between the isotopologues 
 (with NH$_3$ and H$_2$O as good examples) also complicate the analysis, as the beam-filling factors may be very different.
  \item
 \emph{Gaussian decomposition.}
  Errors in our Gaussian decompositions can   introduce errors in the calculated optical depths and the  column densities. 
  \item
 \emph{H$_2$ column densities.}
 The largest source of uncertainty in the abundances in Sect. \ref{section Abundances} is the adopted H$_2$ column densities and the assumption that the H$_2$ emission and the molecule of interest spatially coincides. This can create errors by orders of magnitude.
  \item
  \emph{Observational errors.}
  Errors may exist in the  measured line intensities due to misidentifications, unresolved 
  blends, or pointing and calibration.
  However, the errors must be rather  
  large  to create significant errors
  in the column density  obtained from the rotation diagram and forward model.
 \item
  \emph{Background radiation.}
  The background radiation, e.g. emission from warm dust, 
   may be too large to be neglected. 
  If the condition $T_\mathrm{ex}\gg T_{\mathrm{bg}}$ is violated, shifts along the ordinate in the rotation diagram with a factor 
  of
  $\ln$ (1$-T_{\mathrm{bg}}$/$T_\mathrm{ex}$) will occur. 
  The dust continuum radiation will for example affect the level 
  populations of water, and must be 
  included in accurate calculations for these transitions. 
   \end{enumerate}

\subsection {Determination of relative chemical abundances }

The definition of molecular abundance of a species with respect to H$_2$ is the ratio of the volume densities. 
This  demands detailed knowledge of the geometry of the emitting region. If this is not known,   column densities can be used as a substitute, with the assumption that the two species are well mixed and emitted from the same region, to obtain average abundances along the line of sight

\begin{equation}
X_\mathrm{species} = \frac {N_\mathrm{species}}{N_\mathrm{H_2}}.
\end{equation}

The use of column densities will provide a higher degree of accuracy than the use of volume densities  (Irvine \etal~\cite{Irvine85}).
As long as the molecules are emitted from the same region with approximately the same source size, there is no dependence of beam-filling. In addition, since we have a uniformly calibrated set of data,  uncertainties in calibration will not affect the result. 

The relative isotopologue abundance $R$ can be determined
in the same way
as  the molecular abundances,
\begin{equation}
R = \frac {N_{1}}{N_{2}},
\end{equation}
when we have observations of optically thin isotopologues, or opacity-corrected column densities.

\subsection{Optical depth broadening}

If the intrinsic line shape is Gaussian, a high line opacity will increase the observed line width  (cf. Phillips \etal~\cite{Phillips79}) as 
 \begin{equation} \label{opacity broadening}
\Delta \upsilon \approx \Delta \upsilon_i \, \left( \frac{\ln \tau_p}{\ln 2} \right)^{1/2}
\end{equation}
where $\Delta \upsilon_i$ is the intrinsic velocity width of the line and
$\tau_p$ \mbox{($\gg$1)} is the peak optical depth in the line. 

\clearpage



\clearpage

\clearpage

\begin{figure} 
    \resizebox{\hsize}{!}{\includegraphics{7225fg32.eps}}  
 \caption{CH$_3$OH  with emission from both the CR and HC.  Vibrationally excited transitions are marked by $*$ after the  intensity scale factor  given in the top left 
 corners, with upper state energies given in K and $A$-coefficients in s$^{-1}$.} 
 \label{figureCH3OH}
\end{figure}

\clearpage


\begin{figure}
   \resizebox{\hsize}{!}{\includegraphics{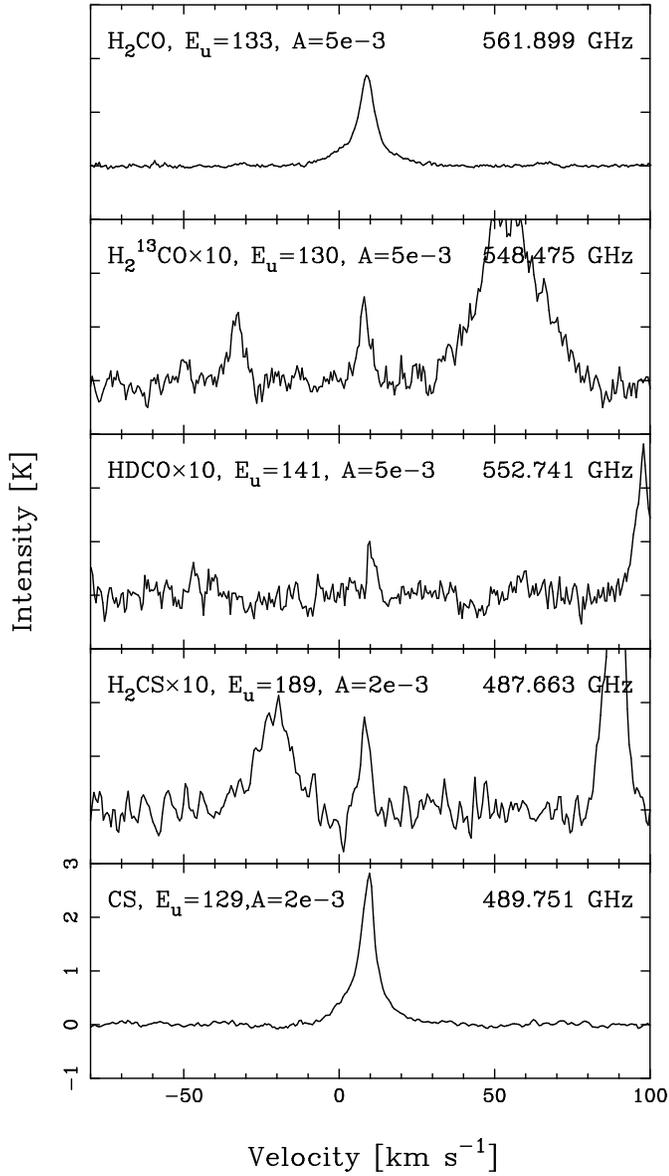}}  
 \caption{The optically thick  line profile of H$_2$CO  show emission from the CR and LVF. The optically thin transitions of
 H$_2^{13}$CO, HDCO, and
 H$_2$CS show only emission from the
 CR. The CS line profile shows emission from the HC, LVF and a narrow component from the CR or ER. An  intensity scale factor is 
 given after the molecular species, the upper state energies are given in K and the $A$-coefficients in 
 s$^{-1}$.}
 \label{H2CO}
\end{figure}
\begin{figure} [h]
    \resizebox{\hsize}{!}{\includegraphics{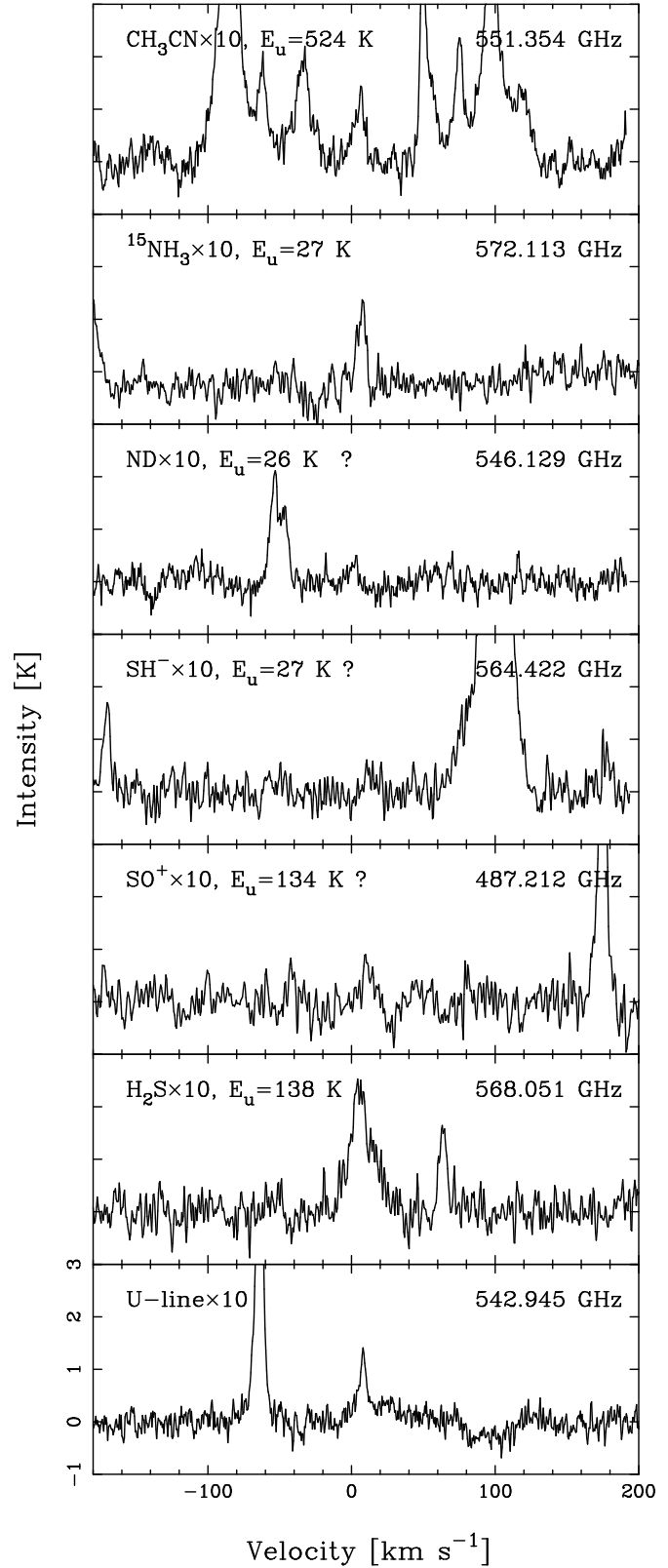}}  
 \caption{Identification suggestions for a few U-lines together with $^{15}$NH$_3$, H$_2$S and CH$_3$CN as comparison lines. The U-line at 542.945 GHz is our strongest non-identified U-line.
 An  intensity scale factor is given after the molecular species.}
 \label{TentativeLines}
\end{figure}

\begin{figure}
       \resizebox{\hsize}{!}{\includegraphics{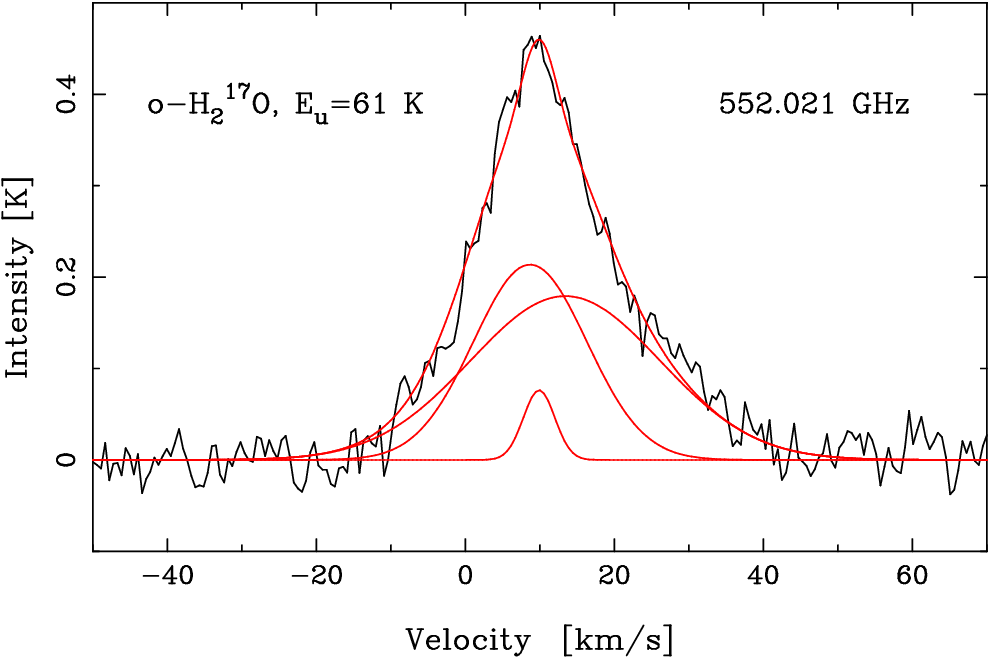}}  
 \caption{A three-component Gaussian fit to H$_2^{17}$O shown together with the individual Gaussians. The line widths are 5~\kms, 18~\kms, and 30~\kms~from the CR, 
 LVF and HVF, respectively.}
 \label{3G fit to h2o17}
\end{figure}

\begin{figure}
       \resizebox{\hsize}{!}{\includegraphics{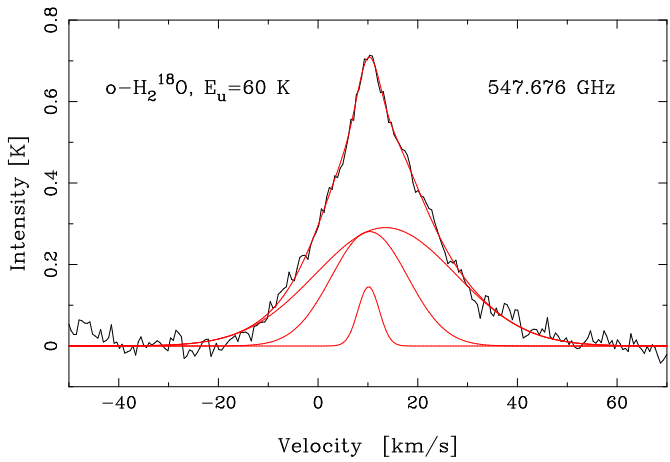}}  
 \caption{A three-component Gaussian fit to H$_2^{18}$O shown together with the individual Gaussians. The line widths are 5~\kms, 18~\kms, and 33~\kms~from the CR, 
 LVF and HVF, respectively.}
 \label{3G fit to h2o18}
\end{figure}

\begin{figure}
 \resizebox{\hsize}{!}{\includegraphics{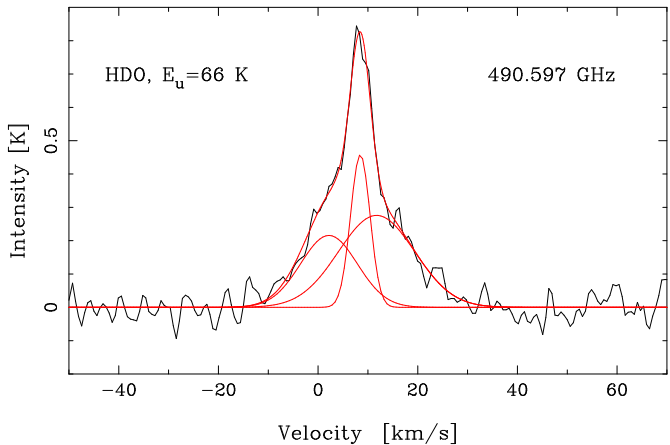}}  
 \caption{A three-component Gaussian fit to HDO shown together with the individual Gaussians. The line widths are 5~\kms, 13~\kms, and 18~\kms~for the CR, HC, and LVF, 
 respectively.}
 \label{3G fit to hdo}
\end{figure}

\begin{figure}
\resizebox{\hsize}{!}{\includegraphics{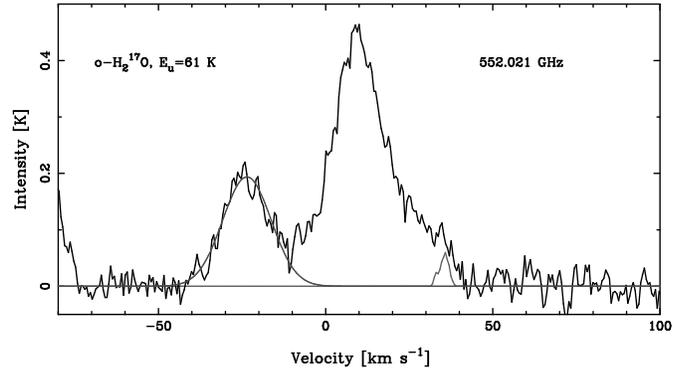}}  
 \caption{H$_2^{17}$O  before removal of blends with reconstructed lines from SO$_2$ and CH$_3$OH shown..}
  \label{H217O with blending lines before removal}
\end{figure}

\begin{figure}
       \resizebox{\hsize}{!}{\includegraphics{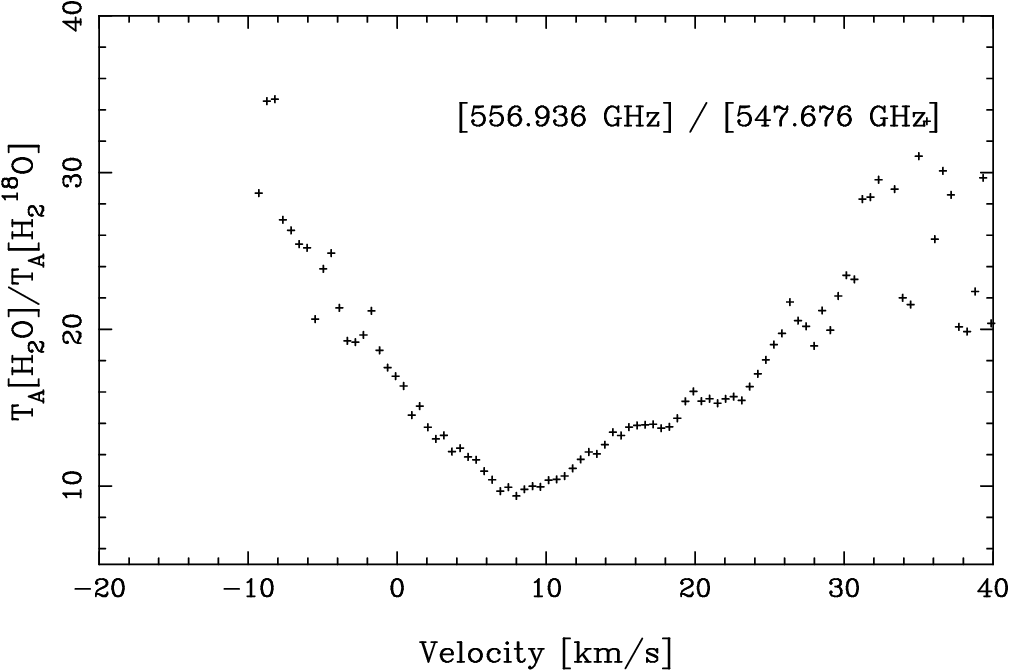}}  
 \caption{Ratio of $o$-H$_2$O over $o$-H$_2^{18}$O.}
 \label{Extra fig h2o_over_h2o18_zoom50_narrow}
\end{figure}

\begin{figure}
       \resizebox{\hsize}{!}{\includegraphics{7225fg40.eps}} 
 \caption{Ratio of $o$-H$_2^{18}$O over $o$-H$_2^{17}$O.}
  \label{fig h2o18overh2o17_zoom_narrow}
\end{figure}

\begin{figure}
       \resizebox{\hsize}{!}{\includegraphics[angle=90]{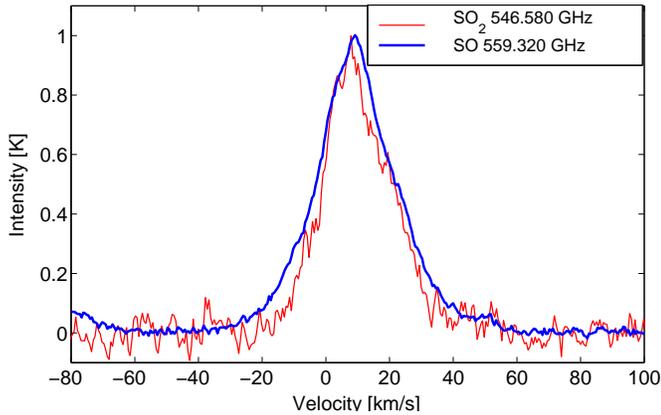}}  
 \caption{Comparison of the SO and SO$_2$ line profiles. Both lines are normalised to unity.}
  \label{fig SO_SO2}
\end{figure}

\begin{figure}
\resizebox{\hsize}{!}{\includegraphics{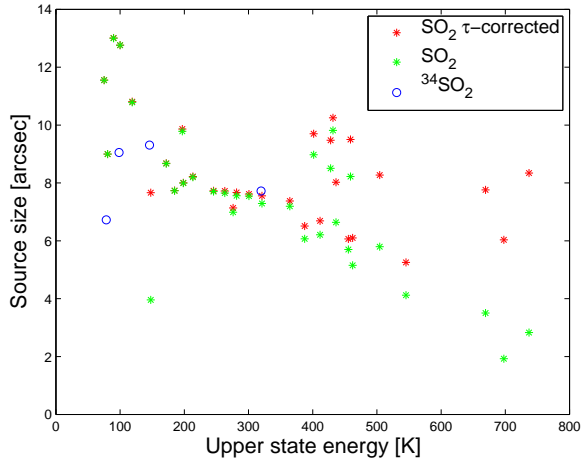}}  
 \caption{Source sizes for opacity-corrected SO$_2$  (calculated with $T_\mathrm{ROT}$\,=\,103 K and source size\,=\,8$\arcsec$) vs. non corrected SO$_2$ together with  $^{34}$SO$_2$  (calculated with $T_\mathrm{ROT}$\,=\,125 K).}
  \label{Figure so2 sizes}
\end{figure}
 
\begin{figure}[h] 
       \resizebox{\hsize}{!}{\includegraphics{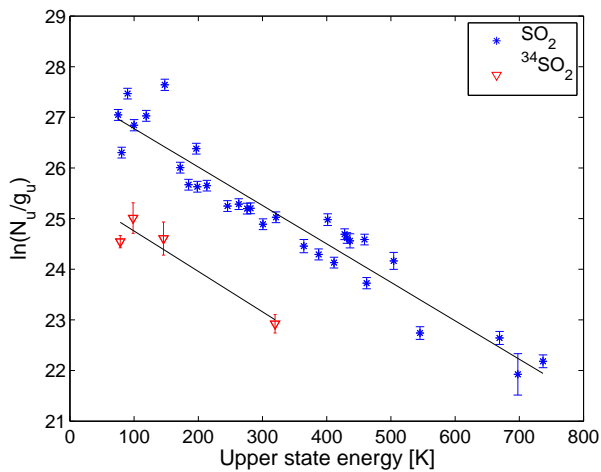}}  
 \caption{Rotation diagram  for SO$_2$,  not corrected for opacity, produces $T_{\mathrm{ROT}}$\,=\,132 K and 
$N_\mathrm{ROT}$\,=\,3.9$\times10^{17}$ cm$^{-2}$ (extended source). The $^{34}$SO$_2$ 
fit give $T_{\mathrm{ROT}}$\,=\,125 K and $N_\mathrm{ROT}$\,=\,5.4$\times10^{16}$ cm$^{-2}$.  }
 \label{rot SO2 not opacity corrected}
\end{figure}

\begin{figure}[h]
       \resizebox{\hsize}{!}{\includegraphics{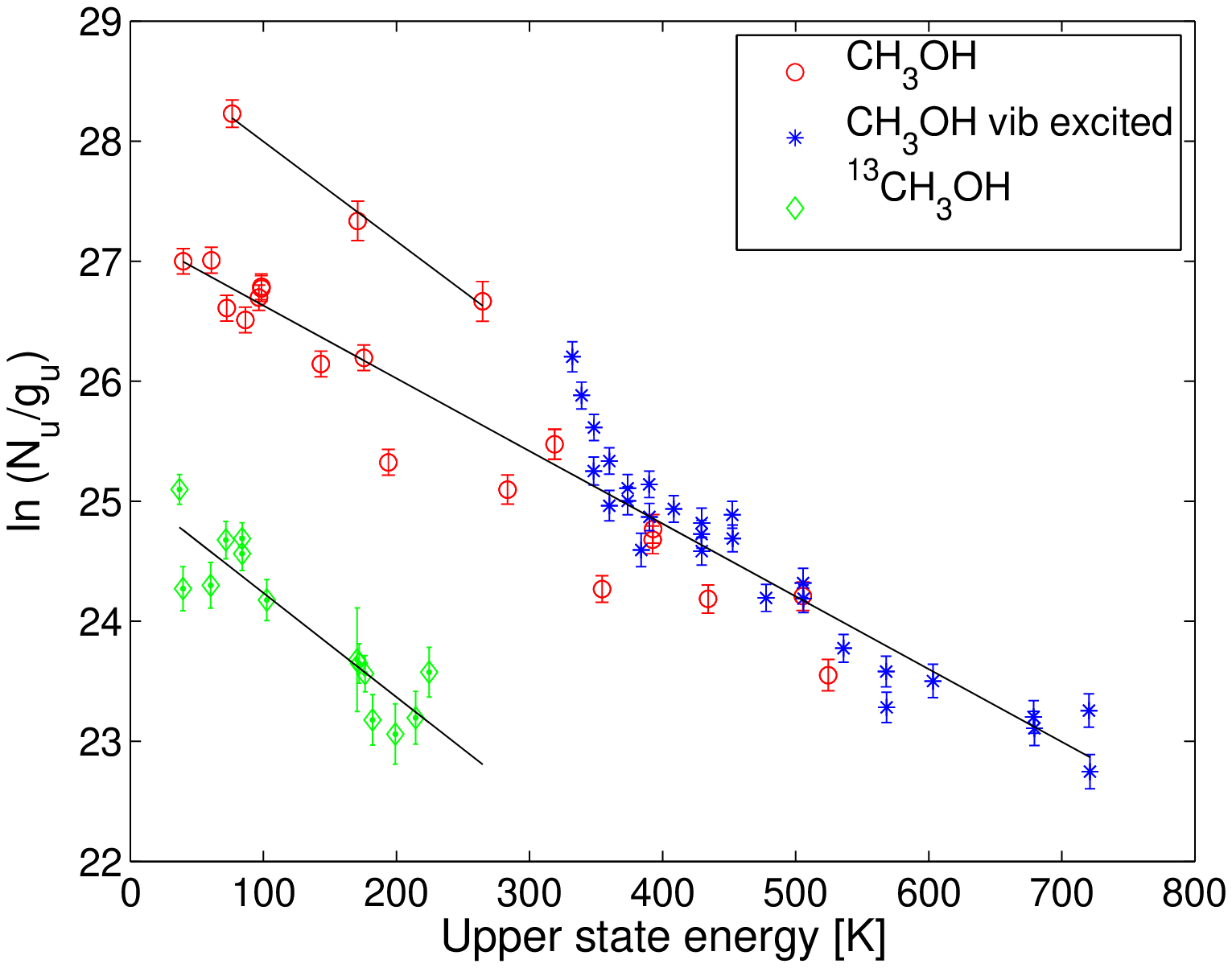}}  
 \caption{Rotation diagrams: CH$_3$OH producing $T_\mathrm{ROT}$\,=\,165 K and $N_\mathrm{ROT}$\,=\,9.3$\times10^{17}$ cm$^{-2}$ (not opacity corrected); three optically thin CH$_3$OH lines produces  $T_\mathrm{ROT}$\,=\,120 K and \mbox{$N_\mathrm{ROT}$\,=\,2.6$\times10^{18}$ cm$^{-2}$}; }$^{13}$CH$_3$OH producing $T_\mathrm{ROT}$\,=\,115 K and $N_\mathrm{ROT}$\,=\,5.9$\times10^{17}$ cm$^{-2}$  (extended source).  
  \label{Rot diagram for non-tau-corrected CH3OH and 3 optically thin ch3oh and 13ch3oh}
\end{figure}

\begin{figure}[h] 
 \resizebox{\hsize}{!}{\includegraphics{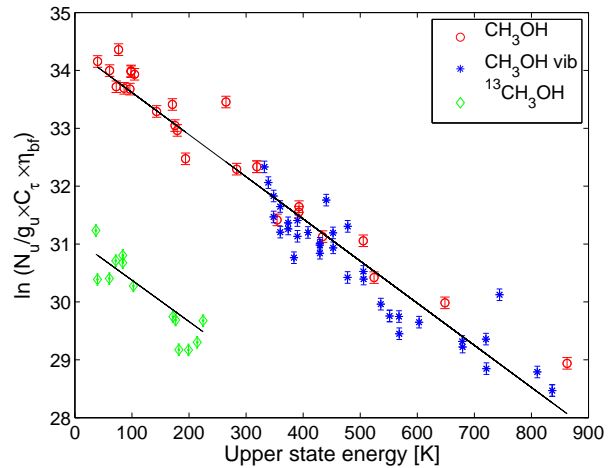}} 
 \caption{CH$_3$OH forward model producing $T_\mathrm{ROT}$\,=\,136$^{+3}_{-4}$~K, $N$\,=\,(1.3$\pm$0.1)$\times10^{18}$ cm$^{-2}$ and a source size of 6$\arcsec^{+0.1}_{-0.3}$. The
 $^{13}$CH$_3$OH forward model use the rotation temperature and source size obtained from CH$_3$OH. }
 \label{rot Albert CH3OH}
\end{figure}


\begin{figure}[h]
       \resizebox{\hsize}{!}{\includegraphics{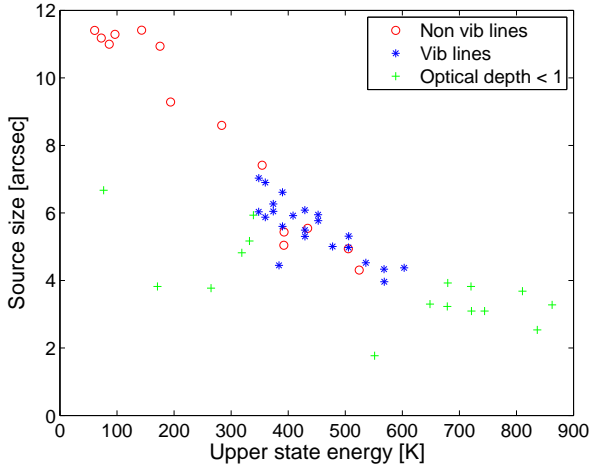}}  
 \caption{Source size variation with energy for CH$_3$OH (no opacity corrections). The low-energy transitions have larger source sizes than higher-energy 
 transitions. Note that all lines with an optical depth less than one, fall below the general trend of decreasing source size with higher energy, since Eq. (\ref{source-size}) is only valid for optically thick lines.}
  \label{SourceSizeCH3OH}
\end{figure}

\begin{figure}[h] 
\resizebox{\hsize}{!}{\includegraphics{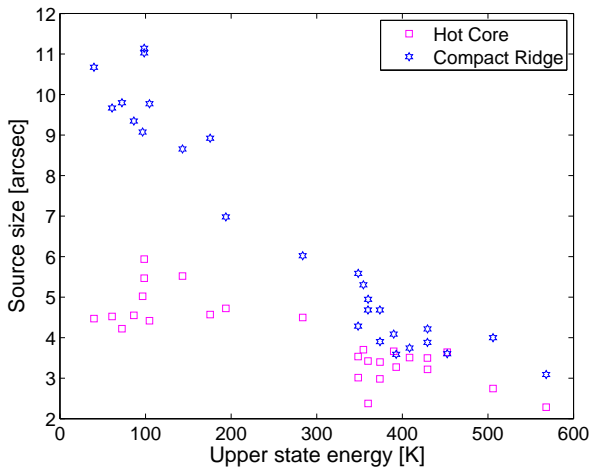}}  
 \caption{Source size variation with energy for two components of CH$_3$OH with no opacity correction.}
 \label{Two component sizes with no tau correction}
\end{figure}

\begin{table} 
\caption{Summary of  all detected species.} 
\label{summary Table}
\begin{tabular} { l r r r}
\hline
\hline
Species  & Number & Upper state & $\int{T^{\,*}_\mathrm{A}  \mathrm{d} \upsilon}$ \\
 & of lines & energy range &    \\
	     &	            &	[K]              &  [K \kms]      \\                           
\hline
CH$_{3}$OCH$_{3}$		&	47	&	106\,--\,448	&	18.3\\ 
SO$_{2}$				&	42	&	75\,--\,737	&	239.9\\  
$^{34}$SO$_{2}$		&	5	&	79\,--\,457	&	2.6	\\  
SO					&	5	&	71\,--\,201	&	181.5\\  
$^{33}$SO			&	3	&	191\,--\,199	&	6.5\\ 
$^{34}$SO			&	2	&	191\,--\,197	&	15.6\\  
CH$_{3}$OH	$v_{\rm t}$=1	&	42	&	332\,--\,836	&		\\  
CH$_{3}$OH			&	34	&	38\,--\,863	&	108.6\\   
$^{13}$CH$_{3}$OH	&	21	&	37\,--\,499	&	6.8\\ 
$^{13}$CH$_{3}$OH $v_{\rm t}$=1	&	2	&	373\,--\,670	&		\\
CH$_{3}$CN			&	17	&	410\,--\,1012	&	9.3\\ 
NO					&	12	&	84\,--\,232	&	11.1\\  
CN					&	8	&	54		&	8.5\\  
H$_{2}$CS			&	5	&	138\,--\,343	&	1.7\\ 
H$_{2}$CO			&	3	&	106\,--\,133	&	38.5\\ 
H$_{2}^{13}$CO		&	1	&	130		&	0.6\\  
HDCO				&	3	&	114\,--\,141	&	2.8\\ 
OCS					&	3	&	604\,--\,658	&	1.3\\ 
$o$-H$_{2}$O			&	1	&	61	&		320.3\\ 
$p$-H$_{2}$O			&	1	&	867	&		2.2\\ 
$o$-H$_{2}^{17}$O		&	1	&	61		&	9.4\\ 
$o$-H$_{2}^{18}$O		&	2	&	60\,--\,430	&	16.2\\
HDO					&	1	&	66		&	10.7\\ 
HC$_{3}$N			&	2	&	648\,--\,799	&	0.6\\ 
CO					&	1	&	83		&	1\,100\\ 
$^{13}$CO			&	1	&	79		&	174.3\\ 
C$^{17}$O			&	1	&	81		&	6.0\\ 
C$^{18}$O			&	1	&	79		&	24.3\\ 
C				&	1	&	24		&	38.7\\ 
NH$_{3}$				&	1	&	27		&	20.6\\ 
$^{15}$NH$_{3}$		&	1	&	27		&	0.7\\ 
HNC					&	1	&	91		&	10.7\\ 
N$_{2}$H$^{+}$		&	1	&	94		&	1.5\\ 
H$_2$S				&	1	&	166		&	4.4\\ 
CS					&	1	&	129		&	24\\ 
$^{13}$CS                       &	1	&	173		& 	v blend\\ 
SiO					&	1	&	190		&	17.6\\ 
$^{29}$SiO			&	1	&	187		&	1.4\\ 
$^{30}$SiO			&	1	&	185		&	0.5\\
HCS$^{+}$			&	1	&	186		&	v blend \\ 
NS					&	1	&	442		&	v blend\\ 
U-line				&	28	&	 		&	9.1\\  
T-line					&	36	&	 		&	7.2\\  
\hline
No. of species  	&		38	&\\
No. of  lines 		&		344	&\\
Total $\int{T^{\,*}_\mathrm{A} } \mathrm{d} \upsilon$ & &&		2\,455 \\
\hline
\end{tabular}
\end{table}

\clearpage

\begin{table*}
\caption{ Isotopologue abundance ratios from our survey, or from the literature, as compared to terrestrial values. The molecular abundance O/S ratios are also included.}
\label{isotope_ratios} 
\centering
\begin{tabular}{r r rr  r r r r r r}
\hline \hline
& [$^{12}$C/$^{13}$C] & [$^{32}$S/$^{34}$S]	 & [$^{34}$S/$^{33}$S]	&[$^{18}$O/$^{17}$O]& [$^{16}$O/$^{18}$O] & [D/H]& [O/S] & [$^{14}$N/$^{15}$N] & [$^{28}$Si/$^{29}$Si]\\
\hline

This survey& 57$\pm$14$^{{\,a}}$ &21$\pm$6$^{{\,b}}$\,--\,23$\pm$7$^{\,c}$ &4.9$^{{\,d}}$  &  3.6$^{{\,e}}$  && 0.001$^{{f}}$--\,0.03$^{{\,g}}$  &  15$^{{\,h}}$--\,20$^{{\,i}}$\\

Terrestrial& 89$^{{j}}$ & 22.5$^{{j}}$   & 5.5$^{{j}}$  &5.5$^{{k}}$&&&35$^v$&273$^{{l}}$&19.6$^{{\,m}}$\\

K88$^j$&47$^{+6}_{-5}$&20.2$^{+2.6}_{-2.1}$&5.7$^{+0.8}_{-0.6}$&&&\\

W\&R94$^n$&77$\pm$7&$\sim$22  & &3.2$\pm$0.2&560$\pm$25 &&&450$\pm$22\\

C96$^o$&75$\pm$21 &35$\pm$10 & 7.53$\pm$0.45 &&&\\

B01$^p$&65.0$\pm$9.2	&&&4.15$\pm$0.59&&\\
P81b$^q$&&&&3.9$\pm$0.2&&\\

Various surveys    &43$\pm$7$^r$&&&4.17$\pm$0.26$^{\,s}$ &330$\pm69^{\,{t}}$& 2$\times10^{-5}$ $^u$ && &16.9$\pm2.0^{{\,m}}$\\   
\hline
\end{tabular}
\begin{list}{}{}
\item$^{{a}}$From $^{12}$CH$_3$OH/$^{13}$CH$_3$OH. $^{{b}}$From $^{32}$SO/$^{34}$SO. $^{{c}}$From $^{32}$SO$_2$/$^{34}$SO$_2$. $^{{d}}$From $^{34}$SO/$^{33}$SO. $^{{e}}$From C$^{18}$O/C$^{17}$O. $^{{f}}$From HDO/H$_2^{17}$O Hot Core. $^{{g}}$From HDO/H$_2^{17}$O Compact Ridge.  $^{{h}}$From H$_2^{13}$CO/H$_2^{12}$CS CR. $^{{i}}$From H$_2^{17}$O/H$_2$S LVF.  $^{{j}}${Kahane \etal~(\cite{Kahane}), the envelope of IRC+10216}. $^{{k}}$Blake \etal~(\cite{Blake}). $^{{l}}$Ho \& Townes (\cite{Ho and Townes}). $^{{m}}$Penzias  (\cite{Penzias81a}). $^{{n}}${Wilson \& Rood (\cite{Wilson and Rood}) and references therein, local ISM.} $^{{o}}${Chin \etal~(\cite{Chin}), towards Orion KL}. $^{{p}}${Bensch \etal~(\cite{Bensch}), towards $\rho$ Ophiuchi Molecular Cloud}. $^{{q}}$Penzias  (\cite{Penzias81b}), towards Orion A. $^{{r}}${Savage \etal~(\cite{Savage}), towards Orion A using CN and $^{13}$CN.}  $^{{s}}${Wouterloot \etal~(\cite{Wouterloot}), towards $\rho$ Ophiuchi Molecular Cloud, value from LTE column densities.}  $^{{t}}$Olofsson (\cite{Olofsson thesis}), from S$^{18}$O observations of molecular cloud cores. $^{{u}}$Neufeld \etal~(\cite{Neufeld06}).  $^{{v}}$ Standard abundances, Grevesse \etal~(\cite{Grevesse}).

\end{list}
\end{table*}


\begin{table*} [!h]
\caption{H$_2$O and isotopologue parameters and derived column densities.} 
\label{result_tableWater} 
\centering
\begin{tabular} { l l rr   rrrrrrr}
\hline
\hline
Species &	Comp$^{a}$&	$\upsilon_{\mathrm{LSR}}$&	$\Delta \upsilon$ &	$T^{\,*}_{\mathrm{A}}$  & $T_\mathrm{k}$& Size	&$N^b$&$\tau$	& $N^c_{\tau\,\mathrm{corr}}$&$N_{\mathrm{ISO}}^d$ \\
	&		&	[km\,s$^{-1}$] &	  [km\,s$^{-1}$]&	  	[K]	  	&[K]	
	&[$\arcsec]$&&cm$^{-2}$]	&cm$^{-2}$]	
	&[cm$^{-2}$]\\              
\hline
$o$-H$_2^{16}$O & Total		&	 	&  	 &   	 &	 	 	 72	& (15)$^{e}$	&& $\sim$1100$^{f}$ &&
1.7$\times10^{18}$	\\
 & CR		&	 	&  &   		 &	 	115	& 	(6)$^{e}$&& $\sim$860$^{f}$ &&
5.6$\times10^{17}$	\\
 			&  LVF	 &	  	&  &   		&	 	72	& (15)$^{e}$	&
 &$\sim$1900$^{f}$ &&8.7$\times10^{17}$   \\
 			&  HVF	 &	5.3	&67.1 & 2.452		 	&	72	&70	 
			&8.7$\times10^{14}$&$\sim$910$^{f}$&8.0$\times10^{17}$&8.8$\times10^{17}$\\
$p$-H$_2^{16}$O&  HC	 & 4.4	&12.0 &0.181		 	&	200 	&10	&1.2$\times10^{19}$&0.3$^g$&  \\
$o$-H$_2^{17}$O& Total	 & 	 	& 	& 		 		&	72	&15	&8.6$\times10^{14}$&0.9$^f$  
			   &1.3$\times10^{15}$ &\\
			   &  CR	 &9.9 	&5.0	&0.076		 		&	115	&6	&3.2$\times10^{14}$&0.7$^f$  
			   &4.4$\times10^{14}$ &\\	
			   &   LVF	&8.7	&18.0&0.214		 	&	72	&15 	& 3.5$\times10^{14}$& 
			   1.5$^f$ &6.7$\times10^{14}$& \\ 
			   &   HVF	& 13.4	&30.1 &0.179 		 				&	72	&15 	& 4.9$\times10^{14}$& 
			   0.7$^f$ &6.8$\times10^{14}$& \\ 
$o$-H$_2^{18}$O   &  Total 		 & 	 	&   &	 	    	&	72	&15	&1.4$\times10^{15}$& 3.4$^f$ 
&5.0$\times10^{15}$&5.0$\times10^{15}$  \\
			&  CR 		 &10.1	 	&5.0  &	0.145	    	&	115	&6	&6.2$\times10^{14}$& 2.6$^f$ 
&1.8$\times10^{15}$&1.7$\times10^{15}$  \\
		       &  LVF	 &10.5	&18.0  &	0.275	 	&	72	&15 	& 4.6$\times10^{14}$
			& 5.9$^f$  & 2.7$\times10^{15}$& 2.6$\times10^{15}$ \\
			&  HVF	 &13.4	&33.4  &	0.298	 	&	72	&15 	& 9.3$\times10^{14}$
			& 2.8$^f$  & 2.8$\times10^{15}$& 2.7$\times10^{15}$ \\	
HDO			&  Total	&	 		&  &	&	72	&15	&9.1$\times10^{15}$	   &$\sim$1.5$^g$&&\\
			&  CR	&	8.5 		&4.6  &	0.459	 	&	115	&6	&1.8$\times10^{16}$	   
			&$\sim$3$^g$&&\\
			&  LVF	 &	 11.8		&18.0  &	0.273		&	72	&15 	&  4.5$\times10^{15}$& 
			$\sim$0.3$^g$&&\\
			&  HC	 &	 2.2		&13.4  &	0.216		&	200	&10 	&  1.5$\times10^{16}$&$\sim$0.5$^g$ 
			&&\\ 
\hline
\end{tabular}
\begin{list}{}{}
\item[$^{a}$] Total=the total integrated intensity is used, CR=Compact Ridge, LVF=Low Velocity Flow, HVF=High Velocity Flow, HC=Hot Core. 
\item[$^{b}$] Corrected for beam-filling. 
\item[$^{c}$]  Corrected for beam-filling, and for optical depth with factor $\tau/(1-e^{-\tau})$
\item[$^{d}$] Column calculated from isotopologue $o$-H$_2^{17}$O, beam-filling and optical depth corrected. 
\item[$^{e}$] Size from isotopologues. The full LVF $o$-H$_2^{16}$O emission may have a larger extent.
\item[$^{f}$] Calculated from the ratio of H$_2^{18}$O and H$_2^{17}$O column densities.
\item[$^{g}$] Calculated with Eq. \ref{tau}. 

\end{list}
\end{table*}

\begin{table*} [!h]
\caption{CO and isotopologue  parameters and derived column densities.} 
\label{result_tableCO} 
\centering
\begin{tabular} { l lrrrrrrrrrrr}
\hline
\hline
Species &	Comp$^{a}$&	$\upsilon_{\mathrm{LSR}}$&	$\Delta \upsilon$ &	$T^{\,*}_{\mathrm{A}}$& $\int{T^{\,*}_\mathrm{A} dv}$ & $T_\mathrm{k}$	&Size&$N$ &$\tau$	& $N_{\mathrm{ISO}}$(C$^{17}$O)& $N_{\mathrm{ISO}}$(C$^{18}$O)	& $N_{\mathrm{ISO}}$($^{13}$CO)\\
	&		&	  &	 &	  	[K]	 & [K km\,s$^{-1}$]	&[K]&[$\arcsec$]	
	&[cm$^{-2}$]	&
	&[cm$^{-2}$]\\              
\hline
CO	&	 PDR$^b$ 	&	 	&	   	&	  	& 	& 100&... &   & & 1.6$\times10^{18}$	& 1.5$\times10^{18}$ 	&1.2$\times10^{18}$\\
	&	 LVF  &	  	&	   &	  	& 	& 100& (30)$^d$ &   &&
2.5$\times10^{19}$	 &2.2$\times10^{19}$ & 2.3$\times10^{19}$   \\
	&	 HVF &	 	&	 &	 	& 	&  100&(70)$^d$&   & &&&	
	2.2$\times10^{18}$	\\
	
$^{13}$CO	&  N$^c$		 &	 8.1	&	 4.8	 &	 21.6		&109.9	&	100	&...&	5.7$\times10^{16}$&&5.4$\times10^{16}$&4.9$\times10^{16}$\\
			&  LVF	 &	 7.8	&	 18.0	&	 2.18		&41.6	&	100	&30&	
			3.9$\times10^{17}$&&4.2$\times10^{17}$&3.7$\times10^{17}$\\
			&  HVF	 &	 6.7	&	 48.7 &	 0.470	&24.2	&	100	&70&	
			5.2$\times10^{16}$&&&\\	
			
C$^{17}$O	&  N$^c$	 	 &	 8.8	&	 4.2	 &	 1.11		&4.95	&	100	&...&	2.5$\times10^{15}$&0.07&&\\
			&  LVF	 &	 9.9	&	 18.0	 &	 0.116	&2.21	&	100	&30&	
		2.0$\times10^{16}$&0.1&&\\	
			
C$^{18}$O	&  N$^c$		 &	 8.7	&	 4.2	 &	 3.81		&17.0	&	100	&...&	8.9$\times10^{15}$&0.3&9.8$\times10^{15}$&\\
			&  LVF	 &	 7.8	&	 18.0  &	 0.367	&7.00	&	100	&30&	
			6.7$\times10^{16}$&0.3& 7.7$\times10^{16}$&\\

H$_2$		&PDR$^{b}$	&	&	&		&			&		&		...            &&& 2.0$\times10^{22}$ & 1.8$\times10^{22}$& 1.5$\times10^{22}$\\
			&ER$^{b}$	&	&	&		&			&		&		...            &&& 2.0$\times10^{22}$ 
			& 1.8$\times10^{22}$& 1.5$\times10^{22}$\\
 			&LVF		&	&	&		&			&		&		(30)$^d$&&& 3.2$\times10^{23}$ 
			&2.8$\times10^{23}$& 2.3$\times10^{23}$\\
 			&HVF		&	&	&		&			&		&		(70)$^d$&&&   & & 
			3.9$\times10^{22}$\\
\hline
\end{tabular}
\begin{list}{}{}
\item[$^{a}$] N=Narrow, LVF=Low Velocity Flow, HVF=High Velocity Flow.
\item[$^{b}$] This is half of the column density obtained from the narrow components of the isotopolouges, since CO narrow component only has emission from the PDR, while the isotopologues have $\sim$ equal emission from  PDR and ER (as discussed in W06).
\item[$^{c}$] Consists of approximately equal emission from PDR and ER.
\item[$^{d}$] Size from CO isotopologues. The size of the full LVF CO emission is larger than the isotopologues and calculated to be 45$\arcsec$.
\end{list}
\end{table*}

\clearpage



\begin{table*}
\caption{SO$_2$ parameters.$^a$} 
\label{SO2 parameters } 
\centering
\begin{tabular} { l r  r  l r l r r r r    l }
\hline
\hline
$\nu_{{ul}}$ & $\Delta \nu_\mathrm{p}$ &  $\Delta \nu_\mathrm{g}$ &    Transition     & $E_u$  &  $A_{{ul}}$  & $T^{\,*}_{\mathrm{A}}$(peak) & Ampl.  & Width & $\int$ $T^{\,*}_{\mathrm{A}}$  d$\upsilon$  &    Note    \\
$[\mathrm{MHz}]$ &	 [MHz] & [MHz] & $J_{{K_a},{K_c}}$&	[K] &	[s$^{-1}$] & 	[mK] & [mK]  & [\kms]  &[K \kms] &\\ 
\hline  
491934.7 &       &      & $7_{4,4}-6_{3,3}$      &  65.0 & 9.49e-4 &       &     &       &       &         blend HDCO\\

541750.9 &   0.9 &   & $14_{3,11}-13_{2,12}$  &  119.0 &   6.31e-4 &   695 &   &   &  15.80   & \\
& &  0.0      &  &  &   &    & 143 &  5.0 &     &3G: CR \\
& & -0.1     &  &  &   &    & 266 &  18.0 &     &3G: LVF \\
& &  3.2      &&   &    && 301 &  34.9  &    &3G: HVF \\

541810.6 &      &     & $30_{6,24}-30_{5,25}$  &   516.7 &   1.34e-3 &      &    &      &      &        weak\\
543413.5 &  $-$2.5 & $-$0.5 & $29_{2,28}-28_{1,27}$  &  401.5 &    1.64e-3 &   457 & 411 &  25.5 &  10.70   & blend  CH$_{3}$OH \& SO$_{2}$\\
543467.7 &      &     & $37_{3,35}-37_{2,36}$  &  664.0 &   7.77e-4 &      &    &      &      &   v blend 
CH$_{3}$OH \& SO$_{2}$\\
 545318.5 &   3.5 &  5.5 & $37_{6,32}-37_{5,33}$  &  736.9 &   1.43e-3 &    66 & 41  &  17.5 & 0.72   & \\
 545517.3 &  $-$0.7 & $-$2.7 & $35_{6,30}-35_{5,31}$  &  669.3 &   1.42e-3 &    99 &  63 &  16.5 & 1.07  & \\
 546579.8 &  $-$1.2 &  2.8 & $19_{3,17}-18_{2,16}$  &  197.0 &    7.40e-4 &   546 & 488 &  25.1 &  12.90  & \\
 547802.2 &   1.2 & $-$1.8 & $28_{6,22}-28_{5,23}$  &  462.2 &   1.36e-3 &   160 & 136 &  16.8 & 2.41   & \\
 548734.3 &  $-$3.7 &$-$2.7 & $31_{6,26}-31_{5,27}$  &  545.3 &   1.40e-3 &   102 &  87 &  11.3 &  1.02   & \\
 548838.9 &      &     & $40_{4,36}-39_{5,35}$  &  808.3 &   5.52e-4 &      &    &      &      &       blend C$^{18}$O\\
549303.3 &   3.3 & 11.3 & $10_{4,6}-9_{3,7}$     &  89.8 &   1.07e-3 &   890 & 857 &  32.8 &  29.20  & blend CH$_{3}$OH ,\\
&&&&&&&&&&  $^{13}$CH$_{3}$OH \& $^{34}$SO\\
549566.4 &  15.4 &  5.4 & $30_{1,29}-29_{2,28}$  &  427.9 &  1.73e-3 &   455 & 369 &  24.7 & 9.38   & blend CH$_{3}$OCH$_{3}$\\
 550946.7 &      &     & $29_{6,24}-29_{5,25}$  &  488.9 &   1.40e-3 &      &    &      &     &        blend $^{13}$CO\\
 551622.9 &  $-$3.1 &     & $38_{2,36}-38_{1,37}$  &  697.7 &   7.94e-4 &    43 &    &      & 0.31  & weak\\
552069.4 & $-$11.6 &     & $34_{1,33}-34_{0,34}$  &  542.8 &  4.43e-4 &      &    &      &      &        blend H$_{2}^{17}$O, SO$_{2}$\\
552078.9 &  $-$2.1 &    0.0 & $26_{6,20}-26_{5,21}$  &  411.4 &   1.37e-3 &   223 & 197 &  17.1 & 3.33   & blend H$_{2}^{17}$O, SO$_{2}$\\
553164.9 &   2.9 &  2.9 & $27_{6,22}-27_{5,23}$  &  436.3 &   1.39e-3 &   224 & 225 &  22.5 &      &        v blend \& 3G CH$_{3}$OH;\\
554212.8 &   7.8 &  1.8 & $31_{1,31}-30_{0,30}$  &  431.5 &   2.34e-3 &   545 & 490 &  21.4 &  10.90   & blend CH$_{3}$OH\\
555121.5 &  $-$2.5 &     & $24_{6,18}-24_{5,19}$  &   364.4 &   1.37e-3 &   264 &    &      & 4.20 &   blend (CH$_{3})_2$O\\
 555204.1 &  $-$6.9 &     & $25_{6,20}-25_{5,21}$  &  387.4 &    1-38e-3 &   223 & 188 &  19.9 & 3.75  & \\
555666.3 &   0.7 & $-$0.7 & $5_{5,1}-4_{4,0}$      &  75.1 &   2.18e-3 &   742 & 677 &  28.1 &    20.00 & blend CH$_{3}$OCH$_{3}$\\
 556959.9 &      &     & $23_{6,18}-23_{5,19}$  &  342.2 &   1.37e-3 &      &    &      &      &        blend H$_{2}$O\\
 557283.2 &  $-$3.8 &  0.2 & $22_{6,16}-22_{5,17}$  &  321.0&  1.36e-3 &   303 & 271 &  22.9 & 6.72   & \\
 558101.2 &      &     & $15_{9,7}-16_{8,8}$    &  308.6 &   1.54e-3 &      &    &      &      &        blend SO\\
558390.9 &  $-$1.1 & $-$1.1 & $21_{6,16}-21_{5,17}$  &  300.8 &   1.35e-3 &   335 & 290 &    20.0 & 5.55  & blend \& 3G CH$_{3}$OH;\\
 558555.8 &   4.8 &  9.2 & $16_{3,13}-16_{0,16}$  &   147.8 &    3.90e-5 &   115 &  80 &  23.8 & 1.93   & \\
 558812.5 &  $-$2.5 &  2.5 & $20_{6,14}-20_{5,15}$  &   281.4 &   1.34e-3 &   327 & 291 &  23.7 & 7.17   & \\
 559500.4 &  $-$0.6 &$-$0.6 & $19_{6,14}-19_{5,15}$  &  263.0 &   1.32e-3 &   352 & 298 &  23.1 & 7.31   & \\
 559882.1 &  $-$1.9 &  0.1 & $18_{6,12}-18_{5,13}$  &  245.5 &   1.31e-3 &   331 & 302 &  21.7 & 6.59   & \\
560318.9 &  $-$1.1 & $-$1.1 & $17_{6,12}-17_{5,13}$  &  229.0 &   1.29e-3 &   393 & 376 &  18.1 &      &  v blend \& 2G  CH$_{3}$OH;\\
560613.5 &  $-$1.5 &  1.5 & $16_{6,10}-16_{5,11}$  &  213.3 &   1.27e-3 &   354 & 342 &  23.4 & 8.52   & blend $^{34}$SO$_{2}$\\
   560891.0 &    $-$3.0 &     & $15_{6,10}-15_{5,11}$  &  198.6 &   1.25 e-3&   378 & 325 &  22.6 & 7.68   & \\
 561094.8 & $-$1.2 & $-$1.2 & $14_{6,8}-14_{5,9}$    &  184.8 &  1.22e-3 &   326 & 304 &    23.0 & 7.30  & \\
561265.6 &  $-$0.4 &  1.4 & $13_{6,8}-13_{5,9}$    &  171.9 &   1.19e-3&   467 & 382 &  23.2 & 9.26   & blend CH$_{3}$OH\\
 561361.4 &      &     & $21_{3,19}-20_{2,18}$  &   234.7 &  8.31e-4 &   522 &    &      &      &        blend SO$_{2}$\\
561392.9 &   0.9 &     & $12_{6,6}-12_{5,7}$    &  160.0 &   1.14e-3 &   436 &    &      &    &        blend CH$_{3}$OH \& SO$_{2}$\\
 561490.5 &   0.5 &     & $11_{6,6}-11_{5,7}$    &   149.0 &  1.09e-3 &   383 &    &      &      &      blend SO$_{2}$\\
 561560.3 &  $-$1.7 &     & $10_{6,4}-10_{5,5}$    &  138.8 &  1.02e-3 &   402 &    &      &      &        v blend SO$_{2}$\\
 561608.6 &  $-$0.4 &     & $9_{6,4}-9_{5,5}$      &  129.7 &   9.37e-4 &   502 &    &      &      &        v blend SO$_{2}$\\
 561639.3 &  $-$0.7 &     & $8_{6,2}-8_{5,3}$      &  121.4 &   8.18e-3 &   564 &    &      &      &        v blend SO$_{2}$\\
 561656.7 &  16.7 &     & $7_{6,2}-7_{5,3}$      &  114.0 &   6.50e-4 &   547 &   &      &      &        v blend SO$_{2}$\\
 561664.2 &  24.2 &     & $6_{6,0}-6_{5,1}$      &  107.6 &   3.99e-4 &   547 &    &      &      &        v blend SO$_{2}$\\
 567592.7 &  $-$1.3 &     & $11_{4,8}-10_{3,37}$   &  100.0 &  1.12e-3 &   822 &    &      &  16.80    & \\
571532.6 &  $-$7.4 & $-$1.4 & $32_{2,30}-31_{3,29}$  & 504.3 &  1.34e-3 &   465 & 171 &  21.3 &     &        blend \& 2G SO$_{2}$;\\
571553.3 &  $-$2.7 &  0.3 & $32_{0,32}-31_{1,31}$  & 459.0 &  2.58 e-3&   465 & 343 &  21.2 &  11.30     & blend \& 2 G SO$_{2}$;\\
574587.8 &  $-$7.2 & $-$3.2 & $23_{3,21}-22_{2,20}$  & 276.0 &  9.46e-4 &   276 & 248 &  21.4 & 5.47      & \\
574807.3 &  $-$2.7 &  3.3 & $6_{5,1}-5_{4,2}$      &  80.7 &   2.07e-3 &   478 & 410 &  23.4 &  9.97   & blend  $^{34}$SO$_{2}$\\
 576042.1 &  $-$5.9 & $-$4.9 & $31_{2,30}- 30_{1,29}$ &  455.6&  2.03e-3 &   200 & 165 &    16.0 &  2.72   & \\   
\hline
Total no. &49&&&&&&&&&\\
\hline
\end{tabular}
\begin{list}{}{}
\item$^{{a}}\nu_{{ul}}$\,=\,rest frequency of the transition $u \to l$; $\Delta \nu_\mathrm{p}$\,=\,the difference between freq and the observed frequency at the peak temperature; $\Delta \nu_\mathrm{g}$\,=\,the difference between freq and the frequency of the Gaussian fit; Transition\,=\,the quantum numbers for the transition; $E_u$\,=\,the upper state energy; $A_{{ul}}$\,=\,the Einstein $A$-coefficient; $T^{\,*}_{\mathrm{A}}$(peak)\,=\,the observed peak temperature of the transition; Ampl.\,=\,the peak temperature of the Gaussian fit; Width\,=\,the width of the transition from the Gaussian fit;  $\int T^{\,*}_{\mathrm{A}}$ d$\upsilon$\,=\,the integrated intensity from the observed spectra; Note: 2G or 3G denotes a two- or three-component Gaussian fit, v blend denotes a visible blend, Total no.\,=\,the total number of transitions of the molecule in the table. All these lines are also marked in the spectra shown in  \mbox{{\bf{paper I}}} Appendix A, and listed in the on-line Table of \mbox{{\bf{paper I}}}. These total numbers of lines  include not visible blends, which are not counted in  
Table \ref{summary Table}. 
\end{list}
\end{table*} 

\begin{table*}
\caption{ $^{34}$SO$_2$ parameters.$^a$} 
\label{34SO2 parameters } 
\centering
\begin{tabular} { l r  r  l r l r r r r    l }
\hline
\hline
$\nu_{{ul}}$ & $\Delta \nu_\mathrm{p}$ &  $\Delta \nu_\mathrm{g}$ &    Transition     & $E_u$  &  $A_{{ul}}$  & $T^{\,*}_{\mathrm{A}}$(peak) & Ampl.  & Width & $\int$ $T^{\,*}_{\mathrm{A}}$  d$\upsilon$  &    Note    \\
$[\mathrm{MHz}]$ &	 [MHz] & [MHz] & $J_{{K_a},{K_c}}$&	[K] &	[s$^{-1}$] & 	[mK] & [mK]  & [\kms]  &[K \kms] &\\ 
\hline 
 542097.7 & $-$0.3 &  4.7   & $11_{6,6}-11_{5,7}$   & 145.8 & 9.83e-4 &      & 102 &    16.0 && from 3G CH$_{3}$OH\\
 542177.4 &     &     & $10_{6,4}-10_{5,5}$   &  135.7 & 9.23e-4 &      &    &      &      &       \\
 542233.5 &     &     & $9_{6,4}-9_{5,5}$     &  126.5 & 8.44e-4 &      &    &      &      &      weak\\
 542270.1 &     &     & $8_{6,2}-8_{5,3}$     &  118.3 & 7.37e-4 &     &    &      &      &       weak\\
 542291.7 &     &     & $7_{6,2}-7_{5,3}$     &   110.9 & 5.86e-4 &      &    &      &      &      weak\\
 542302.3 &     &     & $6_{6,0}-6_{5,1}$     &  104.5 &  3.60e-4 &      &    &      &      &       weak\\
\hline
\end{tabular}
\begin{list}{}{}
\item$^{{a}}$Notation as in Table \ref{SO2 parameters }.
\end{list}
\end{table*}

\begin{table*}
\caption{ SO parameters.$^a$} 
\label{SO parameters } 
\centering
\begin{tabular} { l r  r  l r l r r r r    l }
\hline
\hline
$\nu_{{ul}}$ & $\Delta \nu_\mathrm{p}$ &  $\Delta \nu_\mathrm{g}$ &    Transition     & $E_u$  &  $A_{{ul}}$  & $T^{\,*}_{\mathrm{A}}$(peak) & Ampl.  & Width & $\int$ $T^{\,*}_{\mathrm{A}}$  d$\upsilon$  &    Note    \\
$[\mathrm{MHz}]$ &	 [MHz] & [MHz] & $N_J$&	[K] &	[s$^{-1}$] & 	[mK] & [mK]  & [\kms]& [K \kms]   &\\    
\hline 
487708.5 & 0.5 & $-$1.5 & $7_{7}-6_{7}$     &  71.0 &  1.70e-5 &   206 &  178 &  13.8 & 2.67  & \\
 558087.7 & 2.7 &     & $13_{12}-12_{11}$ &  194.4 & 2.28e-3 &  2094 &     &      &    59.00   & blend SO$_{2}$\\  
  &    &  2.7 & && &      &  160 &  4.7 &      &        3G; CR \\
 &    &  0.7 & && &      & 759 &  18.0 &      &       3G; LVF \\
 &    &  2.7 & && &      & 1178 &  35.5 &      &       3G; HVF \\
 559319.8 & 3.8 &    & $13_{13}-12_{12}$ &  201.1 & 2.30e-3 &  1996 &   &      &  56.90   & \\
   &    &  3.8 & && &      &  165 &  5.0 &      &        3G; CR\\
  &    &  $-$0.2 & &&&      &  674 &  18.0 &      &        3G; LVF \\
    &    &  4.8 & &&&      & 1163 &  35.3 &      &        3G; HVF \\
 560178.7 & 2.7 &    & $13_{14}-12_{13}$ &  192.7 & 2.32e-3 &  2168 &   &   &  61.70   & \\
 &    &  2.7 & &&&      &  213 &  5.0 &      &       3G; CR \\
  &    &  0.7 & && &      & 666 &  18.0 &      &        3G; LVF \\
    &    &  1.7 & && &      & 1285 &  35.6 &      &        3G; HVF \\
 568741.6 & 9.6 &  1.6 & $9_{9}-8_{9}$     & 106.1 &  1.90e-5 &   105 &   78 &  15.9 & 1.24   & \\
\hline
Total no. &5&&&&&&&&&\\
\hline
\end{tabular}
\begin{list}{}{}
\item$^{{a}}$Notation as in Table \ref{SO2 parameters }.
\end{list}
\end{table*}

\begin{table*}
\caption{ $^{33}$SO parameters.$^a$} 
\label{33SO parameters } 
\centering
\begin{tabular}{ l r  r  l r l r r r r   l }
\hline
\hline
$\nu_{{ul}}$ & $\Delta \nu_\mathrm{p}$ &  $\Delta \nu_\mathrm{g}$ &    Transition     & $E_u$  &  $A_{{ul}}$  & $T^{\,*}_{\mathrm{A}}$(peak) & Ampl.  & Width & $\int$ $T^{\,*}_{\mathrm{A}}$  d$\upsilon$  &    Note    \\
$[\mathrm{MHz}]$ &	 [MHz] & [MHz] &$N_J$&	[K] &	[s$^{-1}$] & 	[mK] & [mK]  & [\kms]& [K \kms]  &\\ 
\hline 
 552429.8 & $-$2.2 &  1.8 & $12_{13}-11_{12}$ & 192.5 & 2.18 e-3&   138 & 112 &  20.3 & 2.41  &  \\
 553681.3 & $-$6.7 &$-$2.7 & $13_{13}-12_{12}$ & 199.2 & 2.20e-3 &   113 &  85 &    17.0 &    & Blend 2 U-lines\\
 &&&&&&&&&& \& CH$_3$CN \\
 554555.6 & $-$3.4 & $-$3.4 & $14_{13}-13_{12}$ & 190.8 & 2.24e-3 &   110 &  100 &    20.5 & 2.06 & \\
\hline
Total no. &3&&&&&&&&&\\
\hline
\end{tabular}
\begin{list}{}{}
\item$^{{a}}$Notation as in Table \ref{SO2 parameters }.
\end{list}
\end{table*}

\begin{table*}
\caption{ $^{34}$SO parameters.$^a$} 
\label{34SO parameters } 
\centering
\begin{tabular}{ l r  r  l r l r r r r   l }
\hline
\hline
$\nu_{{ul}}$ & $\Delta \nu_\mathrm{p}$ &  $\Delta \nu_\mathrm{g}$ &    Transition     & $E_u$  &  $A_{{ul}}$  & $T^{\,*}_{\mathrm{A}}$(peak) & Ampl.  & Width & $\int$ $T^{\,*}_{\mathrm{A}}$  d$\upsilon$  &    Note    \\       
$[\mathrm{MHz}]$ &	 [MHz] & [MHz] & $N_J$&	[K] &	[s$^{-1}$] & 	[mK] & [mK]  & [\kms]& [K \kms]  &\\ 
\hline 
 547119.6 & $-$5.4 & $-$2.4 & $12_{13}-11_{12}$ &  190.7 & 2.15e-3 &   380 & 326 &  27.6 & 9.35 & \\
 548389.8 & $-$2.2 & 0.8 & $13_{13}-12_{12}$ & 197.4 & 2.17 e-3&   324 & 304 &  22.8 & 7.31 & \\
 549278.7 &     &    & $14_{13}-13_{12}$ & 189.0 &  2.19e-3 &      &    &      &      &       blend 
 CH$_{3}$OH,  \\
 &&&&&&&&&&SO$_{2}$   \& $^{13}$CH$_{3}$OH\\
 561700.4 &     &    & $9_{9}-9_{8}$     & 104.3 &   1.83e-5 &      &    &      &      &       blend SO$_{2}$ \&  
 C$^{17}$O\\
\hline
Total no. &4&&&&&&&&&\\
\hline
\end{tabular}
\begin{list}{}{}
\item$^{{a}}$Notation as in Table \ref{SO2 parameters }.
\end{list}
\end{table*}

\begin{table*} [!h]
\caption{SiO and isotopologue parameters.$^a$} 
\label{SiO parameters } 
\centering
\begin{tabular}   {l l r  r  l r l r r r r    l }
\hline
\hline
Species & $\nu_{{ul}}$ & $\Delta \nu_\mathrm{p}$ &  $\Delta \nu_\mathrm{g}$ &    Transition     & $E_u$  &  $A_{{ul}}$  & $T^{\,*}_{\mathrm{A}}$(peak) & Ampl.  & Width & $\int$ $T^{\,*}_{\mathrm{A}}$  d$\upsilon$  &    Note    \\
  & $[\mathrm{MHz}]$ &	 [MHz] & [MHz] & $J$&	[K] &	[s$^{-1}$] & 	[mK] & [mK]  & [\kms] &[K \kms] &\\ 
\hline 
SiO	&	564249.2 & $-$2.8 &  & $13- 12$   & 189.6 & 9.66e-3 &   729 &  &   & 17.60   & \\
&&&  2.2      &&&&& 391 & 18.0&&2G; LVF\\
&&&  $-$1.8 &&&&& 315 &  31.5&&2G; HVF\\
$^{29}$SiO &	557184.4 & 2.4 & 1.4 & $13-12$    & 187.2 & 9.31e-3 &    82 &  72 &  21.2 & 1.39&     blend H$_2$O  \\
$^{30}$SiO	&	550605.2 & $-$9.8 & $-$5.8 & $13-12$    & 185.0 & 8.98e-3 &    71 &  59 &   7.5 & 0.45  &? \\
\hline
Total no. &3&&&&&&&&&\\
\hline
\end{tabular}
\begin{list}{}{}
\item$^{{a}}$Notation as in Table \ref{SO2 parameters }.
\end{list}
\end{table*} 

\begin{table*}
\caption{H$_2$S parameters.$^a$} 
\label{H2S parameters} 
\centering
\begin{tabular} { l r  r  l r l r r r r    l }
\hline
\hline
$\nu_{{ul}}$ & $\Delta \nu_\mathrm{p}$ &  $\Delta \nu_\mathrm{g}$ &    Transition     & $E_u$  &  $A_{{ul}}$  & $T^{\,*}_{\mathrm{A}}$(peak) & Ampl.  & Width & $\int$ $T^{\,*}_{\mathrm{A}}$  d$\upsilon$  &    Note    \\
$[\mathrm{MHz}]$ &	 [MHz] & [MHz] &$J_{{K_a},{K_c}}$&	[K] &	[s$^{-1}$] & 	[mK] & [mK]  & [\kms]& [K \kms]   &\\  
\hline     
567079.6 &     &     & $6_{4,2}-6_{3,3}$ & 540.2 & 6.52e-4 &      &   &      &      &       blend NO  \\
568050.7 & $-$7.3 &     & $3_{3,1}-3_{2,2}$ & 165.9 & 3.32e-4 &   247 &    &      & 4.41  &  \\
 &      & $-$7.3 & &&&       & 106 &   7.5 &       &        2G; HC  \\
 &      & $-$0.3 & && &       & 145 &  24.4 &       &         2G; Plateau  \\
\hline
Total no. &2&&&&&&&&&\\  
\hline
\end{tabular}
\begin{list}{}{}
\item$^{{a}}$Notation as in Table \ref{SO2 parameters }.
\end{list}
\end{table*} 

 \begin{table*}
\caption{CH$_3$CN parameters.$^a$} 
\label{CH3CN parameters} 
\centering
\begin{tabular} { l r  r  l r l r r r r    l }
\hline
\hline
$\nu_{{ul}}$ & $\Delta \nu_\mathrm{p}$ &  $\Delta \nu_\mathrm{g}$ &    Transition     & $E_u$  &  $A_{{ul}}$  & $T^{\,*}_{\mathrm{A}}$(peak) & Ampl.  & Width & $\int$ $T^{\,*}_{\mathrm{A}}$  d$\upsilon$  &    Note    \\
 $[\mathrm{MHz}]$ &	 [MHz] & [MHz] & $J_K$&	[K] &	[s$^{-1}$] & 	[mK] & [mK]  & [\kms]& [K \kms] &\\ 
\hline    
550671.7 &          & & $30_{9}-29_{9}$               &  984.6 & 1.34e-2 &    &&&&   blend CH$_{3}$OH \\
550850.0 &              $-$6.0 &     & $30_{8}-29_{8}$               &  864.2 & 1.37e-2 &     &&&&   v blend $^{13}$CO \\ 
551007.4 &             0.4 &     & $30_{7}-29_{7}$               &  758.0 &  1.39e-2&      &&&&  v blend $^{13}$CO\\
551143.9 &            $-$5.1 & $-$5.1 & $30_{6}-29_{6}$               &  665.8 & 1.41e-2 &   145 & 125 &  13.2 &      &       v blend NO; from\\
&&&&&&&&&& 5G CH$_{3}$CN, NO\\
&&&&&&&&&&  \& CH$_{3}$OH\\
\hline
\end{tabular}
\begin{list}{}{}
\item$^{{a}}$Notation as in Table \ref{SO2 parameters }.
$^{\mathrm{*}}$Vibrationally excited transition, $v_{8}=1$.
\end{list}
\end{table*}

\begin{table*} [!h]
\caption{NH$_3$ and  $^{15}$NH$_3$ parameters.$^a$} 
\label{NH3 parameters } 
\centering
\begin{tabular}   {l l r  r  l r l r r r r   l }
\hline
\hline
Species & $\nu_{{ul}}$ & $\Delta \nu_\mathrm{p}$ &  $\Delta \nu_\mathrm{g}$ &    Transition     & $E_u$  &  $A_{{ul}}$  & $T^{\,*}_{\mathrm{A}}$(peak) & Ampl.  & Width & $\int$ $T^{\,*}_{\mathrm{A}}$  d$\upsilon$  &    Note    \\
&  $[\mathrm{MHz}]$ &	 [MHz] & [MHz] & $J_{K}$  &	[K] &	[s$^{-1}$] & 	[mK] & [mK]  & [\kms]  &[K \kms] &\\ 
\hline 
 NH$_3$& 572498.2 & 1.2 &   & $1_{0}-0_{\,0}$ & 27.5 & 1.61e-3 &  2409 &     &      & 20.60   & \\
	&   	&    & 1.2 & &&  &      & 1704 &   4.9 &     &      2G fit; CR \\  
	&	 &    & 2.2 & && &      &  694 &  16.2 &    &      2G fit; HC \\ 
 $^{15}$NH$_3$ &  	572112.8 & -0.2 &   & $1_{0}-0_{\,0}$ & 27.5 & 1.57e-3 &   167 & 150 &     6.7 & 1.02   & \\
 \hline
Total no. &2&&&&\\
\hline
\end{tabular}
\begin{list}{}{}
\item$^{{a}}$Notation as in Table \ref{SO2 parameters }.
\end{list}
\end{table*} 

\begin{table*}
\caption{HC$_3$N parameters.$^a$} 
\label{HC3N parameters} 
\centering
\begin{tabular} { l r  r  l r l r r r r    l }
\hline
\hline
$\nu_{{ul}}$ & $\Delta \nu_\mathrm{p}$ &  $\Delta \nu_\mathrm{g}$ &    Transition     & $E_u$  &  $A_{{ul}}$  & $T^{\,*}_{\mathrm{A}}$(peak) & Ampl.  & Width & $\int$ $T^{\,*}_{\mathrm{A}}$  d$\upsilon$  &    Note    \\
$[\mathrm{MHz}]$ &	 [MHz] & [MHz] & $J$&	[K] &	[s$^{-1}$] & 	[mK] & [mK]  & [\kms]& [K \kms]  &\\ 
\hline    
490955.7 & $-$6.3 &$-$5.3 & $54-53$    & 648.2& 9.50e-3 &    91 &  86 &   6.4 & 0.53   & blend CH$_{3}$OH \\
545417.1 & $-$3.9 &     & $60-59$    &  798.7 & 1.30e-2 &    70 &  57  &  10.3    & 0.60     &  \\
\hline
Total no. &2&&&&&&&&&\\
\hline
\end{tabular}
\begin{list}{}{}
\item$^{{a}}$Notation as in Table \ref{SO2 parameters }.
\end{list}
\end{table*} 

\clearpage

\begin{table*}
\caption{OCS parameters.$^a$} 
\label{OCS parameters} 
\centering
\begin{tabular} { l r  r  l r l r r r r   l }
\hline
\hline
$\nu_{{ul}}$ & $\Delta \nu_\mathrm{p}$ &  $\Delta \nu_\mathrm{g}$ &    Transition     & $E_u$  &  $A_{{ul}}$  & $T^{\,*}_{\mathrm{A}}$(peak) & Ampl.  & Width & $\int$ $T^{\,*}_{\mathrm{A}}$  d$\upsilon$  &    Note    \\
 $[\mathrm{MHz}]$ &	 [MHz] & [MHz] & $J$&	[K] &	[s$^{-1}$] & 	[mK] & [mK]  & [\kms]& [K \kms]   &\\  
\hline    
546859.8 & $-$3.2 &      & $45-44$    &  603.9 & 4.81e-4 &    67 &   &       & 0.31    & blend SO$_{2}$, $^{13}$CH$_{3}$OH  \\
558990.5 & $-$4.5 & $-$2.5 & $46-45$    &  630.7  & 5.14e-4 &    88 &  88 &   6.5 & 0.58    &  \\
571119.7 & $-$2.3 &      & $47-46$    & 658.1  & 5.49e-4 &    68 &     &     & 0.41  & ? \\
\hline
Total no. &3&&&&&&&&&\\
\hline
\end{tabular}
\begin{list}{}{}
\item$^{{a}}$Notation as in Table \ref{SO2 parameters }.
\end{list}
\end{table*}

\begin{table*}
\caption{CH$_3$OH parameters.$^a$} 
\label{CH3OH parameters 1 } 
\centering
\begin{tabular}{ l r  r  l r l r r r r   l }
\hline
\hline
$\nu_{{ul}}$ & $\Delta \nu_\mathrm{p}$ &  $\Delta \nu_\mathrm{g}$ &    Transition     & $E_u$  &  $A_{{ul}}$  & $T^{\,*}_{\mathrm{A}}$(peak) & Ampl.  & Width & $\int$ $T^{\,*}_{\mathrm{A}}$  d$\upsilon$  &    Note    \\
$[\mathrm{MHz}]$ &	 [MHz] & [MHz] & $J_{{K_a},{K_c}}$&	[K] &	[s$^{-1}$] & 	[mK] & [mK]  & [\kms]&[K \kms] &\\ 
\hline  
 486940.9 &             0.9 &     & $4_{2,3}-4_{1,4}$ A     &  60.9 &    5.45e-4 &   884 &   &    & 5.13  & \\
  &                &  0.9 &&&& &  635 &   3.8 &      &      2G; CR \\
 &               &$-$0.1 & &&&      &  241 &  10.1 &            & 2G; HC \\
 487531.9 &             0.9 &   & $10_{1,9}-9_{1,8}$ A    & 143.3 &   5.15e-4 &   855 &   &    & 4.85  & \\
  &                &  0.9 &    & & &      &  509 &   3.3 &      &       2G; CR \\
 &                & $-$0.1 &   & & &      &  359 &   8.1 &      &       2G; HC \\
\hline
\end{tabular}
\begin{list}{}{}
\item$^{{a}}$Notation as in Table \ref{SO2 parameters }.
$^{\mathrm{*}}$Vibrationally excited transition, $v_{t}=1$.
\end{list}
\end{table*} 

\clearpage

\begin{table*}
\addtocounter{table}{-1}
\caption{\emph{continued}} 
\centering
\begin{tabular}{ l r  r  l r l r r r r   l }
\hline
\hline
$\nu_{{ul}}$ & $\Delta \nu_\mathrm{p}$ &  $\Delta \nu_\mathrm{g}$ &    Transition     & $E_u$  &  $A_{{ul}}$  & $T^{\,*}_{\mathrm{A}}$(peak) & Ampl.  & Width & $\int$ $T^{\,*}_{\mathrm{A}}$  d$\upsilon$  &    Note    \\
$[\mathrm{MHz}]$ &	 [MHz] & [MHz] & $J_{{K_a},{K_c}}$&	[K] &	[s$^{-1}$] & 	[mK] & [mK]  & [\kms] &[K \kms] &\\ 
\hline  
551228.6$^*$ & $-$1.4 &     & $10_{1,10}-10_{2,9}$ A  &  452.4 &   8.36e-4 &   237 &     &      &      &       blend  CH$_{3}$OCH$_{3}$,\\
&&&&&&&&&& NO, \& CH$_{3}$CN\\
&      &      & && &       &   90 &   3.3 &       &       5G CH$_{3}$CN,\\
&      &      & && &       &  152 &   9.4 &       &        NO \& CH$_{3}$OH \\
 551736.2$^*$ & $-$0.8 &   & $9_{1,9}-9_{2,8}$ A     &  429.3 &    8.40e-4 &   248 &    &      & 1.33   & \\
&     & $-$0.8 & &&& &  101 &   2.5 &      &        2G; CR \\
\hline
\end{tabular}
\end{table*} 

\clearpage

\begin{table*}
\addtocounter{table}{-1}
\caption{\emph{continued}} 
\centering
\begin{tabular}{ l r  r  l r l r r r r   l }
\hline
\hline
$\nu_{{ul}}$ & $\Delta \nu_\mathrm{p}$ &  $\Delta \nu_\mathrm{g}$ &    Transition     & $E_u$  &  $A_{{ul}}$  & $T^{\,*}_{\mathrm{A}}$(peak) & Ampl.  & Width & $\int$ $T^{\,*}_{\mathrm{A}}$  d$\upsilon$  &    Note    \\
$[\mathrm{MHz}]$ &	 [MHz] & [MHz] &$J_{{K_a},{K_c}}$&	[K] &	[s$^{-1}$] & 	[mK] & [mK]  & [\kms]  &[K \kms] &\\  
\hline  
556603.7 &            $-$1.3 &    & $25_{1,25}-24_{2,22}$ A &  760.8 &   2.66e-4 &    &     &     &     &       v blend CH$_{3}$OH\\
557115.2$^*$ & $-$0.8 &$-$3.8 & $11_{1,10}-11_{2,9}$ A  &  478.2 &    8.50e-4 &   257 &  250 &   9.1 & 2.27   & \\
557147.0$^*$ &   $-$4.0 &     & $11_{9,3}-10_{8,2}$ A   &  836.5 &     1.40e-3 &    86 &   74 &   7.2 & 0.26   & blend CH$_{3}$OH\\
557147.0$^*$ &   $-$4.0 &     & $11_{9,2}-10_{8,3}$ A   &  836.5 &   1.40e-3 &    86 &   74 &   7.2 & 0.26   & blend CH$_{3}$OH\\
557676.6$^*$ & $-$0.4 & & $12_{1,11}-12_{2,10}$ A &  506.0 &   8.46e-4 &   189 &   &    & 1.01   & \\
&     &  0.6 & &&& &  107 &   3.5 &      &        2G; CR \\
\hline
\end{tabular}
\end{table*} 

\clearpage
  
 \begin{table*}
\caption{$^{13}$CH$_3$OH parameters.$^a$} 
\label{13CH3OH parameters 4} 
\centering
\begin{tabular} { l r  r  l r l r r r r   l }
\hline
\hline
$\nu_{{ul}}$ & $\Delta \nu_\mathrm{p}$ &  $\Delta \nu_\mathrm{g}$ &    Transition     & $E_u$  &  $A_{{ul}}$  & $T^{\,*}_{\mathrm{A}}$(peak) & Ampl.  & Width & $\int$ $T^{\,*}_{\mathrm{A}}$  d$\upsilon$  &    Note    \\
$[\mathrm{MHz}]$ &	 [MHz] & [MHz] & $J_{{K_a},{K_c}}$&	[K] &	[s$^{-1}$] & 	[mK] & [mK]  & [\kms] &[K \kms] &\\ 
\hline    
488153.5 &            $-$1.5 &$-$0.5 & $4_{1,4}-3_{0,3}$ A     &  37.0  &   7.46e-4 &   204 & 205 &   4.9 & 1.04   &  \\
488302.6 &             0.6 & $-$0.4 & $4_{2,3}-4_{2,3}$ A     &  60.5 &    5.50e-4 &   154 & 151 &   2.3 & 0.34   &  \\
490299.4 &             0.4 &     & $5_{2,4}-5_{1,5}$ A     &  71.8&   5.75e-4 &   168 &    &      & 0.59  & \\  
491170.0 &               0.0 &    1.0 & $5_{3,3}-4_{2,2}$ A     &  84.0  &   6.52e-4 &   155 & 155 &   4.4 & 0.73  & \\  
491201.3 &             0.3 &  0.3 & $5_{3,2}-4_{2,3}$ A     &  84.0  &   6.53e-4 &   156 & 145 &   4.2 &  0.64   & \\  
491310.1 &                 &      & $11_{1,10}-10_{2,8}$    &  170.6 &    2.15e-4 &    73 &     &       & 0.34   & weak\\
\hline
\end{tabular}
\begin{list}{}{}
\item$^{{a}}$Notation as in Table \ref{SO2 parameters }.
$^{\mathrm{*}}$Vibrationally excited transition, $v_{t}=1$. 
\end{list}
\end{table*} 

\begin{table*}
\caption{(CH$_3)_2$O parameters.$^a$} 
\label{(CH3)2O parameters  } 
\centering
\begin{tabular}{ l r  r  l r l r r r r  l }
\hline
\hline
$\nu_{{ul}}$ & $\Delta \nu_\mathrm{p}$ &  $\Delta \nu_\mathrm{g}$ &    Transition     & $E_u$  &  $A_{{ul}}$  & $T^{\,*}_{\mathrm{A}}$(peak) & Ampl.  & Width & $\int$ $T^{\,*}_{\mathrm{A}}$  d$\upsilon$  &    Note    \\
$[\mathrm{MHz}]$ &	 [MHz] & [MHz] & $J_{{K_a},{K_c}}$&	[K] &	[s$^{-1}$] & 	[mK] & [mK]  & [\kms]& [K \kms]  &\\ 
\hline  
490795.3 &  $-$4.7 &     & $15_{4,11}-14_{3,12}$ AE  & 131.9 &   3.56e-4 & 194 &   &  & 2.05 & blend 3 groups \\   
 490795.3 &     & & $15_{4,11}-14_{3,12}$ EA  & 131.9&  3.56e-4&&&&&   \\   
 490797.5 &     && $12_{5,8}-11_{4,7}$ EA    & 106.2 &   2.26e-4 &    &&&&   \\   
 490798.3 &  && $15_{4,11}-14_{3,12}$ EE  & 131.9 &   3.56e-4 &      &&&&   \\   
 490801.3 &     & & $15_{4,11}-14_{3,12}$ AA  & 131.9 &   3.56e-4 &     &&&&   \\   
 490804.3 &     & & $12_{5,8}-11_{4,7}$ AE    & 106.2 &   4.84e-4 &     &&&&  \\
\hline
\end{tabular}
\begin{list}{}{}
\item$^{{a}}$Notation as in Table \ref{SO2 parameters }.
\end{list}
\end{table*}

\begin{table*}
\addtocounter{table}{-1}
\caption{\emph{continued} } 
\centering
\begin{tabular}{ l r  r  l r l r r r r   l }
\hline
\hline
$\nu_{{ul}}$ & $\Delta \nu_\mathrm{p}$ &  $\Delta \nu_\mathrm{g}$ &    Transition     & $E_u$  &  $A_{{ul}}$  & $T^{\,*}_{\mathrm{A}}$(peak) & Ampl.  & Width & $\int$ $T^{\,*}_{\mathrm{A}}$  d$\upsilon$  &    Note    \\
$[\mathrm{MHz}]$ &	 [MHz] & [MHz] & $J_{{K_a},{K_c}}$&	[K] &	[s$^{-1}$] & 	[mK] & [mK]  & [\kms] &[K \kms] &\\ 
\hline  
 543753.9 &  $-$5.1 &    & $18_{3,15}-17_{2,16}$ AE  &   169.8 &   1.94e-4  &    45 &    &      & 0.20  & entire group \\   
 543753.9 &  $-$5.1 &     & $18_{3,15}-17_{2,16}$ EA  &  &&&&&&   \\   
 543756.9 &  $-$2.1 &     & $18_{3,15}-17_{2,16}$ EE  &   &&&&&&  \\   
 543759.8 &   0.8 &    & $18_{3,15}-17_{2,16}$ AA  &   &&&&&&   \\   
 546827.8 &  $-$4.2 & $-$3.2 & $15_{5,11}-14_{4,10}$ EA  &  144.4 &   5.50e-4  &   116 & 119 &   3.5 & 0.60  & entire group \\   
 546829.2 &  $-$2.8 & $-$1.8 & $15_{5,11}-14_{4,10}$ AE  &   &&&&& \\
\hline
\end{tabular}
\end{table*} 
\clearpage

\begin{table*}
\addtocounter{table}{-1}
\caption{\emph{continued} } 
\centering
\begin{tabular}{ l r  r  l r l r r r r  l }
\hline
\hline
$\nu_{{ul}}$ & $\Delta \nu_\mathrm{p}$ &  $\Delta \nu_\mathrm{g}$ &    Transition     & $E_u$  &  $A_{{ul}}$  & $T^{\,*}_{\mathrm{A}}$(peak) & Ampl.  & Width & $\int$ $T^{\,*}_{\mathrm{A}}$  d$\upsilon$  &    Note    \\
$[\mathrm{MHz}]$ &	 [MHz] & [MHz] & $J_{{K_a},{K_c}}$&	[K] &	[s$^{-1}$] & 	[mK] & [mK]  & [\kms] &[K \kms] &\\ 
\hline  
 554979.1 &  $-$4.9 &     & $23_{10,13}-23_{9,14}$ AA &  391.7 &   7.23e-4 &    90 &    &      & 0.21   & entire 
 group\\
 554979.1 &  $-$4.9 &     & $23_{10,14}-23_{9,15}$ AA   &&&&&&&\\ 
 554980.4 &  $-$3.6 &     & $23_{10,14}-23_{9,15}$ EE    &&&&&&&  \\
 554981.2 &  $-$2.8 &     & $23_{10,13}-23_{9,14}$ EE   &&&&&&& \\
 554981.7 &  $-$2.3 &     & $23_{10,14}-23_{9,15}$ EA  &&&&&&& \\
 554982.5 &  $-$1.5 &     & $23_{10,13}-23_{9,14}$ AE    &&&&&&& \\
\hline
\end{tabular}
\end{table*}

\begin{table*}
\addtocounter{table}{-1}
\caption{\emph{continued}} 
\centering
\begin{tabular}{ l r  r  l r l r r r r   l }
\hline
\hline
$\nu_{{ul}}$ & $\Delta \nu_\mathrm{p}$ &  $\Delta \nu_\mathrm{g}$ &    Transition     & $E_u$  &  $A_{{ul}}$  & $T^{\,*}_{\mathrm{A}}$(peak) & Ampl.  & Width & $\int$ $T^{\,*}_{\mathrm{A}}$  d$\upsilon$  &    Note    \\
$[\mathrm{MHz}]$ &	 [MHz] & [MHz] & $J_{{K_a},{K_c}}$&	[K] &	[s$^{-1}$] & 	[mK] & [mK]  & [\kms]  &[K \kms] &\\ 
\hline  
 555635.9 &      && $16_{10,6}-16_{9,7}$ AA   &   264.2 &   5.66e-4  &&&&& blend SO$_{2}$\\
 555635.9 &     & & $16_{10,7}-16_{9,8}$ AA     &&&&&&&\\
 555637.9 &      && $16_{10,7}-16_{9,8}$ EE   &&&&&&&\\
 555639.4 &      & & $16_{10,6}-16_{9,7}$ EE   &&&&&&&\\
 555639.8 &      & & $16_{10,7}-16_{9,8}$ EA   &&&&&&&\\
 555641.4 &      & & $16_{10,6}-16_{9,7}$ AE   &&&&&&&\\
\hline
\end{tabular}
\end{table*}

\begin{table*}
\addtocounter{table}{-1}
\caption{\emph{continued} } 
\centering
\begin{tabular}{ l r  r  l r l r r r r   l }
\hline
\hline
$\nu_{{ul}}$ & $\Delta \nu_\mathrm{p}$ &  $\Delta \nu_\mathrm{g}$ &    Transition     & $E_u$  &  $A_{{ul}}$  & $T^{\,*}_{\mathrm{A}}$(peak) & Ampl.  & Width & $\int$ $T^{\,*}_{\mathrm{A}}$  d$\upsilon$  &    Note    \\
$[\mathrm{MHz}]$ &	 [MHz] & [MHz] & $J_{{K_a},{K_c}}$&	[K] &	[s$^{-1}$] & 	[mK] & [mK]  & [\kms] &[K \kms] &\\  
\hline  
 556179.4 &  $-$1.6 & $-$0.6 & $31_{0,31}-30_{1,30}$ EA  &   431.3 &  1.50e-4 &    65 &  60 &   3.5 & 0.17   & entire 
 group\\
 556179.4 &  $-$1.6 & $-$0.6 & $31_{0,31}-30_{1,30}$ AE  & &&&&&&  \\
 556179.4 &  $-$1.6 & $-$0.6 & $31_{0,31}-30_{1,30}$ EE  & &&&&&&  \\
 556179.5 &  $-$1.5 & $-$0.5 & $31_{0,31}-30_{1,30}$ AA  &&&&&&& \\
   556212.0 &    & & $31_{1,31}-30_{0,30}$ EA  &   &&&&&&  (Too) weak\\
   556212.0 &     && $31_{1,31}-30_{0,30}$ AE  &  &&&&&&  (Too) weak\\
\hline
\end{tabular}
\end{table*}

\begin{table*}
\caption{H$_2$CO and isotopologue parameters.$^a$} 
\label{H2CO parameters} 
\centering
\begin{tabular} { l l r  r  l r l r r r r    l }
\hline
\hline
Species & $\nu_{{ul}}$ & $\Delta \nu_\mathrm{p}$ &  $\Delta \nu_\mathrm{g}$ &    Transition     & $E_u$  &  $A_{{ul}}$  & $T^{\,*}_{\mathrm{A}}$(peak) & Ampl.  & Width & $\int$ $T^{\,*}_{\mathrm{A}}$  d$\upsilon$  &    Note    \\
& $[\mathrm{MHz}]$ &	 [MHz] & [MHz] &$J_{{K_a},{K_c}}$&	[K] &	[s$^{-1}$] & 	[mK] & [mK]  & [\kms] &[K \kms] &\\ 
\hline    
H$_2$CO & 491968.4 &  2.4 &  1.4 & $7_{1,7}-6_{1,6}$ & 106.3 &  3.44e-3 &  2040 & 1954 &   5.2 & 11.30   & \\
& 561899.3 &  1.3 &     & $8_{1,8}-7_{1,7}$ &  133.3 &   5.20e-3 &  1705 &      &     & 16.3   & \\
 &   &      &  1.3 &   &    &   &       & 1167 &   4.9 &           & 2G; CR  \\
&   &     & $-$0.7 &   &    &    &       &  504 &    19.0 &          & 2G; LVF \\
 & 576708.3 & $-$0.7 &     & $8_{0,8}-7_{0,7}$ & 125.1 & 5.70e-3 &  1116 &     &      & 11.30  & blend  NS\\
 &   &      &   0.3 &   &    &   &       &  742 &   4.8 &      &     2G; CR  \\
&   &      & $-$6.7 &   &    &  &       &  392 &  18.7 &      &       2G; LVF  \\
H$_2^{13}$CO & 548475.2 & 0.2 & 0.2 & $8_{1,8}-7_{1,7}$ & 130.4 & 4.84e-3 &   155 & 135 &   4.3 & 0.62    & \\
HDCO  &  491937.0 & 2.0    & 3.0    & $8_{1,8}-7_{1,7}$ & 114.4  & 3.48e-3 & 370    & 360 & 6.1   & 2.41 &   blend U-line? \\
&	       552740.9 & $-$0.1 & 0.9 & $9_{1,9}-8_{1,8}$ &  141.0 & 4.99e-3 &   101 &  88 &     3.0 & 0.27   & \\
&	       565857.5 &  2.5 &   & $9_{0,9}-8_{0,8}$ & 137.3 & 5.40e-3 &    60 &   &      & 0.19  & weak \\
\hline
Total no. &7&&&&&&&&&\\
\hline
\end{tabular}
\begin{list}{}{}
\item$^{{a}}$Notation as in Table \ref{SO2 parameters }.
\end{list}
\end{table*}

\begin{table*}
\caption{H$_2$CS parameters.$^a$} 
\label{H2CS parameters} 
\centering
\begin{tabular} { l r  r  l r l r r r r   l }
\hline
\hline
$\nu_{{ul}}$ & $\Delta \nu_\mathrm{p}$ &  $\Delta \nu_\mathrm{g}$ &    Transition     & $E_u$  &  $A_{{ul}}$  & $T^{\,*}_{\mathrm{A}}$(peak) & Ampl.  & Width & $\int$ $T^{\,*}_{\mathrm{A}}$  d$\upsilon$  &    Note    \\
$[\mathrm{MHz}]$ &	 [MHz] & [MHz] &$J_{{K_a},{K_c}}$&	[K] &	[s$^{-1}$] & 	[mK] & [mK]  & [\kms]& [K \kms]   &\\  
\hline    
487663.4 &  0.4 &  0.4 & $14_{1,13}-13_{1,12}$ &   188.8 & 1.76e-3 &   172 & 161 &   3.7 & 0.62  & \\
547308.2 & $-$3.8 & $-$2.8 & $16_{0,16}-15_{0,15}$ & 223.7   &  2.51e-3 &    77 &  63 &     3 & 0.19   & \\
549402.4 &  0.4 &  0.4 & $16_{3,14}-15_{3,13}$ &  342.6 & 2.45e-3 &    62 &  55 &     1.8 & 0.15   & \\
549447.5 & $-$1.5 &     & $16_{3,13}-15_{3,12}$ &  342.6 & 2.45e-3 &    98 &     &      &       &        blend U-line\\
557123.2 & 2.2     &      & $16_{1,15}-15_{1,14}$ & 240.6  & 2.64e-3&       &     &       &       &         blend CH$_{3}$OH;\\
&&&&&&&&&&in H$_2$O wing\\
574140.0 &   $-$2.0 &   $-$2.0 & $17_{1,17}-16_{1,16}$ & 261.4  & 2.90e-3 &    130 &  125 &     2.9 & 0.38  & \\
\hline
Total no. &6&&&&&&&&&\\
\hline
\end{tabular}
\begin{list}{}{}
\item$^{{a}}$Notation as in Table \ref{SO2 parameters }.
\end{list}
\end{table*} 

\begin{table*} [!h]
\caption{CS and $^{13}$CS parameters.$^a$} 
\label{CS  parameters }                                        
\centering
\begin{tabular} {l l r  r  l r l r r r r    l }
\hline
\hline
Species	&$\nu_{{ul}}$ & $\Delta \nu_\mathrm{p}$ &  $\Delta \nu_\mathrm{g}$ &    Transition     & $E_u$  &  $A_{{ul}}$  & $T^{\,*}_{\mathrm{A}}$(peak) & Ampl.  & Width & $\int$ $T^{\,*}_{\mathrm{A}}$  d$\upsilon$  &    Note    \\
& $[\mathrm{MHz}]$ &	 [MHz] & [MHz] & $J $&	[K] &	[s$^{-1}$] & 	[mK] & [mK]  & [\kms]& [K \kms]   &\\ 
\hline 
CS	&	489751.1 & 3.1 &    & $10-9$     & 129.3 & 0.00250 &  2814 &     &      &   23.4   &   \\
	&	&    & $-$0.9 &      & &  &     & 424 &   9.0 &     &     HC, 3G fit  \\
	&	 &    & 2.1 &     && &      &  1730 &  4.0 &     &      ER/CR, 3G fit\\
	&	 &    & 1.1 &     && &      &  633 &  18.0 &     &      LVF, 3G fit\\	
$^{13}$CS	&	554726.0 &   & $-$2.0 & $12-11$ & 173.1 & 0.00366 &    &  88 &   4.8 &  & v blend,\\
&&&&&&&&&&  &2G $^{34}$SO$_{2}$\\
\hline
Total no. &2&&&&&&&&&\\
\hline
\end{tabular}
\begin{list}{}{}
\item$^{{a}}$Notation as in Table \ref{SO2 parameters }.
\end{list}
\end{table*} 

 \begin{table*}
\caption{NO parameters.$^a$} 
\label{NO parameters} 
\centering
\begin{tabular} { l r     l r l r r r r   l }
\hline
\hline
$\nu_{{ul}}$ & $\Delta \nu_\mathrm{p}$  &    Transition     & $E_u$  &  $A_{{ul}}$  & $T^{\,*}_{\mathrm{A}}$(peak) & Ampl.  & Width & $\int$ $T^{\,*}_{\mathrm{A}}$  d$\upsilon$  &    Note    \\
$[\mathrm{MHz}]$ &	 [MHz]   &$ J$($F$)&	[K] &	[s$^{-1}$] & 	[mK] & [mK]  & [\kms]& [K \kms]   &\\ 
\hline    
551187.3 & $-$0.7     & $^{2}\Pi_{1/2}\, 5.5(5.5)-4.5(4.5)$ \,e &  84.2 & 2.16e-5 &   540 &&&&   entire group; \\
  &      &     &   &   &       & 265 &   5.4 &       &       5G from CH$_{3}$CN, \\
  &           &   &    &   &       & 284 &  18.6 &       &          NO \& CH$_{3}$OH\\
551187.5 &        & $^{2}\Pi_{1/2}\,5.5(6.5)-4.5(5.5)$\, e &  84.1 &  2.23e-5 &   &&&&  \\
551188.8 &         & $^{2}\Pi_{1/2}\, 5.5(4.5)-4.5(3.5)$\, e &  84.2 & 2.14e-5 &    &&&&   \\
551531.5 &  0.5     & $^{2}\Pi_{1/2}\, 5.5(6.5)-4.5(5.5)$\, f &  84.3 & 2.24e-5 &   644 &    &      & 9.331  & entire group;\\ &&&&&&&&&  blend CH$_{3}$CN\\
551534.0 &    3.0     & $^{2}\Pi_{1/2}\, 5.5(4.5)-4.5(3.5)$\, f &  84.2  &  2.15e-5 &     &&&&  blend CH$_{3}$CN, NO\\
551534.1 &  3.1     & $^{2}\Pi_{1/2}\, 5.5(5.5)-4.5(4.5)$\, f &  84.3 & 2.16e-5 &    &&&&  blend CH$_{3}$CN, NO\\
567064.2 &   $-$11.8     & $^{2}\Pi_{3/2}\, 5.5(6.5)-4.5(5.5)$\, e  & 231.7 & 2.27e-5 &   122 &  97 &  19.1 & 1.805 &   entire group;\\
567069.6 &         & $^{2}\Pi_{3/2}\, 5.5(5.5)-4.5(4.5)$\, e  & 231.7 &  2.63e-5 &    &&&&  blend H$_{2}$S\\
567073.4 &         & $^{2}\Pi_{3/2}\, 5.5(4.5)-4.5(3.5)$\, e  & 231.7 & 1.82e-5 &     &&&&   blend H$_{2}$S\\
567077.9 &         & $^{2}\Pi_{3/2}\, 5.5(6.5)-4.5(5.5)$\, f  &  231.7 & 2.27e-5 &     &&&&   blend  H$_{2}$S\\
567082.7 &         & $^{2}\Pi_{3/2}\, 5.5(5.5)-4.5(4.5)$ \,f  & 231.7 & 2.19e-5 &     &&&&   blend H$_{2}$S\\
567086.6 &         & $^{2}\Pi_{3/2}\, 5.5(4.5)-4.5(3.5)$\, f  &  231.7 & 2.18e-5 &     &&&&    blend H$_{2}$S\\
\hline
Total no. &12 &&&&&&&&\\
\hline
\end{tabular}
\begin{list}{}{}
\item$^{{a}}$Notation as in Table \ref{SO2 parameters }.
\end{list}
\end{table*}

\begin{table*}
\caption{HNC parameters.$^a$} 
\label{HNC parameters} 
\centering
\begin{tabular} { l r  r  l r l r r r r    l }
\hline
\hline
$\nu_{{ul}}$ & $\Delta \nu_\mathrm{p}$ &  $\Delta \nu_\mathrm{g}$ &    Transition     & $E_u$  &  $A_{{ul}}$  & $T^{\,*}_{\mathrm{A}}$(peak) & Ampl.  & Width & $\int$ $T^{\,*}_{\mathrm{A}}$  d$\upsilon$  &    Note    \\
 $[\mathrm{MHz}]$ &	 [MHz] & [MHz] &$J$&	[K] &	[s$^{-1}$] & 	[mK] & [mK]  & [\kms]& [K \kms]   &\\ 
\hline    
543897.6 & 1.6 &     & $6-5$      & 91.4 & 8.04e-3 &   913 &    &      & 10.70 &   v blend U-line   \\
& &  1.6 & && &      & 600 &   3.8 &     &   ER; from  4G  U-line    \\
&& $-$4.4 & && &      & 171 &  9.3 &     &    HC; from 4G  U-line    \\
& &  $-$1.4 &      & & & &224 &  26.8 &     &    Plateau; from 4G U-line  \\
\hline
Total no. &1&&&&&&&&&\\
\hline
\end{tabular}
\begin{list}{}{}
\item$^{{a}}$Notation as in Table \ref{SO2 parameters }.
\end{list}
\end{table*} 

 \begin{table*}
\caption{CN parameters.$^a$} 
\label{CN parameters} 
\centering
\begin{tabular} { l r  r  l r l r r r r r  l }
\hline
\hline
$\nu_{{ul}}$ & $\Delta \nu_\mathrm{p}$ &  $\Delta \nu_\mathrm{g}$ &    Transition     & $E_u$  &  $A_{{ul}}$  & $T^{\,*}_{\mathrm{A}}$(peak) & Ampl.  & Width & $\int$ $T^{\,*}_{\mathrm{A}}$  d$\upsilon$  &    Note    \\
$[\mathrm{MHz}]$ &	 [MHz] & [MHz] & $N$\,($J$, $F$)&	[K] &	[s$^{-1}$] & 	[mK] & [mK]  & [\kms]& [K \kms]   &\\  
\hline    
566729.9 & 1.9 & 0.9 & 5(4.5, 5.5)\,--\,4(3.5, 4.5) & 54.4 &  1.98e-3 &   690 &  &    & 3.345  & entire group;\\ 
&& &&&&&570 & 3.5&& 2G; PDR/ER\\
&&&&&&&122& 10.0&&2G; HC \\
566730.7 &    &    &  5(4.5, 3.5)\,--\,4(3.5, 2.5)  & 54.4 & 1.86e-3 &    &&&&   blend CN \\
566730.8 &   &    &  5(4.5, 4.5)\,--\,4(3.5, 3.5)  & 54.4 & 1.88e-3 &     &&&&   blend CN \\

566946.8 & 1.8 &   &  5(5.5, 5.5)\,--\,4(4.5, 4.5)  & 54.4 &  1.96e-3 &   965 &   &     & 5.15 &   entire group;\\
&&&&&&&&&&  blend CN \\
  &    & 1.8 &  5(5.5, 5.5)\,--\,4(4.5, 4.5)  & 54.4 &  1.96e-3 &      & 736 &   3.8 &           & 3G  CN; PDR/ER \\
 &    & 0.8 &  5(5.5, 5.5)\,--\,4(4.5, 4.5)  & 54.4 &  1.96e-3 &      & 214 &  10.8 &            & v blends; HC \\
566946.9 &  &   &  5(5.5, 6.5)\,--\,4(4.5, 5.5)  & 54.4 & 2.03e-3 &     &&&&   blend CN \\
566947.2 &   &  &  5(5.5, 4.5)\,--\,4(4.5, 3.5)  & 54.4 & 1.95e-3 &    &&&& blend CN \\
566962.0 &   1.0 & 0.0   &  5(5.5,4.5)\,--\,4(4.5,4.5)        & 54.4 &    8.00e-5 &      & 100 &     2 &            & entire group;\\
&&&&&&&&&&  3G  CN \\
566963.7 &    &    & 5(5.5,5.5)\,--\,4(4.5,5.5)        & 54.4 &  6.7e-5 &     &&&&  \\
\hline
Total no. &8&&&&&&&&&\\
\hline
\end{tabular}
\begin{list}{}{}
\item$^{{a}}$Notation as in Table \ref{SO2 parameters }.
\end{list}
\end{table*}

\begin{table*}
\caption{N$_2$H$^+$  parameters.$^a$} 
\label{N2H+ parameters} 
\centering
\begin{tabular} { l r  r  l r l r r r r  l }
\hline
\hline
$\nu_{{ul}}$ & $\Delta \nu_\mathrm{p}$ &  $\Delta \nu_\mathrm{g}$ &    Transition     & $E_u$  &  $A_{{ul}}$  & $T^{\,*}_{\mathrm{A}}$(peak) & Ampl.  & Width & $\int$ $T^{\,*}_{\mathrm{A}}$  d$\upsilon$  &    Note    \\
$[\mathrm{MHz}]$ &	 [MHz] & [MHz] & $J$&	[K] &	[s$^{-1}$] & 	[mK] & [mK]  & [\kms]& [K \kms]   &\\  
\hline     
558966.6 & 4.6 & 1.6 & $6-5$      & 93.9 & 1.08e-2 &   315 & 300 &   4.8 & 1.46   & \\
\hline
Total no. &1&&&&&&&&&\\
\hline
\end{tabular}
\begin{list}{}{}
\item$^{{a}}$Notation as in Table \ref{SO2 parameters }.
\end{list}
\end{table*}

\begin{table*}
\caption{HCS$^+$ parameters.$^a$} 
\label{HCS+ parameters} 
\centering
\begin{tabular} { l r  r  l r l r r r r    l }
\hline
\hline
$\nu_{{ul}}$ & $\Delta \nu_\mathrm{p}$ &  $\Delta \nu_\mathrm{g}$ &    Transition     & $E_u$  &  $A_{{ul}}$  & $T^{\,*}_{\mathrm{A}}$(peak) & Ampl.  & Width & $\int$ $T^{\,*}_{\mathrm{A}}$  d$\upsilon$  &    Note    \\
$[\mathrm{MHz}]$ &	 [MHz] & [MHz] & $J$&	[K] &	[s$^{-1}$] & 	[mK] & [mK]  & [\kms]& [K \kms]   &\\ 
\hline     
554576.6 &   &  & $13-12$    & 186.3 & 3.66e-3 &      &  57  & 3.4     &     &      v blend $^{33}$SO \\
&&&&&&&&&&  2G  $^{33}$SO\\
\hline
Total no. &1&&&&&&&&&\\
\hline
\end{tabular}
\begin{list}{}{}
\item$^{{a}}$Notation as in Table \ref{SO2 parameters }.
\end{list}
\end{table*}

\begin{table*}
\caption{NS parameters.$^a$} 
\label{NS parameters} 
\centering
\begin{tabular} { l r  r  l r l r r r r    l }
\hline
\hline
$\nu_{{ul}}$ & $\Delta \nu_\mathrm{p}$ &  $\Delta \nu_\mathrm{g}$ &    Transition     & $E_u$  &  $A_{{ul}}$  & $T^{\,*}_{\mathrm{A}}$(peak) & Ampl.  & Width & $\int$ $T^{\,*}_{\mathrm{A}}$  d$\upsilon$  &    Note    \\
$[\mathrm{MHz}]$ &	 [MHz] & [MHz] & $J$&	[K] &	[s$^{-1}$] & 	[mK] & [mK]  & [\kms]  &[K \kms] &\\ 
\hline     
576720.2 &   &  &  $^2\Pi_{1/2}$ $12.5-11.5$   & 187.36 &  &      &    &      &     &     v blend H$_2$CO \\
\hline
Total no. &1&&&&&&&&&\\
\hline
\end{tabular}
\begin{list}{}{}
\item$^{{a}}$Notation as in Table \ref{SO2 parameters }.
\end{list}
\end{table*}

\begin{table*}  
\caption{CO  and C parameters.$^a$} 
\label{CO parameters } 
\centering
\begin{tabular}  {l l r  r  l r l r r r r    l }
\hline
\hline
Species & $\nu_{{ul}}$ & $\Delta \nu_\mathrm{p}$ &  $\Delta \nu_\mathrm{g}$ &    Transition     & $E_u$  &  $A_{{ul}}$  & $T^{\,*}_{\mathrm{A}}$(peak) & Ampl.  & Width & $\int$ $T^{\,*}_{\mathrm{A}}$  d$\upsilon$  &    Note    \\
&$[\mathrm{MHz}]$ &	 [MHz] & [MHz] & $J $&	[K] &	[s$^{-1}$] & 	[mK] & [mK]  & [\kms]& [K \kms]  &\\ 
\hline 
CO&576267.9 & $-$0.1 &      & $5-4$      & 83.0 & 1.22e-5 & 62200 &      &      & 785.50   & \\
 $^{13}$CO  &  550926.3 & 0.3 &     & $5-4$        & 79.3 & 1.10e-5 & 24300 &     &      & 174.30   &  \\
&&&  0.3 & &&&      & 21600 &   4.8 &    &  3G; N    \\
&& & $-$0.7 & &&& &  2180 &    18.0 &     &  3G; LVF   \\
&& & $-$0.7 & &&& &  470 &    48.7 &     &  3G;HVF   \\

C$^{17}$O  & 561712.8 & 1.8 & 1.8 & $5-4$      & 80.9 & 1.14e-5 &  1169 &  &    & 6.03   & blend \\
&&&&&&&&&&&  $^{34}$SO, SO$_{2}$\\
&&&  1.8 & &&&      & 1070 &   4.2 &    &  3G; N    \\
&& & 3.8 & &&& &  116 &    18.0 &     &  3G; LVF   \\

C$^{18}$O  & 548831.0 &  2.0 &   & $5-4$      & 79.0 & 1.06e-5 &  4219 &     &     & 24.30  &  \\
&&&  1.0 & &&&      & 3810 &   4.2 &    &  3G; N    \\
&& & 0.0 & &&& &  367 &    18.0 &     &  3G; LVF   \\

C  & 492160.7 & &   & $1-0$ & 23.6 &7.99e-8 & 7800     &7794  & 4.5  & 38.7       & \\
\hline
Total no. &5&&&&&&&&&\\
\hline
\end{tabular}
\begin{list}{}{}
\item$^{{a}}$Notation as in Table \ref{SO2 parameters }.
\end{list}
\end{table*}

\begin{table*}
\caption{Water parameters.$^a$ } 
\label{water parameters} 
\centering
\begin{tabular}{l l r  r  l r l r r r r    l }
\hline
\hline
Species &$\nu_{{ul}}$ & $\Delta \nu_\mathrm{p}$ &  $\Delta \nu_\mathrm{g}$ &    Transition     & $E_u$  &  $A_{{ul}}$  & $T^{\,*}_{\mathrm{A}}$(peak) & Ampl.  & Width & $\int$ $T^{\,*}_{\mathrm{A}}$  d$\upsilon$  &    Note    \\
&$[\mathrm{MHz}]$ &	 [MHz] & [MHz] & $J_{{K_a},{K_c}}$&	[K] &	[s$^{-1}$] & 	[mK] & [mK]  & [\kms]  &[K \kms] &\\ 
\hline  
$o-$H$_2$O 	& 	556936.0 & $-$7.0 &    & $1_{1,0}-1_{0,1}$ & 61.0 & 3.46e-3 &  7600 &    &      & 320.30   & \\
$p-$H$_2$O & 488491.1 & $-$0.9 & $-$5.9 & $6_{2,4}-7_{1,7}$ & 867.3 & 1.22e-5 &   196 & 181 &    12.0 & 2.24   & \\

$o-$H$_2^{17}$O  & 552021.0 &  2.0 &   & $1_{1,0}-1_{0,1}$ & 60.7 & 3.37e-3 &   464 &    &    & 9.44  & blend  SO$_{2}$ \\
&&&&&&&&&&&  \&CH$_{3}$OH  \\ 
&    &&  4.0 &  &   &   &      & 76&   5.0 &      &      3G; CR\\
&   &  &  1.0 &    &  &    &      & 214 &  18.0 &      &  3G; LVF \\
\hline
\end{tabular}
\begin{list}{}{}
\item$^{{a}}$Notation as in Table \ref{SO2 parameters }.
\end{list}
\end{table*}

\begin{table*}
\caption{U-line parameters.$^a$} 
\label{U-line parameters} 
\centering
\begin{tabular} { l r r   r r     l }
\hline
\hline
$\nu_{{ul}}$  &   $T^{\,*}_{\mathrm{A}}$(peak)  & Ampl.  & Width & $\int$ $T^{\,*}_{\mathrm{A}}$  d$\upsilon$   &   Note    \\
$[\mathrm{MHz}]$ &	  	[mK]  &[mK]   & [\kms]&  [K \kms] & \\ 
\hline  
 487209 &    91 &     &       & 0.673 &   SO$^{+}$ ?\\
 488598 &    94 &  73 &   3.4 & 0.256 &  CH$_{3}$OCHO?\\
 488633 &    83 &  76 &   4.3 & 0.313 &   CH$_{3}$CHO?\\
 491496 &   104 &     &       & 1.438 &  CH$_{3}$OCHO?\\
 491892 &   157 &     &       &  0.64 &   tripple peaked\\
 541981 &   109 &  87 &   4.5 &  0.38 &   CH$_{3}$OCHO?\\
\hline
\end{tabular}
\begin{list}{}{}
\item$^{{a}}$Notation as in Table \ref{SO2 parameters }.
\end{list}
\end{table*}

\clearpage
\begin{table*}
\caption{T-line parameters.$^a$} 
\label{T-line parameters} 
\centering
\begin{tabular} { l r r   r r     l }
\hline
\hline
$\nu_{{ul}}$  &   $T^{\,*}_{\mathrm{A}}$(peak)  & Ampl.  & Width & $\int$ $T^{\,*}_{\mathrm{A}}$  d$\upsilon$   &   Note    \\\
 $[\mathrm{MHz}]$ &	  	[mK]  &[mK]   & [\kms] &[K \kms]&\\ 
\hline    
 486845 &    80 &   83  &  3.3     & 0.275   & SO$^{+}$ ?\\
 487507 &    70 &   77  &   2.7    & 0.213   & CH$_{3}$CHO?\\
 488477 &    49 &  47 &   3.2 & 0.147 &   \\
 489193 &   107 & 114    &   1.7    & 0.412   & \\
 489709 &       &     &       &            & SiS?; v blend CS\\
 541926 &   104 &  93 &   2.7 & 0.292 &  \\
\hline
\end{tabular}
\begin{list}{}{}
\item$^{{a}}$Notation as in Table \ref{SO2 parameters }. 
\end{list}
\end{table*}

\clearpage


\begin{thebibliography}{}



\bibitem[2005]{Araya}     
Araya, E., Hofner, P., Kurtz, S., Bronfman, L., and DeDeo, S.  2005,
     ApJS, 157, 279
 
\bibitem[1983]{Batrla83}      
Batrla, W., Wilson, T.L., Bastien, P., and Ruf, K. 1983,
    A\&A, 128, 279      

\bibitem[2001]{Bensch}      
Bensch, F., Pak, I., Wouterloot, J.G.A., Klapper, G., and Winnewisser G. 2001,
    ApJ, 562, L185     
 
\bibitem[2005]{Beuther}  
Beuther, H., Zhang, Q., Greenhill, L.J.,  \etal~2005,
     ApJ, 632, 355

\bibitem[1969]{Bevington}  
Bevington, P.R., 1969, Data Reduction and Error Analysis for the Physicas Sciences, McGraw-Hill book company

\bibitem[1986]{Blake etal 86}
Blake, G.A., Sutton, E.C., Masson, C.R., and Phillips, T.G. 1986,
   ApJS, 60, 357

\bibitem[1987]{Blake}  
Blake, G.A., Sutton, E.C., Masson, C.R.,  and Phillips, T.G.  1987,
     ApJ, 315, 621 \mbox{{\bf B87}}

\bibitem[1988]{Brown88}
Brown, P.D., Charnley, S.B., and Millar, T.J. 1988,
    MNRAS, 231, 409     
     
\bibitem[1993]{Caselli93}
Caselli, P., Hasegawa, T.I., and Herbst, E. 1993,
    ApJ, 408, 548     
     
\bibitem[2006]{Cernicharo} 
Cernicharo, J., Goicoechea, J.R., Daniel, F.,  \etal~2006,
     ApJ, 649, 33   

\bibitem[2007]{Chang Cuppen and Herbst 07}
Chang, Q., Cuppen, H.M., and Herbst, E. 2007,
    A\&A, 469, 973


\bibitem[1968]{Cheung68}
Cheung, A.C., Rank, D.M., Townes, C.H., Thornton, D.D., and Welch, W.J. 1968,
  \prl, 21, 1701

 
\bibitem[1996]{Chin}  
Chin, Y.-N., Henkel, C., Whiteoak, J.B., Langer, N., and Churchwell, E.B. 1996,
   A\&A, 305, 960    
 
\bibitem[2005]{Comito} 
Comito, C., Schilke, P., Phillips, T., G.,  \etal~2005,
     ApJS, 156, 127, {\bf C05 }     

     
\bibitem[1984]{Friberg84} 
Friberg, P. 1984,
   A\&A, 132, 265

\bibitem[2006]{Garrod and Herbst06}
Garrod, R.T., and Herbst, E. 2006,
    A\&A, 457, 927

\bibitem[2007]{Garrod Wakelam and Herbst 07}
Garrod, R.T., Wakelam, V., and Herbst, E. 2007,
   A\&A, 467, 1103

   
\bibitem[1996]{Gensheimer} 
Gensheimer, P.D., Mauersberger, R., and Wilson, T.L. 1996,
   A\&A, 314, 281
     
\bibitem[1981]{Genzel81} 
Genzel, R., Reid, M.J., Moran, J.M., and Downes, D. 1981, 
     ApJ, 244, 884    
   
\bibitem[1982] {Genzel82} 
Genzel, R., Downes, D., Ho, P.T.P., and Bieging, J. 1982,
    ApJ, 259, L103        
           
\bibitem[1989]{Genzel and Stutzki} 
Genzel, R. \& Stutzki, J., 1989,
    ARA\&A, 27, 41        

\bibitem[2005]{Geppert} 
Geppert, W., Hellberg, F., \"Osterdahl, \etal~2005,
      IAU Symposium  231, Eds. Lis, D.C., Blake, G.A., and Herbst, E., Cambridge University Press, 117 

\bibitem[2000]{Gibb00} 
Gibb, E.L., Whittet, D.C.B., Schutte, W.A.,  \etal~2000,
    ApJ, 536, 347  
    
\bibitem[2004]{Gibb04} 
Gibb, E.L, Whittet, D.C.B., Boogert, A.C.A., and Tielens, A.G.G.M. 2004,
   ApJSS, 151, 35              

\bibitem[2006]{Goicoechea}
Goicoechea, J.R., Cernicharo, J., Lerate, M.R., \etal~2006, 
    ApJ, 641, L49   
            
\bibitem[1997]{Goldsmith97} 
Goldsmith, P.F., Bergin, E.A.,  and Lis, D.C. 1997,
     ApJ, 491, 615    
    
\bibitem[1999]{Goldsmith Langer} 
Goldsmith, P.F. \& Langer, W.D. 1999,
      ApJ, 517, 209      

\bibitem[1998]{Greenhill98}  
Greenhill, L.J., Gwinn, C.R., Schwartz, C., Moran, J.M., and  Diamond, P.J. 1998,
    \nat, 396, 650      
 
\bibitem[1996]{Grevesse}  
Grevesse, N., Noels, A., Sauval, A.J. 1996,
    ASP Conf. Ser., 99, 117    

\bibitem[1998]{Groner98}
 Groner, P.,  Albert, S.,  Herbst, E., and de Lucia, F.C.  1998,
 \apj, 500, 1059 


\bibitem[1985]{Hermsen85}   
Hermsen, W., Wilson, T.L., Walmsley, C.M., and Batrla, W. 1985,
      A\&A, 146, 134


\bibitem[1988a]{Hermsen}   
Hermsen, W., Wilson, T.L., and Bieging, J.H. 1988a,
      A\&A, 201, 276      

\bibitem[1988b]{Hermsen88b}   
Hermsen, W., Wilson, T.L., Walmsley, C.M., and Henkel C. 1988b,
      A\&A, 201, 285 


\bibitem[2005]{Hjalmarson05} 
Hjalmarson, \AA.,  Bergman, P., Biver, N., \etal~2005,    
        Adv. in Space Res., 36, 1031  
	
\bibitem[1983]{Ho and Townes}  
Ho, P.T.P. \& Townes, C.H. 1983
    ARA\&A, 21, 239      

\bibitem[1983]{Hollis 83}
Hollis, J.M., Lovas, F.J.,  Suenram, R.D., Jewell, P.R., and Snyder, L.E. 1983,
   ApJ, 264, 543

\bibitem[1999]{Ikeda} 
Ikeda, M., Maezawa, H., Ito, T., \etal~1999,
    ApJ, 527, L59   

\bibitem[1985]{Irvine85}   
Irvine, W.M., Schloerb, F.P., Hjalmarson, \AA., and Herbst, E. 1985,
in Protostars and Planets II, Eds. Black, D.C. and Matthews, M.S.,
University of Arizona Press, 579

\bibitem[1987]{Irvine87} 
 Irvine, W.M.,  Goldsmith, P.F., Hjalmarson, \AA. 1987,  
in Interstellar Processes,
Eds. Hollenbach, D.J. and Thronson, Jr., H.A., Reidel Publishing Co, 561
                        
\bibitem[1984]{Johansson84}     
Johansson, L.E.B., Andersson, C., Elld\'er, J.,  \etal~1984,
      A\&A, 130, 227     
 
\bibitem[1988]{Kahane} 
Kahane, C., Gomez-Gonzalez, J., Cernicharo, J., and Gu\'elin, M. 1988,
   A\&A, 1988, 190, 167      
 
\bibitem[1983]{Keene}  
Keene, J., Blake, G.A., and Phillips, T.G. 1983,
     ApJ, 271, L27      
     
\bibitem[1976]{Lampton} 
Lampton, M., Margon, B., and Bowyer, S. 1976,
    ApJ, 208, 177    

\bibitem[2003]{Larsson}  
Larsson, B., Liseau, R., Bergman, P.,  \etal~2003,
    A\&A, 402, L69      
 
\bibitem[2006] {Lerate} 
Lerate, M.R., Barlow, M.J., Swinyard, B.M.,  \etal~2006,
    MNRAS, 370, 597    
          
\bibitem[2003]{Liseau} 
Liseau, R., Larsson, B., Brandeker, A.,  \etal~2003,
    A\&A, 402, L73      
 
\bibitem[2003]{Lovas}  
Lovas, F.J. 2003, ''Spectral Line Atlas for Interstellar Molecules (SLAIM$\emptyset$3) Ver.1'',
    private communication of a CD


\bibitem[1990]{Mangum90}
Mangum, J.G., Wootten, Al, Loren, R.B., and Wadiak, E.J. 1990,
   \apj, 348, 542
 
\bibitem[1987]{Masson}    
Masson, C.R., Lo, K.Y., Phillips, T.G., \etal~1987,
     ApJ, 319, 446      
 
\bibitem[2000]{Melnick}  
Melnick, G.J., Stauffer, J.R., Ashby, M.L.N., \etal~2000,  
    ApJ, 539, L77   
 
\bibitem[1986]{Menten86}    
Menten, K.M., Walmsley, C.M., Henkel, C.,  \etal~ 1986,
      A\&A, 169, 271

\bibitem[1988]{Menten}    
Menten, K.M., Walmsley, C.M., Henkel, C.,  and  Wilson, T.L. 1988,
      A\&A, 198, 253
 
\bibitem[1989]{Migenes89}
Migenes, V., Johnston, K.J., Pauls, T.A., and Wilson, T.L. 1989,
ApJ, 347, 294


\bibitem[2005]{Millar}
Millar, T.J. 2005, IAU Symposium 231, Eds. Lis, D.C., Blake, G.A., and Herbst, E.,  
Cambridge University Press, 77

\bibitem[1984]{Mitchell84}
Mitchell, G.F. 1984,
   ApJ, 287, 665
      
\bibitem[2001]{Muller} 
M\"uller, H.S.P., Thorwirth, S., Roth, D.A., and Winnewisser, G. 2001,
     A\&A, 370, L49      
 
\bibitem[1979]{Nagai} 
Nagai, T., Kaifu, N., Nagane, K., and Akaba, K. 1979,
     PASJ, 31, 317      
 
\bibitem[1995]{Neufeld95}  
Neufeld, D.A., Lepp, S., and Melnick, G.J. 1995,
ApJS, 100, 132
 
\bibitem[2006]{Neufeld06}  
Neufeld, D.A., Green, J.D., Hollenbach,  \etal~2006,
    ApJ, 647, L33
 
\bibitem[2003]{Nordh}  
Nordh, H.L, von Sch\'eele, F., Frisk, U.,  \etal~2003
A\&A, 402, L21   
               
\bibitem[1998]{Nummelin98}  
Nummelin, A., Dickens, J.E., Bergman, P.,  \etal~1998, 
     A\&A, 337, 275 
 
\bibitem[2000]{Nummelin2000}  
Nummelin, A.,  Bergman, P., Hjalmarson, {\AA},  \etal~2000, 
     ApJS, 128, 213     
 
\bibitem[1981]{HansOlofsson81}  
Olofsson, H., Hjalmarson, \AA., and Rydbeck, O.E.H. 1981,
    A\&A, 100, L30      
 
\bibitem[1982]{HansOlofsson82}  
Olofsson, H., Elld\'er, J., Hjalmarson, \AA., and Rydbeck, O.E.H. 1982,
    A\&A, 113, L18      

\bibitem[1984]{HansOlofsson84}
Olofsson, H.   1984,
A\&A, 134, 36O 
     
\bibitem[2003a]{Olofsson03}   
Olofsson, A.O.H., Olofsson, G., Hjalmarson, \AA.,   \etal~2003a,
     A\&A, 402, L47     
 
\bibitem[2003b]{Olofsson thesis} 
Olofsson, A.O.H.  2003b,   Thesis, Chalmers University of Techonology, ISBN 91-7291-341-X   
              
\bibitem[2006]{Olofsson}   
Olofsson, A.O.H., Persson, C.M., Koning, N., \etal~2006,
      A\&A \mbox{{\bf Paper I}}

     
 \bibitem[2001]{Pardo}   
 Pardo, J., R., Cernicharo, J., Herpin, F., \etal~2001,
    ApJ, 562, 799    

\bibitem[1983]{Pauls 1983}
Pauls, T.A., Wilson, T.L., Bieging, J.H., and Martin, R.N. 1983,
    A\&A, 124, 23
 
\bibitem[1981a]{Penzias81a}    
Penzias, A.A. 1981a,
   ApJ, 249, 513     
    
\bibitem[1981b]{Penzias81b}  
Penzias, A.A. 1981b,
   ApJ, 249, 518      
 
\bibitem[1979]{Phillips79}  
Phillips, T.G., Wannier, P.G., Scoville, N.Z., and Huggins, P.J. 1979,
    ApJ, 231, 720

\bibitem[1998]{Pickett}  
Pickett, H.M., Poynter, R.L., Cohen, E.A., \etal~1998,
J.Quant.Spectrosc. \& Rad.Transfer, 60, 883    

\bibitem[1993]{Pineau des Foretes93}  
Pineau des For\^ets, G., Roueff, E., Schilke, P., and Flower, D.R. 1993,
     MNRAS, 262, 915   

 
\bibitem[2000]{Plume}  
Plume, R., Bensch, F., Howe, J.E., \etal~2000,
   ApJ, 539, L133
 
\bibitem[1998]{Rodriguez-Franco}  
Rodr\'iguez-Franco, A., Mart\'in-Pintado, J., and Fuente, A.  1998,
    A\&A, 329, 1097
 
\bibitem[2001]{Rodriguez-Franco01}  
Rodr\'iguez-Franco, A., Wilson, T.L., Mart\'in-Pintado, J., and Fuente, A.  2001,
    ApJ, 559, 985    
 
\bibitem[1999]{Salas}  
Salas, L., Rosando., M., Cruz-Gonz\'alez, I., \etal~1999, 
         ApJ, 511, 822

\bibitem[2002]{Savage}     
Savage, C., Apponi, A.J., Ziurys, L.M., and Wyckoff, S. 2002,
    ApJ, 578, 211             
 

\bibitem[2005]{Schoier}  
Sch\"oier, F.L., van der Tak, F.F.S., van Dishoeck E.F., and Black, J.H.  2005,
    A\&A, 432, 369 
 
\bibitem[1992]{Schilke92}
Schilke, P., G\"usten, R., Schulz, A., Serabyn, E., and Walmsley, C.M. 1992,
A\&A, 261, L5


\bibitem[2001]{S01}  
Schilke, P., Benford, D.J., Hunter, T.R., Lis, D.C., and Phillips, T.G.  2001,
     ApJS, 132, 281, {\bf S01  }   
 
\bibitem[1995]{Serabyn and Weisstein}
Serabyn, E., and Weisstein, E. 1995,
   ApJ, 451, 238

\bibitem[2004]{Stantcheva and Herbst 04} 
Stantcheva, T., and Herbst, E. 2004,
   A\&A, 423, 241

\bibitem[1986]{Sutton 86}
Sutton, E.C., Blake, G.A., Genzel, R., Masson, C.R., and Phillips, T.G. 1986,
    ApJ, 311, 921 

\bibitem[1995]{Sutton}   
Sutton, E.C., Peng, R., Danchi, W.C.,  \etal~1995,
     ApJS, 97, 455, \textbf{S95}
 
\bibitem[1995]{Tauber}    
Tauber, J.A., Lis, D.C., Keene, J., Schilke, P., and B\"uttgenbach, T.H.  1995,
    A\&A, 297, 567


\bibitem[1990]{Turner90}
Turner, B.E. 1990,
\apj, 362, L29



\bibitem[1991]{Turner 91}
Turner, B.E. 1991,
  ApJS, 76, 617

     
\bibitem[1997]{Ungerechts97}
Ungerechts, H., Bergin, E.~A., Goldsmith, \etal~1997, 
\apj, 482, 245

\bibitem[2006]{Wada Mochizuki and Hiraoka  06}
Wada, A., Mochizuki, N., and Hiraoka, K. 2006,
   ApJ, 644, 300


\bibitem[1985]{Watson}   
Watson, D.M., Genzel, R., Townes, C.H., and Storey, J.W.V. 1985,
   ApJ, 298, 316    
     
\bibitem[2003]{White}                       
White, G.J., Araki, A., Greaves, J.S.,  Ohishi, M., and Higginbottom, N.S. 2003,   
    A\&A, 407, 589, {\bf W03 }

\bibitem[1994]{Wilner 94}
Wilner, D.J., Wright, M.C.H., and Plambeck, R.L. 1994,
   ApJ, 422, 642

\bibitem[1979]{Wilson 1979}
Wilson, T.L., Downes, D, and Bieging, J. 1979,
    A\&A, 71, 275


\bibitem[1992] {Wilson  Matteucci }     
Wilson, T.L., \& Matteucci, F. 1992,  
    A\&ARv., 4, 1     
 
\bibitem[1986] {Wilson86}  
Wilson, T.L., Serabyn, E., Henkel, C., and Walmsley, C.M. 1986,
    A\&A, 158, L1  
 
\bibitem[1994] {Wilson and Rood}   
Wilson, T.L. \& Rood, R. 1994,
   ARA\&A, 32, 191       

\bibitem[2000]{Wilson}  
Wilson, T.L., Gaume, R.A., Gensheimer, P., and Johnston, K.J. 2000,
    ApJ, 538, 665    

\bibitem[2006]{Wirstrom}   
Wirstr\"om, E.S., Bergman, P.,  Olofsson, A.O.H., \etal~2006,
    A\&A, 453, 979, {\bf  W06 }

\bibitem[1990]{Womack90} 
Womack, M., Ziurys, L.~M., Wyckoff, S., and Sage, L. 1990,
\baas, 22, 1329    
    
\bibitem[2005]{Wouterloot}  
Wouterloot, J.G.A., Brand, J., and Henkel C. 2005,
    A\&A, 430, 549
 
\bibitem[1996]{Wright}  
Wright, M.C.H., Plambeck, R.L., Wilner, D.J. 1996,
    ApJ, 469, 216             
 
\bibitem[2000]{Wright 2000}  
Wright, C.M.,  van Dishoeck, E.F., Black, J.H., \etal~2000,
   A\&A, 358, 689

\end{thebibliography}
\end{document}